\pdfoutput=1
\documentclass[a4paper,11pt]{article}
\PassOptionsToPackage{hyperfootnotes=false}{hyperref}
\PassOptionsToPackage{dvipsnames}{xcolor}
\usepackage{jinstpub}
\usepackage{lineno}
\usepackage[capitalise]{cleveref}
\usepackage{booktabs}
\usepackage{tabularx}
\usepackage{multirow}
\usepackage{capt-of}
\usepackage{placeins}
\usepackage{float}
\usepackage{siunitx}
\usepackage{pdfpages}

\usepackage{xcolor}
\usepackage{orcidlink}
\usepackage[normalem]{ulem}

\makeatletter
\gdef\@fpheader{Published in \href{https://doi.org/10.1088/1748-0221/21/10/P10041}{JINST 21 {\upshape(2026)} P10041}}
\makeatother


\title{\boldmath Transferable Fast Calorimeter Shower Generation \\ via Multi-Geometry Pre-training}

\author[a, b, d]{Thorsten~Buss\,\orcidlink{0000-0002-1717-2138},}

\author[b]{Henry~Day-Hall\,\orcidlink{0000-0002-7881-2506},}

\author[b]{Frank~Gaede\,\orcidlink{0000-0002-7055-9200},}

\author[a]{Gregor~Kasieczka\,\orcidlink{0000-0003-3457-2755},}

\author[b]{Katja~Kr\"uger\,\orcidlink{0000-0002-1956-6608},}

\author[c]{Peter~McKeown\,\orcidlink{0009-0006-9722-2233},}

\author[a,1]{Lorenzo~Valente\,\orcidlink{0009-0007-0080-8738}\note{Corresponding author. Authorship is alphabetical.}}
\emailAdd{lorenzo.valente@uni-hamburg.de}

\affiliation[a]{Institut für Experimentalphysik, Universität Hamburg,\\
Luruper Chaussee 149, 22607 Hamburg, Germany}
\affiliation[b]{Deutsches Elektronen-Synchrotron DESY,\\
Notkestr. 85, 22607 Hamburg, Germany}
\affiliation[c]{CERN, 1211 Geneva 23, Switzerland}
\affiliation[d]{Institute for Theoretical Particle Physics and Cosmology, RWTH Aachen University,\\
Sommerfeldstraße 16, 52056 Aachen, Germany}

\abstract{
Detailed \texttt{Geant4} simulation of calorimeter showers dominates the computing budget of high-energy physics experiments. 
Deep generative surrogates reduce this cost,
but they have remained tied to the detector they were trained on, so each new geometry needs a large in-domain dataset.
We study whether a single point cloud shower generator can be pre-trained on multiple detectors and transferred to unseen calorimeters. The pre-training geometries come from \emph{synthetic} geometric variation rather than real-detector data.
We introduce \textsc{SimpleBox}, a family of $10^{4}$ box calorimeters spanning the plane of sampling fraction and longitudinal segmentation, and benchmark it against pre-training on realistic detectors.
On a calorimeter unseen in pre-training, with $10^{3}$ target showers for fine-tuning, the two priors reduce the aggregated sliced Wasserstein distance to \texttt{Geant4} by factors of $5.2$ (synthetic) and $8.0$ (realistic) relative to training from scratch. At larger target sizes the synthetic prior performs better than the realistic one. Geometric diversity alone is therefore a practical way to pre-train a transferable shower generator.
}


\arxivnumber{2608.18233}

\begin{document}
\maketitle
\raggedbottom

\newpage
\section{Introduction}
\label{sec:intro}

The upcoming High-Luminosity Large Hadron Collider (HL-LHC) requires roughly an order of magnitude more simulated events than current generation experiments produce~\cite{HEPSoftwareFoundation:2017ggl}.
Producing them is the largest consumer of computing resources~\cite{ATLAS:2021pzo}, and the detailed \texttt{Geant4}~\cite{GEANT4:2002zbu,Allison:2006ve,Allison:2016lfl} modelling of calorimeter showers dominates that cost, so fast surrogate models are needed to fit within the available computing budget.
Deep generative models are now the leading surrogates of this kind, and existing approaches include generative adversarial networks (GANs)~\cite{Paganini:2017hrr,Paganini:2017dwg,deOliveira:2017rwa,Erdmann:2018kuh,Erdmann:2018jxd,Musella:2018rdi,Belayneh:2019vyx,Buhmann:2020pmy,Butter:2020qhk,ATLAS:2021pzo,FaucciGiannelli:2023fow,Simsek:2024zhj}, variational autoencoders (VAEs)~\cite{Buhmann:2021caf,ATLAS:2022jhk,Cresswell:2022tof,Hoque:2023zjt,Diefenbacher:2023prl,Liu:2024kvv}, autoregressive transformers~\cite{Birk:2025wai,Birk:2026udp,Granger:2026vfd,Cardona-Giraldo:2026rnn}, as well as diffusion and flow models~\cite{Krause:2021ilc,Krause:2021wez,Mikuni:2022xry,Krause:2022jna,Buhmann:2023bwk,Buhmann:2023kdg,Xu:2023xdc,Buckley:2023daw,Acosta:2023zik,Mikuni:2023tqg,Amram:2023onf,Pang:2023wfx,Ernst:2023qvn,Jiang:2025pil,Kobylianskii:2024ijw,Schnake:2026iec,Schnake:2024mip,Du:2024gbp,Jiang:2024bwr,Favaro:2024rle,Raikwar:2025fky,Majerz:2025ykn,Buss:2025kiu,Buss:2025cyw,Buss:2025bec,Buss:2026yrf,Nguyen:2026wsv,Jiang:2026zov,Huang:2026irl}.
For a review, see Ref.~\cite{Hashemi:2023rgo}.
Their development is organised and benchmarked across the community through shared efforts such as the CaloChallenge~\cite{Krause:2024avx}, now being extended to the CMS HGCal~\cite{Amram:CHEP2026}.
These models are also entering experimental production: ATLAS simulates a large fraction of its Run-3 calorimeter showers with AtlFast3~\cite{ATLAS:2021pzo,ATLAS:2024vdo}, whose GAN component was among the first generative models deployed in the production chain of a large experiment. CMS is exploring FlashSim, an end-to-end analysis-level fast simulation based on normalising flows~\cite{Vaselli:2858890,CMS-DP-2024-080}.

Generative calorimeter models fall into two broad families depending on their native shower representation: fixed-grid models, tied to the segmentation of one particular detector, and point cloud models, which describe a shower as a set of energy deposits whose number varies from shower to shower. A point cloud representation presupposes no readout segmentation, so it can describe detectors with complex geometries, although the deposits themselves are still shaped by the detector that produces the shower. We therefore adopt a point cloud representation.
Point cloud models, including CaloClouds and the models that followed it~\cite{Buhmann:2023bwk,Buhmann:2023kdg,Buss:2025kiu,Buss:2025cyw}, as well as the \textsc{AllShowers} backbone of Ref.~\cite{Buss:2026yrf} that we build on, generate the energy deposits directly.
A separate line of work uses a cylindrical grid representation rather than a point cloud, aiming at a single model pre-trained across multiple geometries and then adapted to a new, unseen geometry.
Examples are CaloDiT-2~\cite{Raikwar:2025fky}, a diffusion transformer that extends the single-detector CaloDiT~\cite{Raikwar:2024peb} to multiple geometries, and the universal vision transformer of Ref.~\cite{Favaro:2026awn}, which similarly extends the single-detector study of Ref.~\cite{Favaro:2025ift}.
Conditioning can also be made continuous within a single detector: ParaFlow~\cite{Erdmann:2025tsq} learns photon showers in a fixed toy calorimeter as a smooth function of the upstream passive-material configuration. This is interpolation within the trained family, not transfer to an unseen calorimeter.
In each of these learned approaches, the trained model stays bound to the detectors it saw during training: a new geometry still requires fine-tuning on a sizeable in-domain dataset.
A more recent approach, BRICKS~\cite{Hildebrandt:2026lyj}, instead composes physics-motivated, Markovian radiation--matter kernels for zero-shot simulation across material configurations, that is, simulation of a configuration unseen in training without any further training data. It has not yet been shown to work as a full multi-geometry shower generator. A single generator that transfers efficiently to unseen detector geometries thus remains an open problem.

The closest precursor, the Cross-Geometry Transfer Learning study of Ref.~\cite{Gaede:2025shc}, showed that a shower generator pre-trained on a single calorimeter transfers to a new geometry with less data than training from scratch. Independent evidence from analysis tasks goes in the same direction: pre-training on fast simulation reduces the statistics needed for classification and regression benchmarks \cite{Schott:2026czu}.
That study drew its pre-training pool from a single real detector, which limits both the diversity and the size of the pool to what has already been designed and simulated.
We instead build the pre-training pool from \emph{synthetic geometric variation} rather than real-detector data. We sample the space of simple calorimeter geometries directly, which is inexpensive and sets the geometric coverage of the prior by construction.
A target detector inside the pre-training region can then be simulated with little or no fine-tuning, while one outside it can be reached with far fewer target showers.

This work is a step towards foundation models for calorimetry~\cite{Hallin:2025ywf,Cardona-Giraldo:2026rnn}: a transferable, geometry-general prior for a single task (shower generation), although not a general-purpose multi-task model.
Such a prior reduces the cost and effort currently required to create a new fast simulation application: adapting to a new or evolving detector with little target data, and iterating over candidate geometries in design studies without simulating a full dedicated dataset for each variant.
The main contributions of this work are:
(i) We introduce \textsc{SimpleBox}, a synthetic family of $10^{4}$ box calorimeter geometries densely covering the plane spanned by the sampling fraction $f_{\rm s}$ and the longitudinal segmentation $N_{\rm layers}$. These two axes fix the energy-deposition scale and the longitudinal sampling rather than the full geometry; the geometry-dependent lateral structure is left to be recovered during fine-tuning.
(ii) A re-simulated, point cloud version of the five realistic \textsc{LEMURS}~\cite{McKeown:2025gtw} detectors is provided, extracted on a $1\,\mathrm{mm}\times1\,\mathrm{mm}$ grid instead of the cell-aggregated Universal Grid Representation of the public release.
(iii) \textsc{AllShowers} is equipped with a geometry-aware conditioning, which makes transfer across detectors possible (\cref{sec:method}).
(iv) We benchmark \textsc{SimpleBox} against LEMURS pre-training at downstream data scales $D\in\{10^{2},10^{3},10^{4},10^{5}\}$ on the held-out FCCee-ALLEGRO calorimeter, a noble-liquid (LAr) detector whose considerably higher sampling fraction sits outside the pre-training region. We further test \textsc{SimpleBox} fine-tuning on each of the Si/Sci LEMURS detectors, and probe interpolation within the synthetic family through a zero-shot test on a held-out, in-range \textsc{SimpleBox} configuration.
(v) The compute and memory budget of the two strategies is examined to estimate the break-even point $N^{\star}$ past which pre-training followed by fine-tuning becomes cheaper than training $N$ detectors from scratch. All artefacts of this work, datasets, code and pre-trained weights, are publicly released (\cref{app:code-data}).

\Cref{sec:datasets} presents the \textsc{SimpleBox} and LEMURS datasets; \cref{sec:method} sets out the \textsc{AllShowers} architecture, the geometry-aware conditioning, and the pre-training and fine-tuning protocol; \cref{sec:results} reports the pre-training quality, the transfer results, and the compute-budget analysis; and \cref{sec:conclusions} concludes with limitations and outlook.

\section{Datasets}
\label{sec:datasets}

We use two complementary datasets of photon-induced EM showers, generated with a common simulation chain and exported under a shared point cloud schema. In both cases the detector is described in \texttt{DD4hep}~\cite{Frank:2014zya,Gaede:2020tui}, with \texttt{Geant4}~11.2.2~\cite{GEANT4:2002zbu,Allison:2006ve,Allison:2016lfl} run through the \texttt{ddsim} driver within the \texttt{Key4hep} stack~\cite{Carceller:2025ydg}, and the response is written in the \texttt{EDM4hep} data model~\cite{Gaede:2022leb}. The two datasets differ only in the detector geometry: \textsc{SimpleBox} (\cref{ssec:datasets-simplebox}) uses a custom box geometry of alternating planar Si/W sampling layers, whereas the five realistic LEMURS calorimeters (\cref{ssec:datasets-lemurs-calos}) use their published geometries, among them the \texttt{OpenDataDetector}~\cite{OpenDataDetector} (v5.0.0) for the ODD calorimeter. Both datasets are generated through the \texttt{ddfastsim} package~\cite{ddfastsim}. The \texttt{EDM4hep} hit collections are converted to a lightweight HDF5 point cloud format. In both datasets the deposits are clustered on a $1~\mathrm{mm}\times1~\mathrm{mm}$ transverse grid per layer; we call these grid elements cells throughout. Cells below $10~\mathrm{keV}$ are discarded. All datasets are released in a single record~\cite{mgg_datasets} (\cref{app:code-data}).
Each shower is a variable-cardinality set of points: $\mathbf{p}_{i}
  = \bigl(x^{(i)},\; y^{(i)},\; \ell^{(i)},\; E^{(i)}\bigr)$, where $(x,y)$ are the transverse displacements with respect to the impact point (continuous), $\ell\in\{0,\dots,N_{\rm layers}-1\}$ is the longitudinal (depth) layer index (integer) and $E$ is the cell-summed deposited energy.

\Cref{ssec:datasets-simplebox} describes \textsc{SimpleBox}, a synthetic family of $10^{4}$ box-shaped geometries used to
test interpolation across changes in sampling fraction $f_{\rm s}$ and
longitudinal segmentation.
\Cref{ssec:datasets-lemurs-calos} describes \textsc{LEMURS}, a step-level re-simulation of the five realistic barrel ECALs of Ref.~\cite{McKeown:2025gtw}, with the four Si/Sci detectors used for pre-training and the noble-liquid (LAr) FCCee-ALLEGRO calorimeter reserved as the downstream transfer target, the same held-out detector used by CaloDiT-2~\cite{Raikwar:2025fky}.

\subsection{SimpleBox: synthetic geometric variations}
\label{ssec:datasets-simplebox}

\begin{figure}[htbp]
\centering
\begin{minipage}[b]{0.50\linewidth}
    \includegraphics[width=\linewidth]{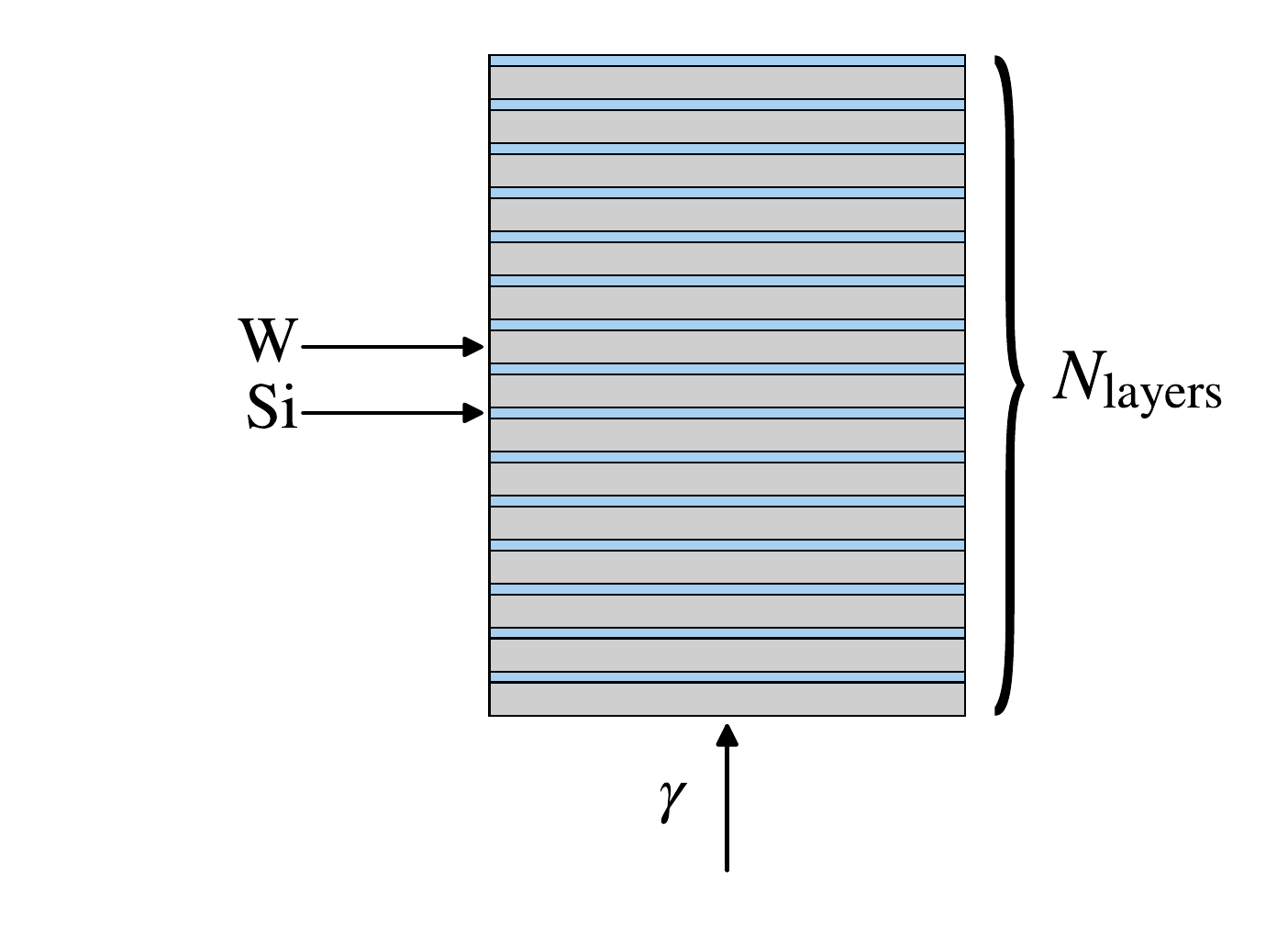}
\end{minipage}\hspace{0.01\linewidth}%
\begin{minipage}[b]{0.48\linewidth}
    \includegraphics[width=\linewidth]{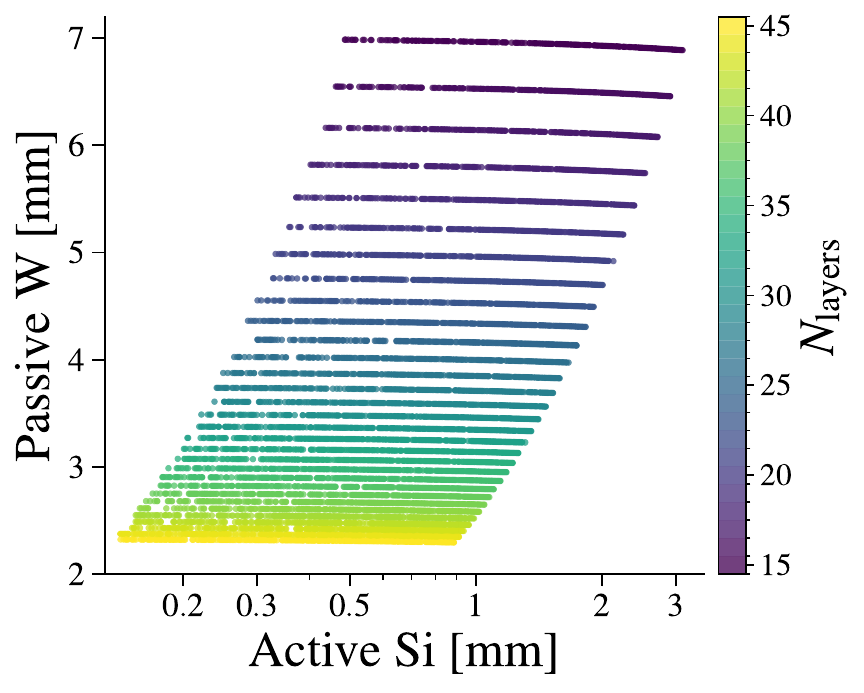}
\end{minipage}
\caption{Left: schematic of a \textsc{SimpleBox} calorimeter with alternating passive W (grey) and active Si (blue) layers, the absorber first along the photon path.
Right: distribution of layer thicknesses across the $10^{4}$ configurations, coloured by $N_{\rm layers}$.}
\label{fig:simplebox-datasets}
\end{figure}
\begin{figure}[htbp]
    \centering
    \includegraphics[width=1\linewidth]{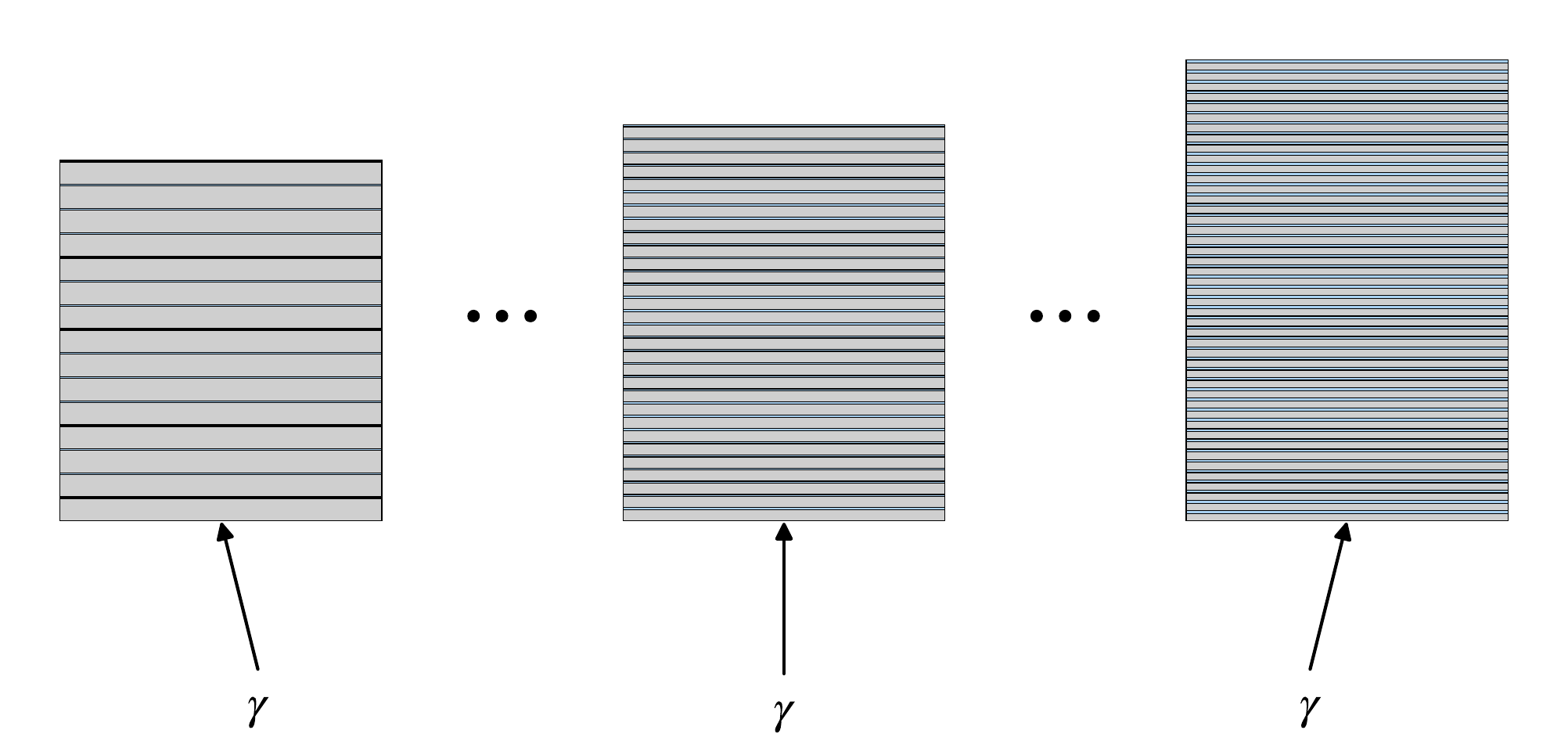}
    \caption{Representative \textsc{SimpleBox} geometries from the multi-geometry pre-training set. From left to right $N_{\rm layers}$ increases: the active Si (blue) and passive W (grey) layers grow thinner and more numerous, with the photon $\gamma$ entering each block from below. The full set comprises $10^{4}$ configurations sampled with $f_{\rm s}^{\rm target}\sim\mathcal{U}[0.01,0.05]$ and $N_{\rm layers}\sim\mathcal{U}\{15,\dots,45\}$.}
    \label{fig:simplebox_multigeo}
\end{figure}

We introduce \textsc{SimpleBox}, a family of box-shaped sampling calorimeters made of
alternating passive tungsten (W) and active silicon (Si) layers, the absorber
first along the photon path
(\cref{fig:simplebox-datasets}, left).
A configuration is specified by
the number of layers $N_{\rm layers}$ and the active fraction
$a\in[0,1]$, defined as the fraction of total radiation lengths contributed by
the active material,
\begin{equation}
  a \;\equiv\;
  \frac{t_{\rm Si}/X_0^{\rm Si}}{t_{\rm Si}/X_0^{\rm Si}+t_{\rm W}/X_0^{\rm W}}.
  \label{eq:simplebox-a}
\end{equation}
It should not be confused with the sampling fraction $f_{\rm s}$, the fraction of the deposited energy that is registered in the active layers.
At fixed total depth\footnote{The value is chosen so that EM showers up to $100~\mathrm{GeV}$ are contained to $\gtrsim 99\%$.} $n_{X_0}=30\,X_0$, the
layer thicknesses are:
\begin{equation}
  t_{\rm Si}=\frac{n_{X_0}X_0^{\rm Si}\,a}{N_{\rm layers}},\qquad
  t_{\rm W} =\frac{n_{X_0}X_0^{\rm W}\,(1-a)}{N_{\rm layers}},
  \label{eq:simplebox-thickness}
\end{equation}

with $X_0^{\rm Si}=93.7~\mathrm{mm}$ and
$X_0^{\rm W}=3.5~\mathrm{mm}$~\cite{Workman:2022ynf}. The right panel
of \cref{fig:simplebox-datasets} shows the distribution of $(t_{\rm Si},t_{\rm W})$ pairs over the final simulated configurations.

The sampling fraction $f_{\rm s}$ has no closed form in terms of the active
fraction $a$ and the number of layers $N_{\rm layers}$, so targeting a
prescribed $f_{\rm s}$ requires a two-stage procedure.
\begin{enumerate}
    \item Calibration: $a$ is drawn uniformly per $N_{\rm layers}\in\{15,\dots,45\}$, the geometries are simulated, and a degree-$4$ polynomial $\hat{a}(f_{\rm s},N_{\rm layers})$ is fit to invert the $f_{\rm s}$--$a$ relation (closure RMSE $4\times10^{-5}$, $\sim0.1\%$; \cref{app:simplebox}).
    \item Generation: $f_{\rm s}^{\rm target}\sim\mathcal{U}[0.01,0.05]$ and $N_{\rm layers}\sim\mathcal{U}\{15,\dots,45\}$ are sampled, mapped to $a_{\rm pred}$, and the resulting geometries are simulated.
    Photons are fired at the box face
    \begin{itemize}
        \item with incidence angle up to $45^{\circ}$ from its normal,
        \item with a direction uniform in solid angle over that cone ($\cos\theta\sim\mathcal{U}[\cos 45^{\circ},1]$ with $\theta$ measured from the normal, $\phi\sim\mathcal{U}[0,2\pi)$),
        \item and with impact energy $E_{\rm inc}\sim\mathcal{U}[1,100]~\mathrm{GeV}$.
    \end{itemize}
    We produce $400$ showers for each configuration.
\end{enumerate}
\Cref{fig:simplebox_multigeo} shows the resulting geometries.

We simulate $4\times10^{6}$ \textsc{SimpleBox} showers in total, matching the size of the four-detector LEMURS pre-training pool (\cref{ssec:datasets-lemurs-calos}) so that the two pre-training strategies are trained on the same number of showers.
We now consider how to divide this budget between the number of geometries $N_{\rm g}$ and the number of showers per geometry $N_{\rm s}$.
The statistical error in the shower distribution the model learns has two sources: the variation of shower behaviour from one geometry to another, as the sampling fraction $f_{\rm s}$ and the number of layers $N_{\rm layers}$ change, and the shower-to-shower fluctuations within a single geometry.
The within-geometry fluctuations scale as $1/(N_{\rm g}N_{\rm s})$ and are therefore already fixed by the total budget, while the geometry-to-geometry variation scales as $1/N_{\rm g}$ and is reduced only by simulating more distinct geometries.
We therefore spend the budget on as many geometries as possible, taking $N_{\rm g}=10^{4}$ rather than the $10^{2}$ or $10^{3}$ of a more shower-heavy split, so that the $(f_{\rm s},N_{\rm layers})$ plane is densely covered.
Each geometry still needs enough showers for the model to learn the distribution it produces, and $N_{\rm g}=10^{4}$ leaves $N_{\rm s}=400$.
This budget allocation concerns the statistical error in that learned distribution rather than the final quality of the model. That quality is assessed on held-out configurations in \cref{ssec:results-pretrain}.
$N_{\rm s}=400$ is not a strict lower bound. To quantify the simulation cost that synthetic geometric pre-training requires, we also build \textsc{SimpleBox}-mini, a random $2.5\%$ subsample of the full pool ($10^{5}$ showers); it beats training from scratch at every fine-tuning size on the held-out FCCee-ALLEGRO target, while giving up part of the full pool's advantage at the smallest sizes (\cref{app:simplebox-mini}).
We analyse the resulting compute cost, and the point at which pre-training becomes cheaper across downstream detectors, in \cref{ssec:results-breakeven}.
\FloatBarrier
\subsection{LEMURS point cloud calorimeters}
\label{ssec:datasets-lemurs-calos}

\begin{figure}[hbp]
    \centering
    \includegraphics[width=.8\linewidth]{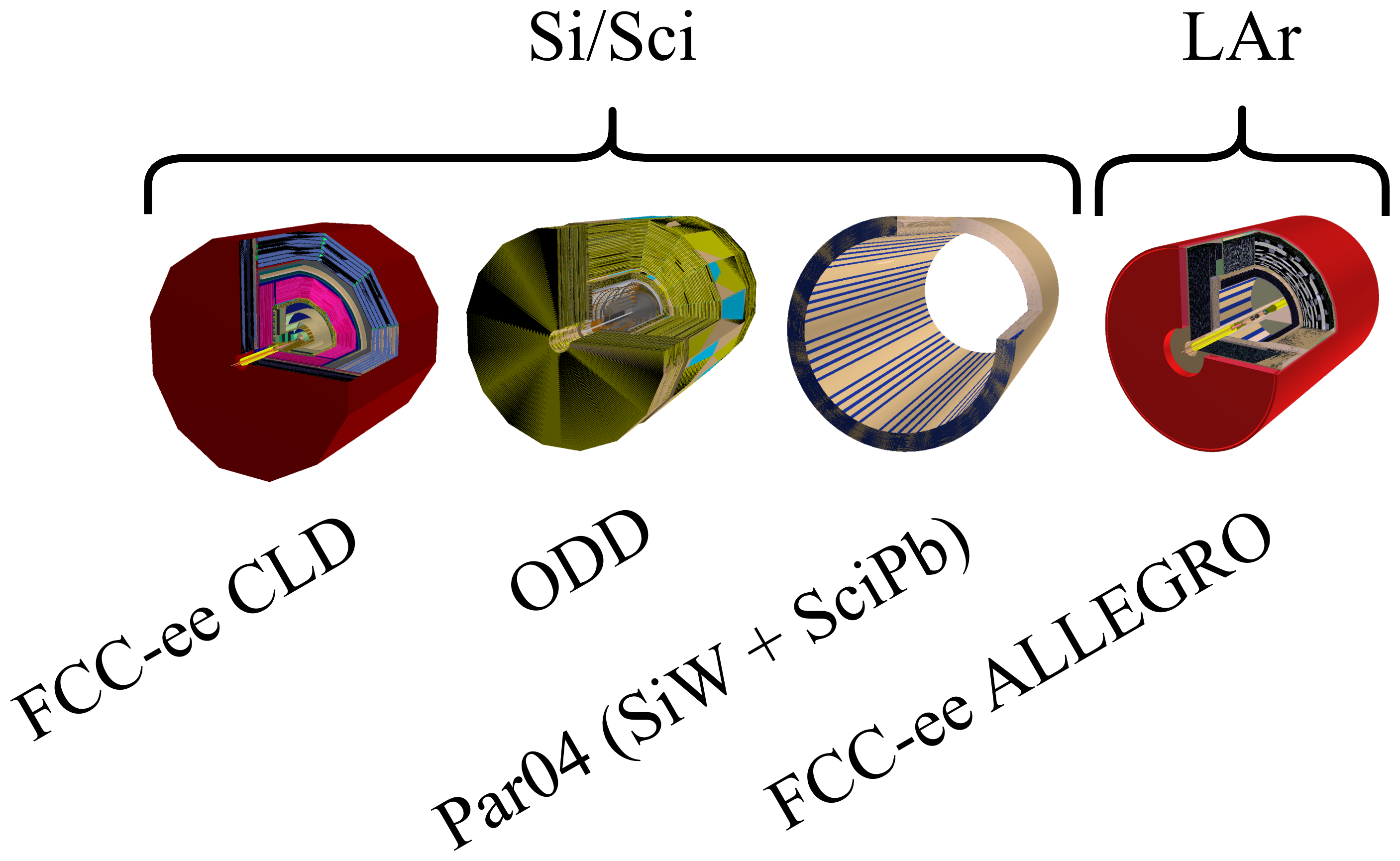}
    \caption{The realistic barrel ECALs that make up the LEMURS dataset: the Si/Sci sampling calorimeters (FCCee-CLD, ODD, Par04-SiW, Par04-SciPb) used for pre-training, and the noble-liquid (LAr) FCCee-ALLEGRO calorimeter reserved as the downstream transfer target. Adapted with permission from Ref.~\cite{Raikwar:2025fky}. CC BY-SA 4.0.}
    \label{fig:lemurs-detectors}
\end{figure}

The LEMURS dataset is a step-level \texttt{Geant4} re-simulation~\cite{Zaborowska:2025nsv}\footnote{The public
release ships only the voxelised Universal Grid Representation showers.} of the five barrel ECALs of Ref.~\cite{McKeown:2025gtw} (\cref{fig:lemurs-detectors}).
The five detectors split into four Si/Sci sampling calorimeters with $f_{\rm s} < 0.04$ (Par04-SiW, Par04-SciPb, ODD, FCCee-CLD~\cite{Bacchetta:2019fmz}) and one noble-liquid calorimeter\footnote{ALLEGRO is described in version \texttt{o1\_v03} (11 layers), in place of the \texttt{v02} (12 layers) used by the original LEMURS dataset~\cite{McKeown:2025gtw}.} with $f_{\rm s}\simeq 0.16$ (FCCee-ALLEGRO, LAr/Pb)~\cite{Pekkanen:2024zbe}; geometric parameters are listed in \cref{tab:detectors}.

\begin{table}[htbp]
\centering
\caption{Geometric and sampling parameters of the five LEMURS ECALs.
Layers are radial; ``Depth'' is the active-stack radial extent;
$\langle f_{\rm s}\rangle$ is estimated from $3\times 10^{4}$
randomly selected showers per detector. The Active/Passive column gives the thicknesses of the two dominant materials only, not the full per-layer pitch.}
\label{tab:detectors}
\sisetup{table-format=4.1}
\begin{tabular}{l | S S c S[table-format=3.0] S[table-format=1.4]}
\toprule
Detector
  & {$R_{\min}$ [mm]} & {$N_{\rm layers}$} & {Active / Passive [mm]}
  & {Depth [mm]} & {$\langle f_{\rm s}\rangle$} \\
\midrule
Par04-SiW       &  800.0 & 90 & Si(0.3) / W(1.4)   & 153 & 0.0324 \\
Par04-SciPb     &  800.0 & 45 & Sci(1.2) / Pb(4.4) & 252 & 0.0334 \\
ODD              & 1250.0 & 48 & Si(0.5) / W(1.9)   & 242 & 0.0268 \\
FCCee-CLD       & 2150.0 & 40 & Si(0.5) / W(1.9)   & 202 & 0.0259 \\
\midrule
FCCee-ALLEGRO   & 2172.8 & 11 & LAr(1.2) / Pb(2.0) & 405 & 0.1619 \\
\bottomrule
\end{tabular}
\end{table}

Showers are produced with \texttt{Geant4} via \texttt{ddsim} with \texttt{HitCreationMode}\,$=2$, in which every \texttt{Geant4} step in the sensitive detector elements is recorded. 
Photons are fired from a \texttt{ddsim} particle gun placed on the inner barrel surface at $\mathbf{x}_{\rm gun}=(R_{\rm g}\cos\phi,R_{\rm g}\sin\phi, R_{\rm g}\cot\theta)$, with $R_{\rm g}=R_{\min}-\epsilon$ and $\epsilon=10^{-8}~\mathrm{mm}$.
The momentum points outward, along the line from the interaction point through the gun position, and the impact energy is $E_{\rm inc}\sim\mathcal{U}[1,100]~\mathrm{GeV}$ for all calorimeters.
The angular distribution is common, $\cos\theta\sim\mathcal{U}[\cos 2.27,\cos 0.87]$, $\phi\sim\mathcal{U}[0,2\pi)$, covering roughly $50^{\circ}$--$130^{\circ}$ from the beam axis; sampling uniformly in $\cos\theta$ rather than in $\theta$ keeps the incidence isotropic in solid angle over this band.
We generate $10^{6}$ showers per detector, giving $4\times10^{6}$ pre-training samples and a further $10^{6}$ for ALLEGRO.

Each shower is projected into a particle-centric cylindrical frame, assigned to one of $N_{\rm layers}$ equal-width radial shells, and clustered on a $1~\mathrm{mm}\times 1~\mathrm{mm}$ transverse grid per shell; cells below $E_{\rm thr}=10~\mathrm{keV}$ are discarded.\footnote{Well below the MIP scale of any of the active materials.} 
The exact local-frame equations and acceptance windows are given in \cref{app:lemurs-preprocessing} and illustrated in \cref{fig:lemurs-frame}.
The output is the point cloud representation together with the conditioning vector $\mathbf{c}$, identical in schema to \textsc{SimpleBox}. 
Mean point cloud cardinality across the $4\times10^{6}$ Si/Sci pre-training showers is $\sim\!3.2\times10^{3}$ (max $\sim\!7.8\times10^{3}$); ALLEGRO is markedly denser, with mean $\sim\!1.3\times10^{4}$ and tails reaching $\sim\!2.5\times10^{4}$. Its sampling fraction is about five times that of the Si/Sci detectors (\cref{tab:detectors}), so a far larger share of each shower's energy is deposited in the active medium and more cells pass the fixed $10~\mathrm{keV}$ threshold.
Because the architecture accepts point clouds of variable cardinality (\cref{ssec:method-arch}), the same model and conditioning layers are reused in the downstream stage without modification.

\section{Method}
\label{sec:method}

We adapt the \textsc{AllShowers} point cloud generator~\cite{Buss:2026yrf} to the multi-geometry setting, so that a single backbone can be pre-trained once and transferred to new calorimeter geometries.
\Cref{ssec:method-arch} describes the architecture and the geometry-aware conditioning that lets one model span detectors with different sampling fractions and segmentations.
\Cref{ssec:method-protocol} then defines the pre-training and fine-tuning protocol and the two pre-training pools, and \cref{ssec:method-metrics} defines the observables and metrics used throughout.

\subsection{AllShowers architecture}
\label{ssec:method-arch}

We build on the \textsc{AllShowers} of Ref.~\cite{Buss:2026yrf}, a two-stage conditional flow matching model in which a transformer vector field regresses the point cloud coordinates and a \textsc{PointCountFM} network predicts the per-layer point multiplicities.
The architectural backbone is inherited unchanged from the public release.
Our contributions are confined to the conditioning interface and to a post-hoc bias correction on the multiplicity stage.

In the original work, each shower is conditioned on its incident energy $E_{\rm inc}$, the unit vector of its incident momentum $\hat{\mathbf{n}}$, and a learnable embedding of the particle species.
Since we restrict ourselves to electromagnetic showers from photons, the species embedding is constant and is dropped.
The generator must instead distinguish between detector geometries that share the same kinematic input.
We replace the particle-type slot with two scalar geometry conditioners, the sampling fraction $f_{\rm s}$ and the number of active layers $N_{\rm layers}$.
Each passes through its own learnable normalisation and is concatenated to the kinematic block: $ \mathbf{c} = (E_{\rm inc},\;\hat{\mathbf{n}},\;f_{\rm s},\;N_{\rm layers})$.
The two scalars encode the absolute energy-deposition scale and the longitudinal granularity of each detector.
They do not encode the transverse (Moli\`ere) scale, which is set by the absorber material and is instead adapted during fine-tuning.

We retain the global energy-conditioning rescaling of Ref.~\cite{Buss:2026yrf} ($E_{\rm inc}\!\to\!1.033\,E_{\rm inc}$) on the transformer stage that generates the point coordinates.
The multiplicity stage, however, requires a separate calibration.
\textsc{PointCountFM} (PCFM) regresses per-layer point counts in a log-transformed space. The inverse $\exp(\cdot)$ then amplifies any residual latent noise, and by Jensen's inequality this leaves a systematically positive bias on the per-layer means at inference.
We absorb this effect by per-layer multiplicative factors $b_\ell = \langle n_\ell^{\rm real}\rangle / \langle n_\ell^{\rm gen}\rangle$,
defined on the active layers and averaged over five incident-energy percentile bins.
Each generated count is replaced by $n_\ell\,b_\ell$ rounded to the nearest non-negative integer at sampling time.
To avoid data leakage, the factors are fitted on the training partition alone, separately at every training-data size of the scaling study.
Such a per-layer correction is needed in our setting because the residual bias is no longer uniform across layers, as it was in the single-detector, single-energy-scale configuration of Ref.~\cite{Buss:2026yrf}. With several geometries differing in $N_{\rm layers}$ and active-stack depth, a single global scalar would leave per-layer residuals in the longitudinal profile.
 The implementation of both stages is public~\cite{mgg_code_allshowers,mgg_code_pcfm} (\cref{app:code-data}).

\subsection{Pre-training and fine-tuning protocol}
\label{ssec:method-protocol}

We consider two pre-training pools, both introduced in \cref{sec:datasets}: the synthetic \textsc{SimpleBox} family, which provides a dense coverage of $(f_{\rm s},N_{\rm layers})$, and the four Si/Sci LEMURS detectors, which occupy the realistic-detector edge of that distribution. Their sampling fractions fall within the \textsc{SimpleBox} range, while their segmentations extend beyond the synthetic $\{15,\dots,45\}$, with $N_{\rm layers}$ up to $90$, so transfer to them already probes extrapolation in $N_{\rm layers}$.
Both pools are kept disjoint from the downstream targets.

\begin{figure}[htbp]
    \centering
    \includegraphics[width=.9\linewidth]{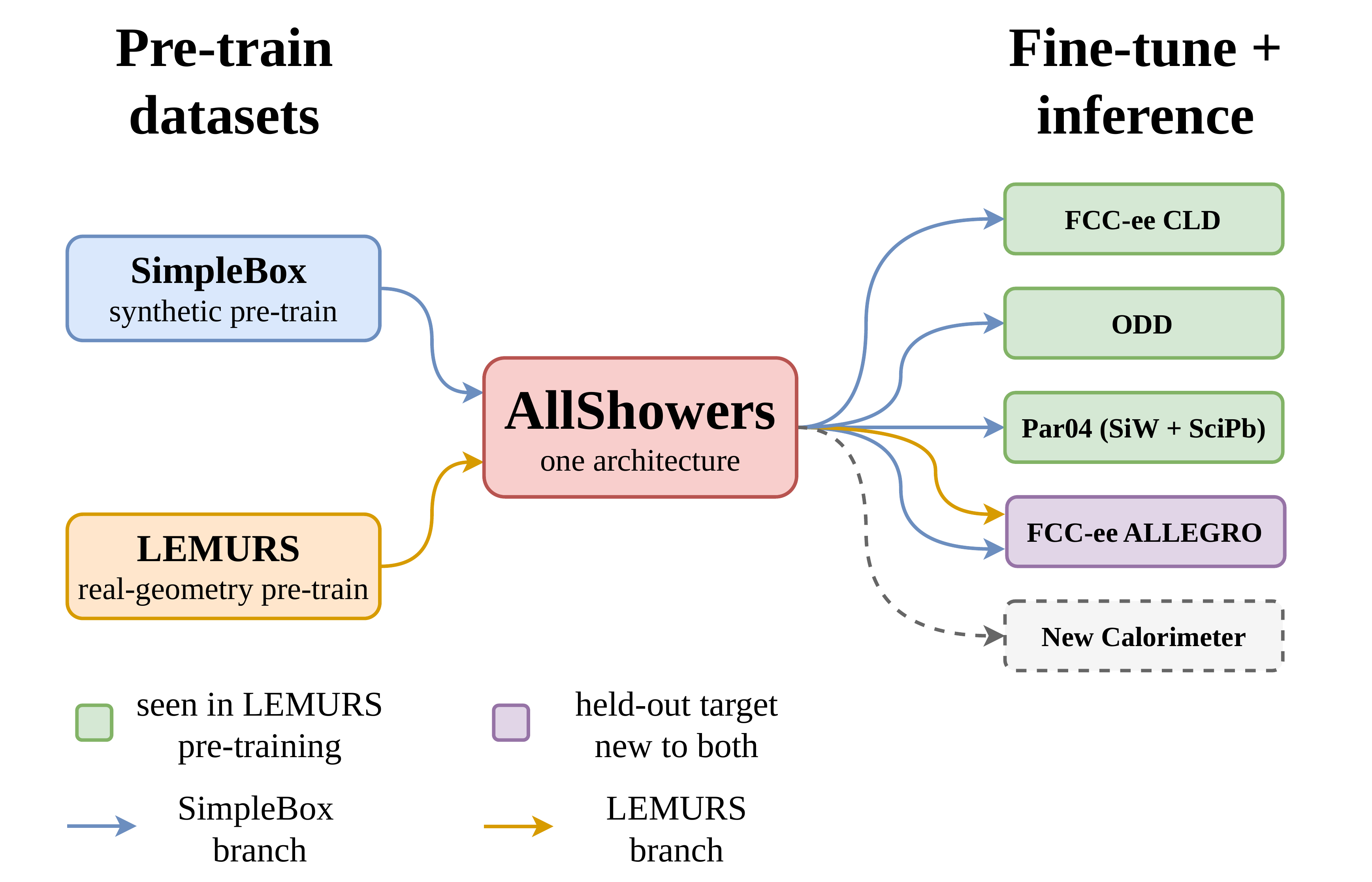}
    \caption{Transfer pipeline. A single \textsc{AllShowers} architecture is pre-trained on two alternative pools: the synthetic \textsc{SimpleBox} family (blue branch) and the four real-geometry Si/Sci LEMURS detectors (orange branch). The \textsc{SimpleBox} branch is then fine-tuned on the realistic Si/Sci calorimeters (FCCee-CLD, ODD, Par04-SiW, Par04-SciPb) and on FCCee-ALLEGRO; the LEMURS branch, which already contains the Si/Sci detectors in pre-training (green), is fine-tuned only on FCCee-ALLEGRO. ALLEGRO (purple) is held out from both pools and is where the two pre-trainings are compared against a from-scratch baseline. The dashed arrow marks the extension to any unseen geometry.}
    \label{fig:transfer-pipeline}
\end{figure}

The primary downstream target is FCCee-ALLEGRO, with $f_{\rm s}\!\simeq\!0.16$ and $N_{\rm layers}\!=\!11$.
It is held out from both pools, lies outside the pre-training $(f_{\rm s},N_{\rm layers})$ region, and is where the two pre-trainings are compared directly (\cref{fig:transfer-pipeline}).
It is therefore an extrapolation stress test along three axes: a noble-liquid (LAr) active medium absent from the pre-training pools, a sampling fraction about three times the highest value covered in pre-training ($f_{\rm s}\simeq 0.05$), and a segmentation ($N_{\rm layers}=11$) below every source geometry.
The \textsc{SimpleBox}-pretrained backbone is additionally fine-tuned on the four realistic Si/Sci LEMURS detectors (FCCee-CLD, ODD, Par04-SiW and Par04-SciPb), and evaluated zero-shot on a held-out \textsc{SimpleBox}-like configuration that lies inside the pre-training range but is never seen during training. This zero-shot test probes how well the pre-trained backbone interpolates within the synthetic family.
Every fine-tuned configuration is compared against a from-scratch\footnote{Models are initialised with random weights, the conventional approach in which each detector geometry requires a full training of its own.} baseline trained on the same downstream budget.

To warm-start a target with $N_{\rm layers}^{\rm target}\!\neq\!N_{\rm layers}^{\rm source}$, we copy the overlapping slice of each pre-trained tensor whose shape depends on $N_{\rm layers}$ into the freshly initialised target tensor.
The remainder keeps the target module's own random initialisation, not zero.
A single backbone is warm-started from a pre-trained source onto each downstream target, including ALLEGRO, whose $N_{\rm layers}\!=\!11$ is smaller than that of any source detector.
The two Si/Sci targets with the most layers, ODD ($N_{\rm layers}=48$) and Par04-SiW ($N_{\rm layers}=90$), are likewise warm-started from the $45$-layer \textsc{SimpleBox} pre-training through the same overlapping-slice copy.
All other training ingredients (Ranger optimiser, cosine schedule, batch size, training-time optimal-transport coupling) are inherited unchanged from Ref.~\cite{Buss:2026yrf}; the complete learning-rate schedules, step budgets and the \textsc{PointCountFM} calibration are listed in \cref{app:hyperparams}.

The data-scaling study spans four train sizes, $D\!\in\!\{10^{2},10^{3},10^{4},10^{5}\}$, drawn at random from each training partition.
At each $D$ we compare, on ALLEGRO, four strategies on the same evaluation suite: from-scratch, fine-tuning from \textsc{SimpleBox}-pretrain, fine-tuning from LEMURS-pretrain, and fine-tuning from a reduced \textsc{SimpleBox}-mini pre-training ($10^{5}$ showers, $2.5\%$ of the \textsc{SimpleBox} pool). We refer to these four training strategies as the arms of the comparison throughout. On the Si/Sci detectors LEMURS-pretrain does not apply, as they belong to its pre-training pool, so the comparison there is from-scratch versus \textsc{SimpleBox}-pretrain.
On ALLEGRO, each $(D,\mathrm{strategy})$ pair is repeated over five random seeds, and we report the mean over seeds with a band of one standard deviation; the Si/Sci scaling runs use a single training run per $(D,\mathrm{strategy})$. The validation set, from which the evaluation reference is drawn, is kept fixed across $D$ and strategies to remove split variance from the scaling axis.

\subsection{Evaluation metrics}
\label{ssec:method-metrics}

We assess generation quality by comparing generated and \texttt{Geant4} distributions of three observables: the longitudinal profile (the energy per calorimeter layer), the radial profile (the energy-weighted distribution of hit distances from the per-shower energy-weighted barycentre), and the cell-energy spectrum.
For each observable, we evaluate two complementary metrics.
The first is the Wasserstein-1 (earth-mover) distance normalised by the standard deviation of the \texttt{Geant4} reference, $W_1/\sigma_{\rm ref}$, which measures the bulk displacement in observable space on a scale common to all observables; its per-detector values on the LEMURS targets are tabulated in \cref{app:percalo-kl} (\cref{tab:sb-lemurs-w1}).
The second is a Kullback--Leibler divergence, with the expectation taken over the \texttt{Geant4} reference and evaluated on bins set by its equiprobable quantiles, so that the bin edges follow the probability mass and the sparsely populated tails stay adequately sampled.

For each metric, the three per-observable values are combined into a single aggregate score by an unweighted geometric mean over the observables,
\begin{equation}
\label{eq:dbar}
\bar{d} = \Bigl(\textstyle\prod_{o=1}^{3} d_o\Bigr)^{1/3},
\end{equation}
where $d_o$ is the distance for observable $o$, so that no observable with a larger numerical scale dominates. We report both the per-observable values and the aggregate $\bar{d}$, and each comparison uses $10^{4}$ showers per sample.

Both metrics above compare the pooled distribution of an observable, with the entries from all showers pooled into a single histogram. A generator can reproduce a pooled distribution and its mean profile while misrepresenting how that observable fluctuates from shower to shower and how its bins correlate within a shower. The transfer comparison of \cref{ssec:results-allegro-transfer} turns on that distinction, so there we add a third metric that keeps the showers separate: each observable becomes a per-shower vector, the energy in each layer for the longitudinal profile and the corresponding binned vectors for the radial profile and the cell-energy spectrum. The two clouds of vectors are then compared with the sliced Wasserstein distance (SWD)~\cite{JMLR:v22:20-451}.
It averages the first-order Wasserstein distance over $256$ random one-dimensional projections of the vectors. Each projection mixes the bins, so the distance is sensitive to correlations that a pooled comparison cannot see. We normalise it by the root-mean-square dispersion of the \texttt{Geant4} cloud about its own mean, which keeps it on the same dimensionless scale as $W_1/\sigma_{\rm ref}$.
The per-observable values are aggregated with the same geometric mean of \cref{eq:dbar}, and these comparisons use $5\times10^{3}$ showers per sample. As a resolution scale, we also quote the same distance evaluated between two disjoint halves of the \texttt{Geant4} reference. This is the statistical resolution of the metric at the sample size used, not a hard lower bound: a generated sample that reaches it is indistinguishable from \texttt{Geant4} by this metric at the available statistics, and a sample close to convergence can scatter around it.
It appears as a dashed line labelled `statistical resolution' in the transfer figures.

\section{Results}
\label{sec:results}

We evaluate the method in four steps.
\Cref{ssec:results-pretrain} first shows that each pre-training stage reproduces the \texttt{Geant4} physics across the geometries it covers.
\Cref{ssec:results-sb-to-lemurs} then transfers the \textsc{SimpleBox} pre-trained model to the four realistic Si/Sci detectors, and \cref{ssec:results-allegro-transfer} to the held-out ALLEGRO calorimeter, where the \textsc{SimpleBox} and LEMURS pre-trainings are compared against training from scratch as a function of the available data.
Finally, \cref{ssec:results-breakeven} examines the associated compute budget for an ALLEGRO-like downstream detector.

\subsection{Pre-training quality}
\label{ssec:results-pretrain}

\begin{figure}[htbp]
    \centering
    \includegraphics[width=1\linewidth]{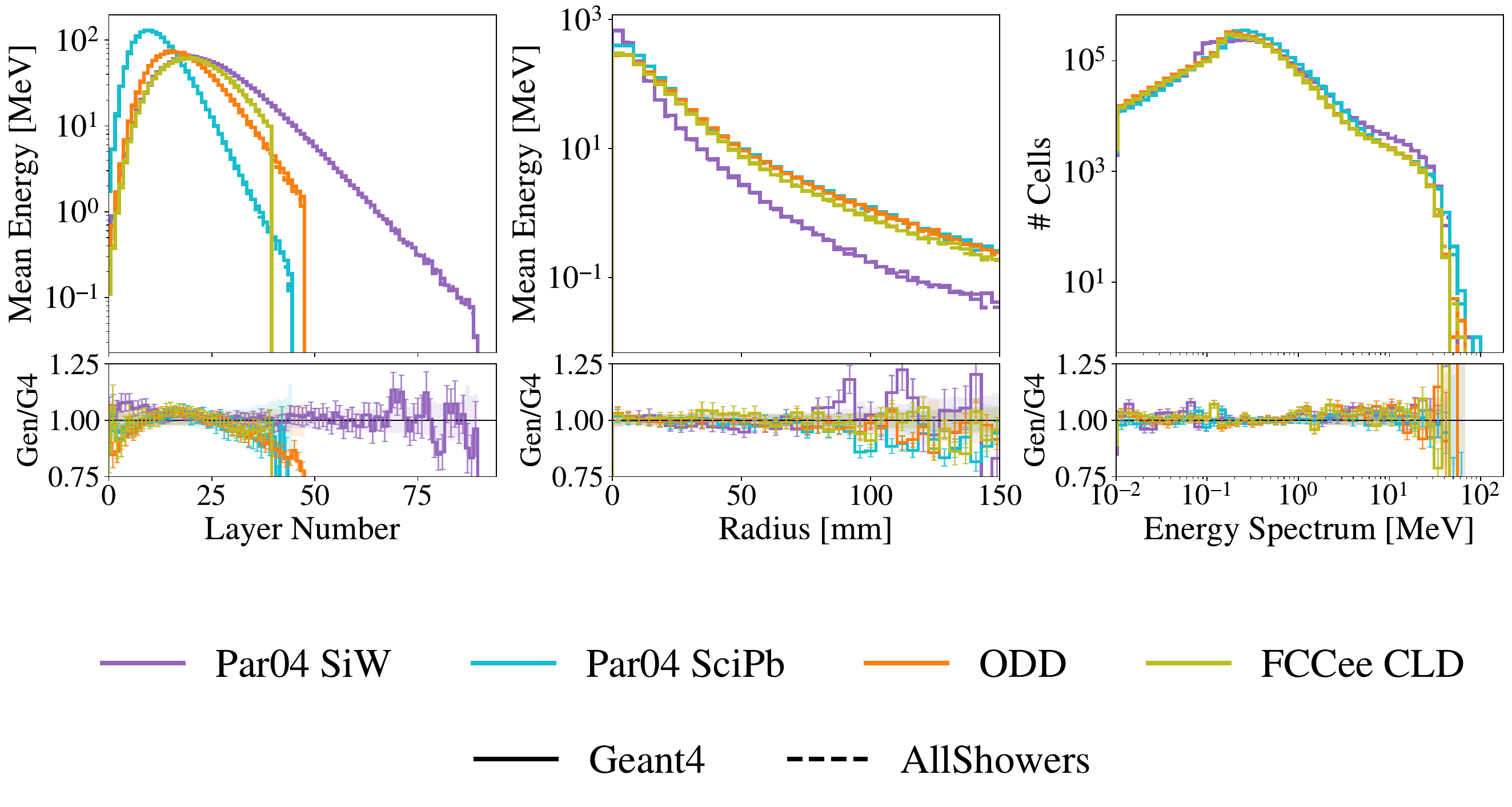}
    \caption{LEMURS pre-training quality. Distributions of the longitudinal profile (left), the radial profile (centre) and the cell-energy spectrum (right), overlaid for the four Si/Sci detectors (Par04-SiW, Par04-SciPb, ODD, FCCee-CLD).
    Lower panels show the generated-to-\texttt{Geant4} ratio. Here and in the following ratio panels, the shaded band around unity is the statistical uncertainty of the \texttt{Geant4} reference, tinted per detector where several references share a panel; the bars carry that of the generated sample. Showers are the independent statistical unit. A single backbone matches all four geometries simultaneously, with the ratio staying close to unity across the populated range.}
    \label{fig:lemurs_pretrain_observables}
\end{figure}

\begin{figure}[htbp]
    \centering
    \includegraphics[width=1\linewidth]{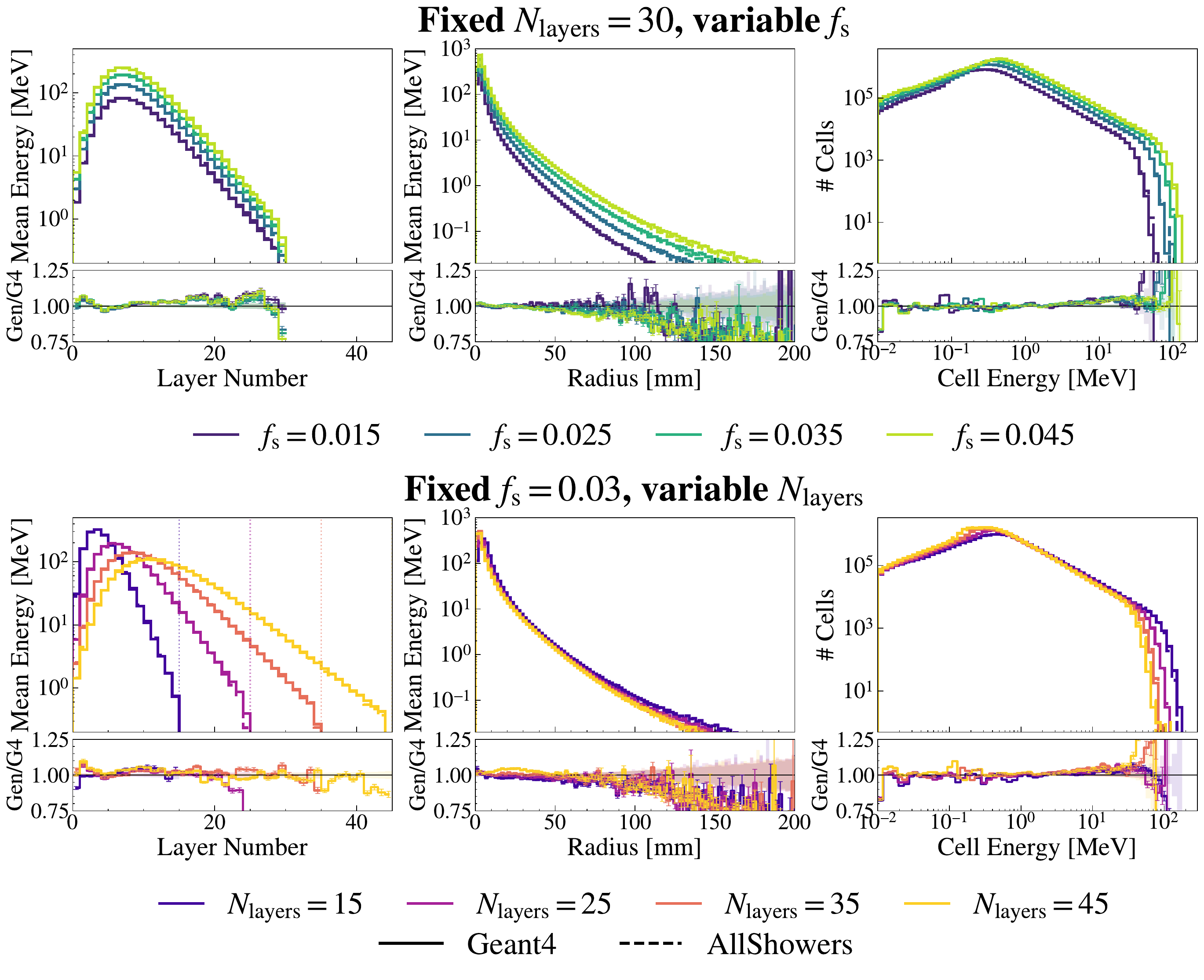}
    \caption{\textsc{SimpleBox} pre-training quality is evaluated on eight held-out geometries spanning the conditioning grid.
    The first row fixes $N_{\rm layers}=30$ and varies the sampling fraction ($f_{\rm s}=[0.015$, $0.045]$); the second row fixes $f_{\rm s}=0.03$ and varies the number of layers ($N_{\rm layers}=[15$, $45]$).
    Columns show the longitudinal profile (left), the radial profile (middle) and the cell-energy spectrum (right), with generated samples (dashed) overlaid on \texttt{Geant4} (solid) and the generated-to-\texttt{Geant4} ratio in the sub-panels. 
    The model reproduces the systematic shift of each observable as the two conditioning variables are varied.
    The \textsc{SimpleBox}-mini counterpart is shown in \cref{fig:simplebox_mini_pretrain_observables}.}
    \label{fig:simplebox_pretrain_observables}
\end{figure}

\begin{figure}[htbp]
    \centering
    \includegraphics[width=1\linewidth]{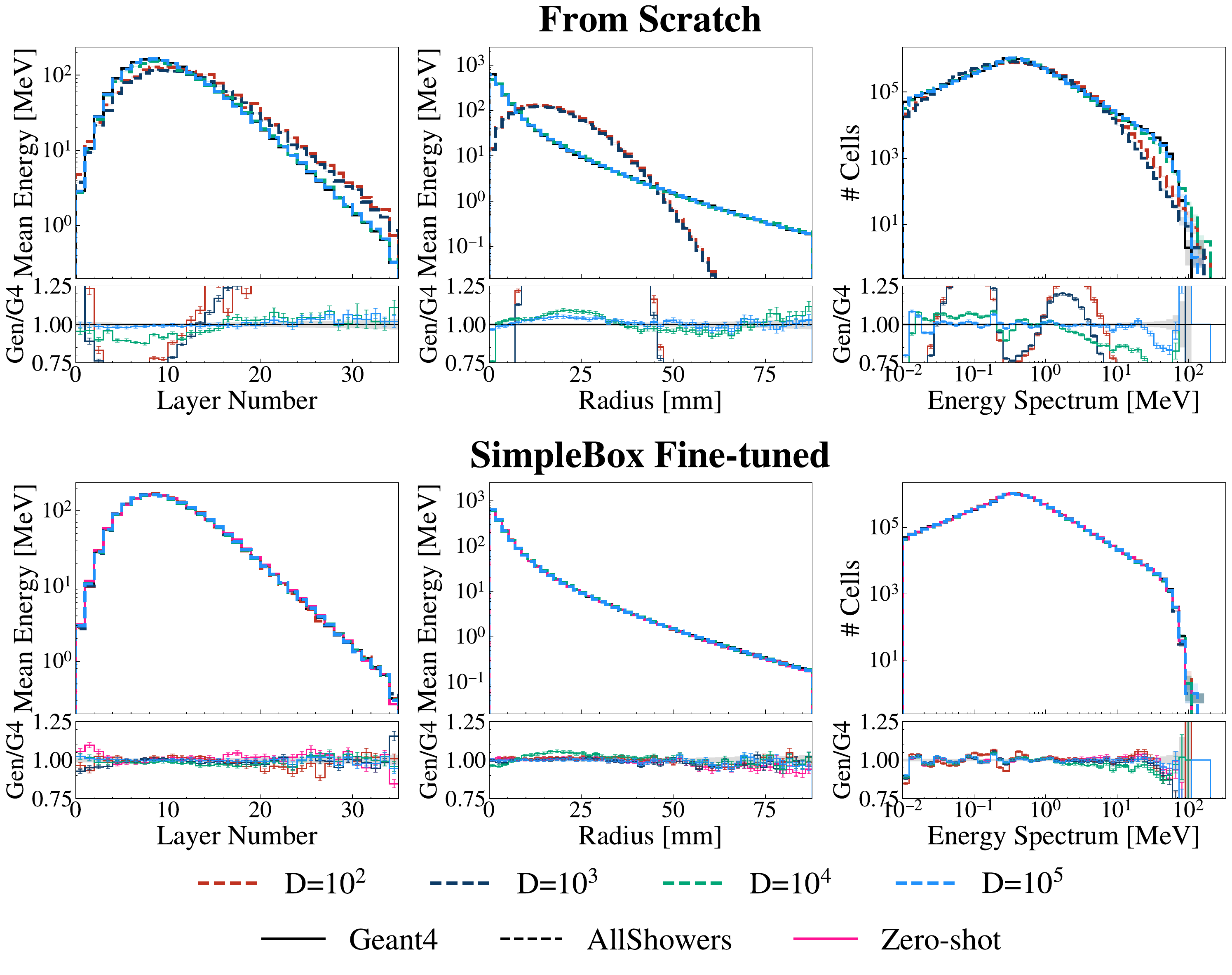}
    \caption{Transfer to a held-out \textsc{SimpleBox}-like detector ($f_{\rm s}=0.035$, $N_{\rm layers}=35$) that lies inside the pre-training $(f_{\rm s},N_{\rm layers})$ grid but is absent from the training set.
    The \textsc{SimpleBox}-pretrained model, fine-tuned at four fine-tuning dataset sizes $D$, is compared against from-scratch training (top and bottom rows) on the three evaluation observables. 
    Generated distributions (dashed) are overlaid on the \texttt{Geant4} (solid) results, together with the zero-shot pre-trained model (pink), evaluated without any fine-tuning. 
    The close agreement, reached already at zero-shot, is consistent with interpolation within the synthetic family.}
    \label{fig:simplebox_pretrain_zeroshot}
\end{figure}

Before investigating transfer, we verify each pre-training stage on the geometries it covers.
For LEMURS, this requires matching four physically distinct detectors with a single set of weights; for \textsc{SimpleBox}, covering a continuous grid of $(f_{\rm s},N_{\rm layers})$ configurations.
Throughout this section, generated and \texttt{Geant4} samples are compared on the three observables of \cref{ssec:method-metrics}.

\Cref{fig:lemurs_pretrain_observables} shows the LEMURS pre-training result.
A single backbone reproduces the three observables for all four Si/Sci detectors simultaneously, despite their different longitudinal extents, with the generated-to-\texttt{Geant4} ratio flat across the bulk of every distribution.
The largest deviations appear in the sparsely populated high-radius and high-energy tails, where the reference statistics themselves are limited.

For \textsc{SimpleBox}, we probe the conditioning grid directly.
We choose eight geometries as a dedicated validation set, simulated independently with \texttt{Geant4} ($10{,}000$ showers each, $80{,}000$ in total) at chosen points of the $(f_{\rm s},N_{\rm layers})$ grid.
Their exact configurations are absent from the training set, although its dense sampling of the same grid contains nearby values, so the comparison probes interpolation rather than memorisation.
\Cref{fig:simplebox_pretrain_observables} evaluates the pre-trained model on these held-out configurations: the first row varies the sampling fraction at fixed segmentation, the second the number of layers at fixed sampling fraction.
The generated distributions track \texttt{Geant4} along both conditioning axes: as $f_{\rm s}$ increases, the overall energy scale rises, and as $N_{\rm layers}$ increases, the longitudinal profile stretches. The model follows both trends without retraining.

Finally, \cref{fig:simplebox_pretrain_zeroshot} probes how the \textsc{SimpleBox} prior transfers to a held-out \textsc{SimpleBox}-like detector ($f_{\rm s}=0.035$, $N_{\rm layers}=35$).
Both \cref{fig:simplebox_pretrain_observables} and \cref{fig:simplebox_pretrain_zeroshot} therefore rely on held-out configurations; the difference is what they test.
The former evaluates the prior without any adaptation across eight points of the grid, which probes its quality directly. The latter instead treats a single in-range configuration as a downstream target and takes it through the full fine-tuning protocol, including a from-scratch baseline.
The zero-shot pre-trained model (pink) reproduces the \texttt{Geant4} longitudinal profile, radial profile, and cell-energy spectrum without any fine-tuning, consistent with interpolation between nearby training configurations at this in-range point.
The zero-shot prediction is accurate here, so fine-tuning preserves the \texttt{Geant4}-level agreement instead of rebuilding it. This holds across every fine-tuning dataset size $D$ shown, down to the smallest $D=10^2$: the generated observables show no visible dependence on $D$ over this range.
A model trained from scratch on the same detector lacks this prior and behaves differently. It is equally poor at $D=10^2$ and $D=10^3$ and only reaches agreement with the reference once $D=10^4$, a data-efficiency threshold rather than a gradual improvement with $D$.
The tail of the radial distribution makes the contrast concrete: at $D=10^{2}$ and $10^{3}$ the from-scratch model flattens the core of the profile and truncates it near $50\,\mathrm{mm}$, well short of the range covered by \texttt{Geant4}. The pre-trained model reproduces the full radial range at every $D$, and already at zero-shot. The two pre-trained models evaluated here are released~\cite{mgg_weights_allshowers,mgg_weights_pcfm} (\cref{app:code-data}).

\FloatBarrier

\subsection{Transfer of the SimpleBox prior across the LEMURS Si/Sci calorimeters}
\label{ssec:results-sb-to-lemurs}

With pre-training quality established, we now turn to the central transfer claim: a model pre-trained on the synthetic \textsc{SimpleBox} family can be adapted, with limited target data, to realistic detectors it never saw during pre-training.
The four Si/Sci LEMURS calorimeters differ in absorber material, sampling fraction, and longitudinal segmentation.

\begin{figure}[p]
    \centering
    \includegraphics[width=0.75\linewidth]{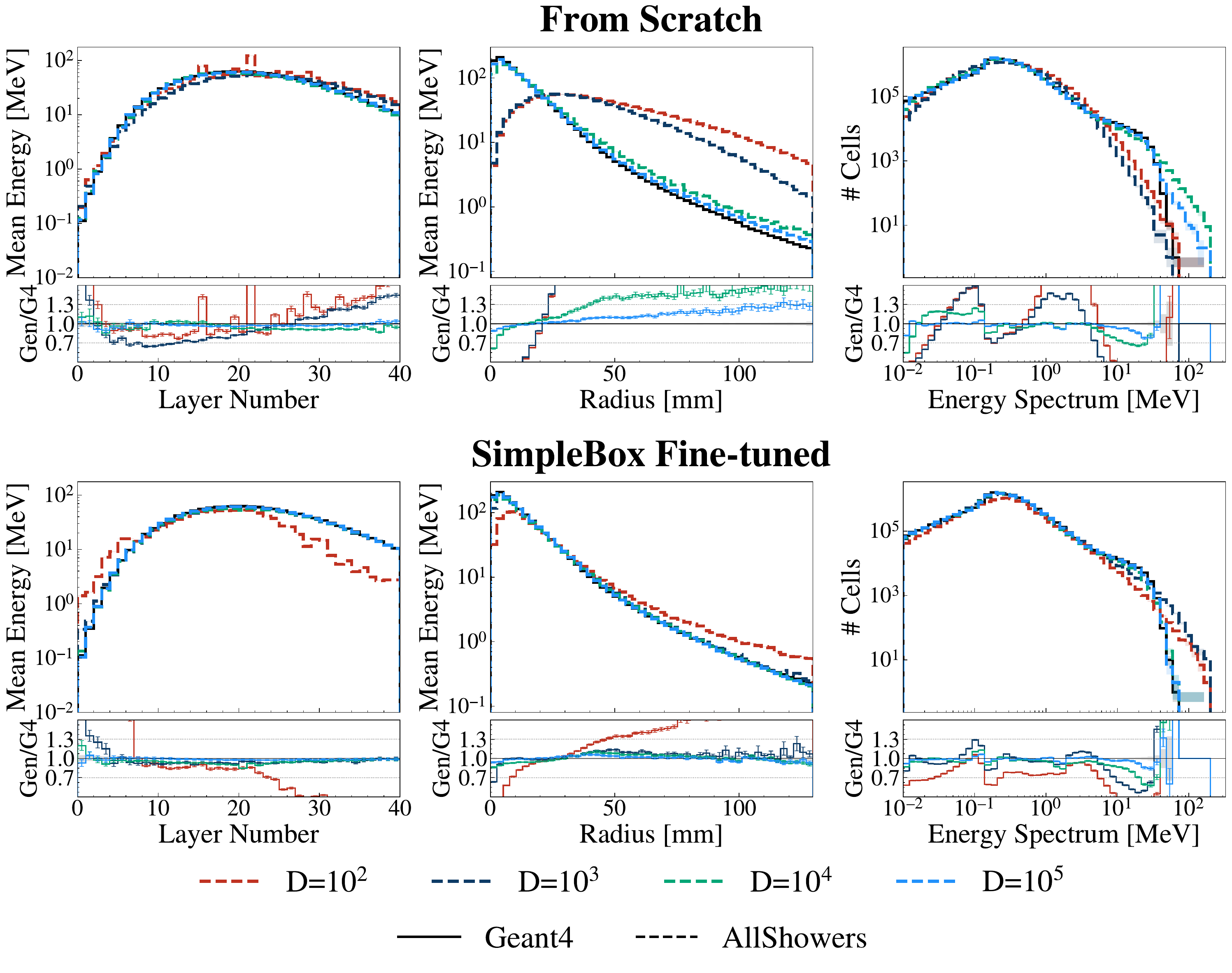}
    \caption{Transfer to FCCee-CLD ($N_{\rm layers}=40$): longitudinal profile (left), radial profile (centre) and cell-energy spectrum (right) for the from-scratch baseline (top) and the \textsc{SimpleBox}-pretrained model fine-tuned on the target (bottom), at fine-tuning sizes $D=10^{2}$--$10^{5}$, with generated distributions (dashed) overlaid on the \texttt{Geant4} reference (solid) and the generated-to-\texttt{Geant4} ratio below each panel. The no-PCFM counterpart, with per-layer point counts obtained from \texttt{Geant4} rather than \textsc{PointCountFM}, is shown in \cref{fig:simplebox_finetune_cld_g4}.}
    \label{fig:simplebox_finetune_cld}
\end{figure}

\begin{figure}[p]
    \centering
    \includegraphics[width=0.75\linewidth]{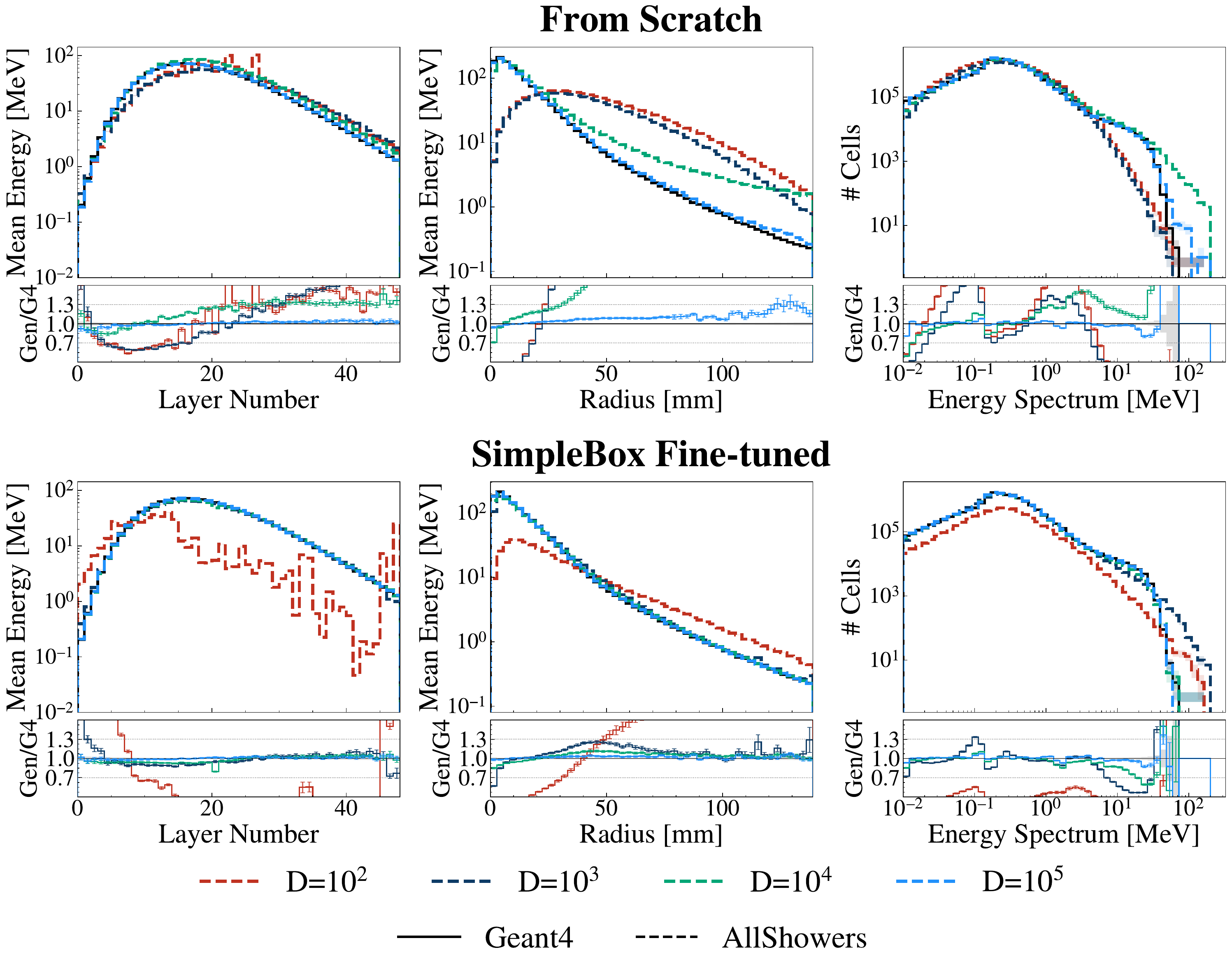}
    \caption{Transfer to ODD ($N_{\rm layers}=48$), with the same panel layout as \cref{fig:simplebox_finetune_cld}. The no-PCFM counterpart is shown in \cref{fig:simplebox_finetune_odd_g4}.}
    \label{fig:simplebox_finetune_odd}
\end{figure}

\begin{figure}[p]
    \centering
    \includegraphics[width=0.8\linewidth]{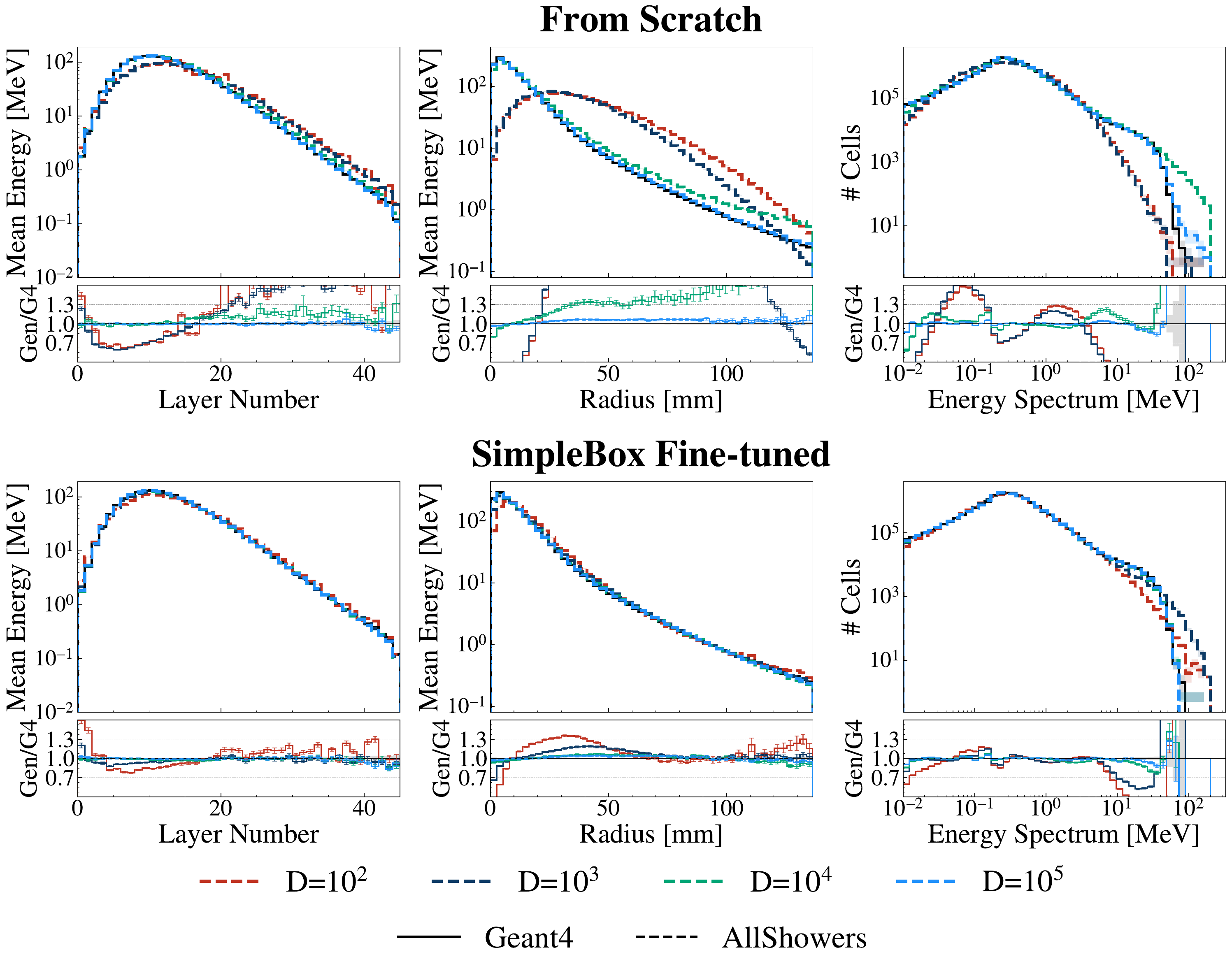}
    \caption{Transfer to Par04-SciPb ($N_{\rm layers}=45$), with the same panel layout as \cref{fig:simplebox_finetune_cld}. The no-PCFM counterpart is shown in \cref{fig:simplebox_finetune_par04scipb_g4}.}
    \label{fig:simplebox_finetune_par04scipb}
\end{figure}

\begin{figure}[p]
    \centering
    \includegraphics[width=0.8\linewidth]{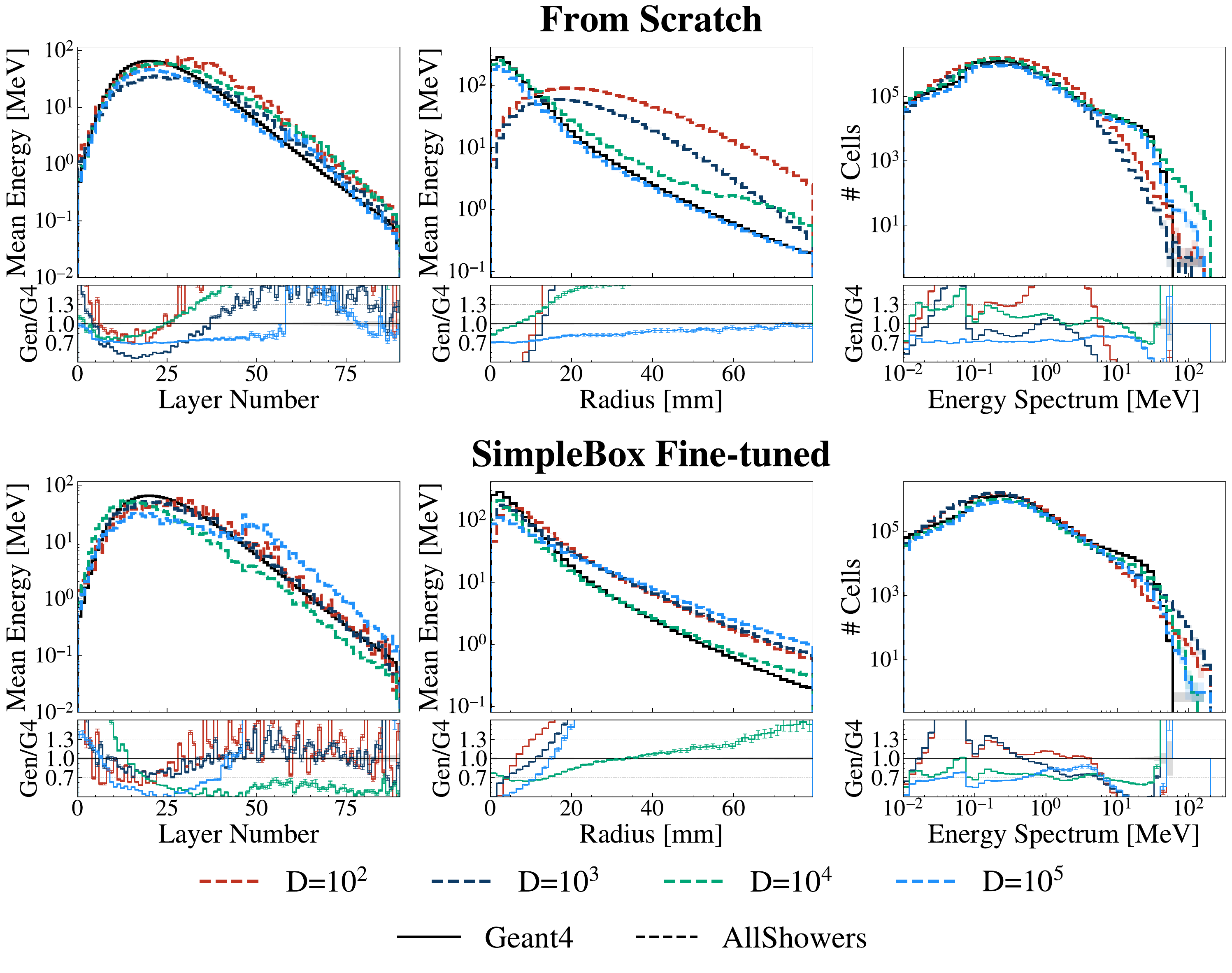}
    \caption{Transfer to Par04-SiW ($N_{\rm layers}=90$, the largest segmentation of the four targets), with the same panel layout as \cref{fig:simplebox_finetune_cld}. The no-PCFM counterpart is shown in \cref{fig:simplebox_finetune_par04siw_g4}.}
    \label{fig:simplebox_finetune_par04siw}
\end{figure}

\Cref{fig:simplebox_finetune_cld,fig:simplebox_finetune_odd,fig:simplebox_finetune_par04scipb,fig:simplebox_finetune_par04siw} show the longitudinal profile, radial profile, and cell-energy spectrum, one detector per figure, across the fine-tuning dataset sizes $D$ of the scaling study. Each overlays the \textsc{SimpleBox}-pretrained model fine-tuned on the target geometry (bottom rows) on the \texttt{Geant4} reference and on a from-scratch baseline trained on the same budget (top rows). The corresponding per-observable $W_1/\sigma_{\rm ref}$ distances to \texttt{Geant4} are tabulated in \cref{app:percalo-kl} (\cref{tab:sb-lemurs-w1}).
The geometric variation spanned synthetically by \textsc{SimpleBox} transfers to realistic detectors of different materials and segmentations. It reduces the per-detector \texttt{Geant4} sample needed to reach comparable agreement, with the detector-dependent exceptions discussed below.

The advantage of the prior is largest at small $D$: at $D=10^{3}$ the \textsc{SimpleBox}-pretrained model already reproduces the three observables on every target, where the from-scratch baseline is visibly poorer, and the two approaches become hard to separate by eye as $D$ grows. At $D=10^{2}$ the picture is detector dependent, as discussed below. The low-$D$ gain therefore comes from the prior rather than from the target data alone.


The generator produces the point coordinates and the per-layer point multiplicities in two separate stages (\cref{ssec:method-arch}); the figures in this section apply the \textsc{PointCountFM} (PCFM) correction to the multiplicity stage, while \cref{app:no-pcfm} repeats the same detectors with the per-layer counts taken directly from \texttt{Geant4}.
Comparing the two isolates the multiplicity stage as the less stable component of the generator: with the counts taken from \texttt{Geant4} the agreement degrades smoothly and monotonically as $D$ decreases. With PCFM, the degradation is larger and less regular at the smallest $D$. This is consistent with PCFM being trained only on the per-layer point counts of each shower.

The behaviour is detector dependent. Par04-SiW, with the largest segmentation ($N_{\rm layers}=90$), is the most demanding for the multiplicity stage. Its samples are generated without the per-layer bias correction (\cref{app:hyperparams}), and it shows the clearest low-$D$ decline in \cref{fig:simplebox_finetune_par04siw}, while its no-PCFM counterpart in \cref{fig:simplebox_finetune_par04siw_g4} remains close to the \texttt{Geant4} reference. The coordinate model therefore transfers well, and the residual sits in the multiplicity stage. 
In aggregate, this is also the target where pre-training most clearly fails to help at the largest $D$: the fine-tuned $\bar{d}$ rises well above the from-scratch value at $D=10^{5}$ (\cref{tab:sb-lemurs-w1}). The no-PCFM counterpart (\cref{fig:simplebox_finetune_par04siw_g4}) shows no such degradation, so we attribute the rise to the \textsc{PointCountFM} multiplicity stage rather than to the transfer itself.
Par04-SciPb is stable even with PCFM down to $D=10^{2}$, whereas FCCee-CLD and ODD show a PCFM decline at $D=10^{2}$ but are reproduced at every $D$ once the per-layer counts are taken from \texttt{Geant4} (\cref{fig:simplebox_finetune_cld_g4,fig:simplebox_finetune_odd_g4}).

\FloatBarrier

\subsection{Comparison of pre-training strategies on FCCee-ALLEGRO}
\label{ssec:results-allegro-transfer}

\begin{figure}[p]
    \centering
    \includegraphics[width=0.94\linewidth]{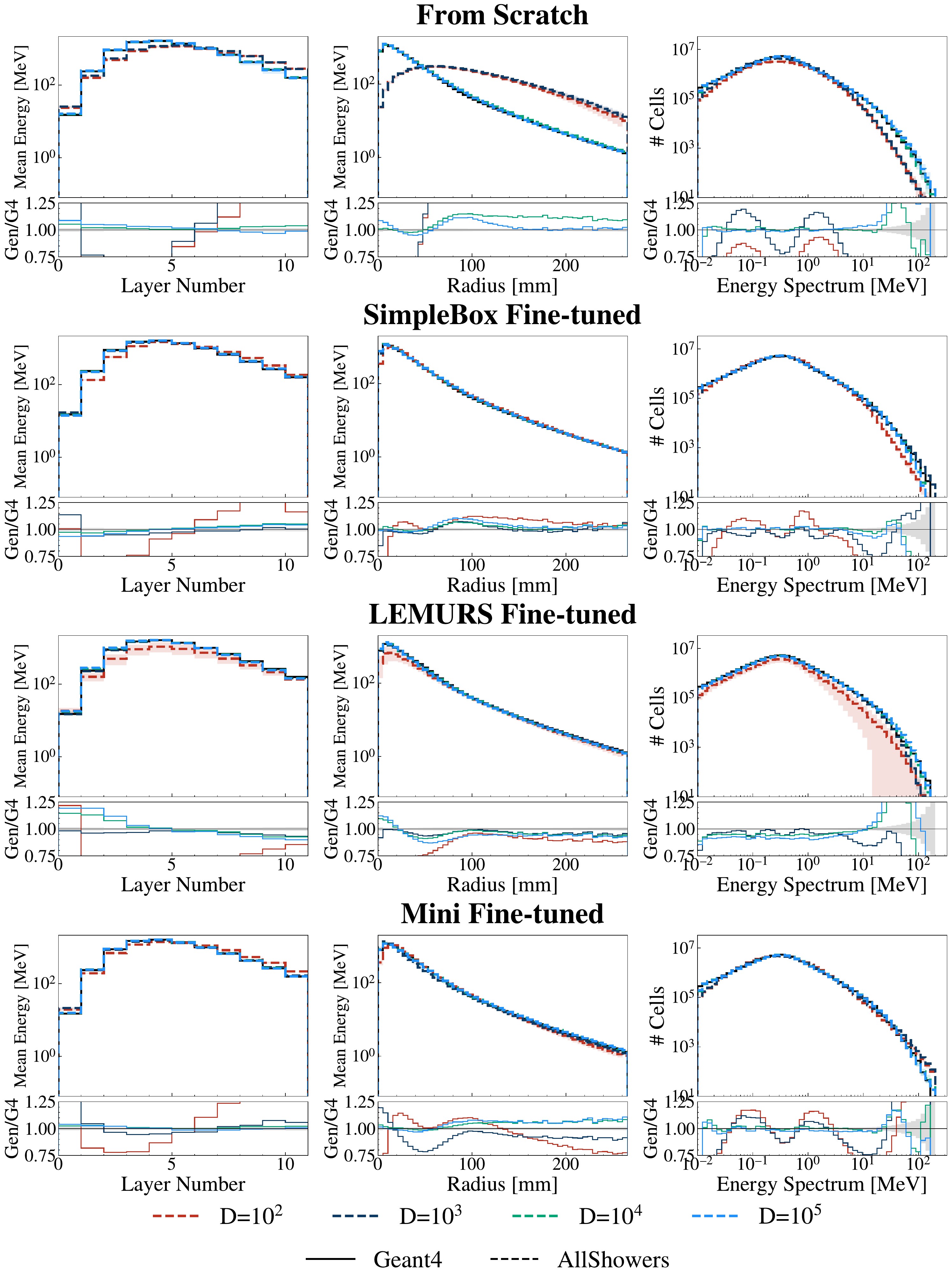}
    \caption{Transfer to FCCee-ALLEGRO ($N_{\rm layers}=11$): longitudinal profile (left), radial profile (centre) and cell-energy spectrum (right), one row per training strategy, with generated distributions overlaid on the \texttt{Geant4} reference; bands span the mean $\pm$ one standard deviation over five seeds, and the generated-to-\texttt{Geant4} ratio is shown below each panel. The no-PCFM counterpart, with per-layer point counts obtained from \texttt{Geant4} rather than \textsc{PointCountFM}, is shown in \cref{fig:observables_allegro_g4}.}
    \label{fig:observables_allegro}
\end{figure}

\begin{figure}[htbp]
    \centering
        \includegraphics[width=1\textwidth]{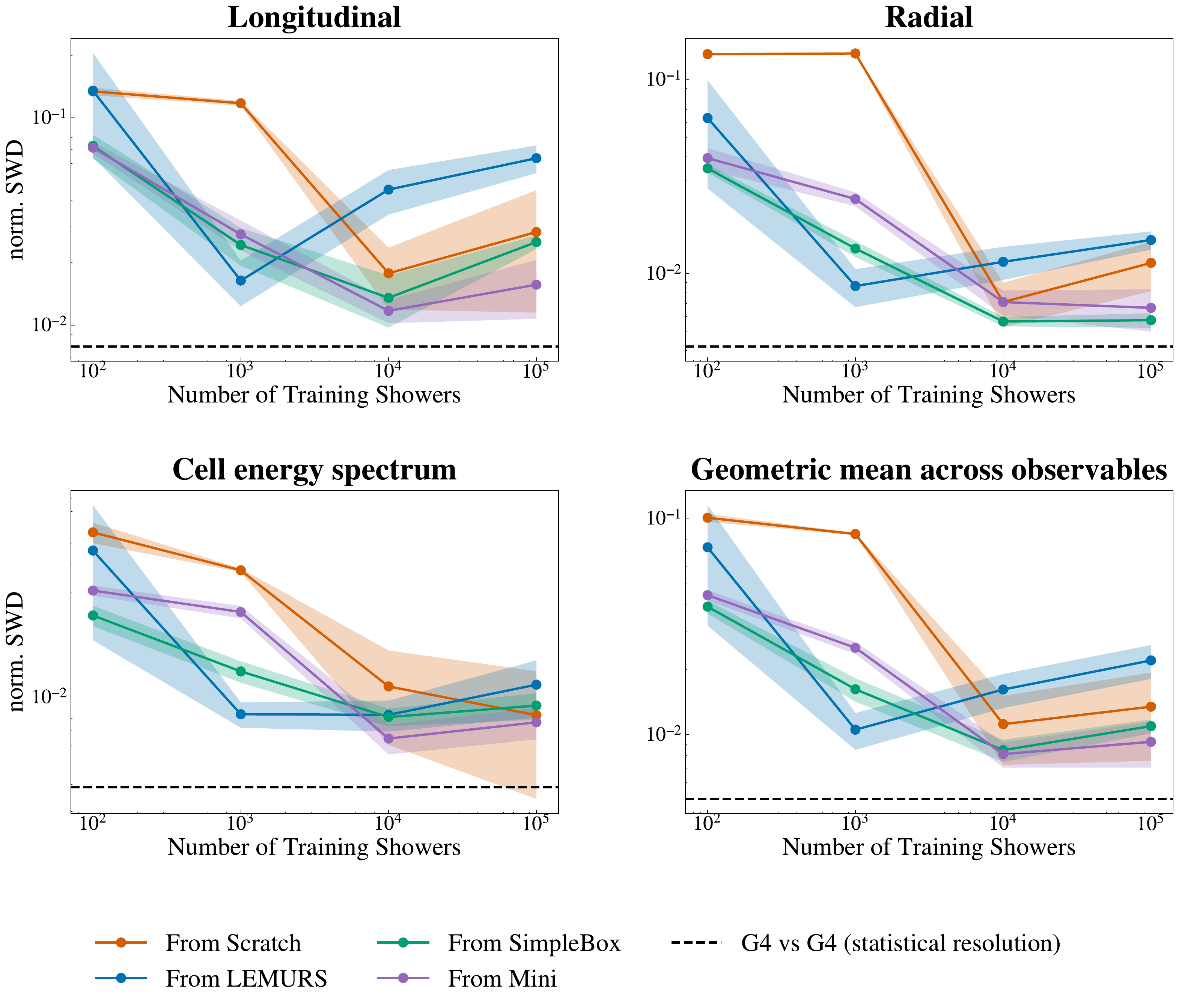}
    \caption{Sample efficiency of the strategies on FCCee-ALLEGRO, measured by the sliced Wasserstein distance (SWD) to \texttt{Geant4} (\cref{ssec:method-metrics}) as a function of the fine-tuning dataset size $D$. Panels show the three observables and their geometric mean $\bar{d}$. Bands span the mean $\pm$ one standard deviation over five seeds; the dashed line is the \texttt{Geant4}-versus-\texttt{Geant4} statistical resolution (\cref{ssec:method-metrics}). A Kullback--Leibler counterpart is given in \cref{fig:kl-allegro-pcfm}.}
    \label{fig:sample-efficiency}
\end{figure}

\begin{figure}[htbp]
    \centering
        \includegraphics[width=1\textwidth]{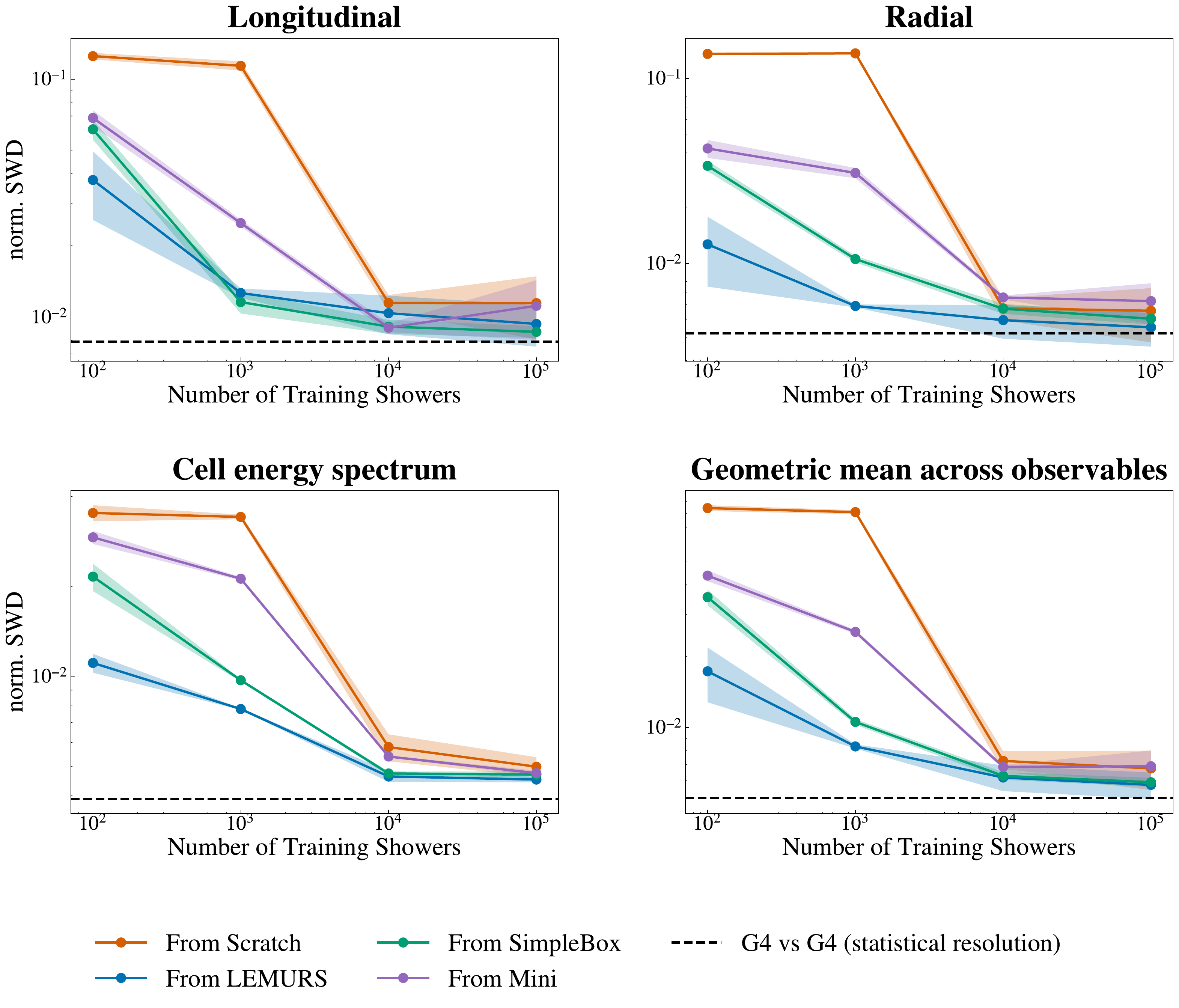}
    \caption{The sliced Wasserstein distances of \cref{fig:sample-efficiency}, recomputed with the per-layer point multiplicities taken from \texttt{Geant4} instead of from \textsc{PointCountFM}. Panels, bands and resolution line follow \cref{fig:sample-efficiency}. A Kullback--Leibler counterpart is given in \cref{fig:kl-allegro-g4cond}.}
    \label{fig:g4cond_ablation}
\end{figure}

\begin{figure}[htbp]
    \centering
        \includegraphics[width=0.85\textwidth]{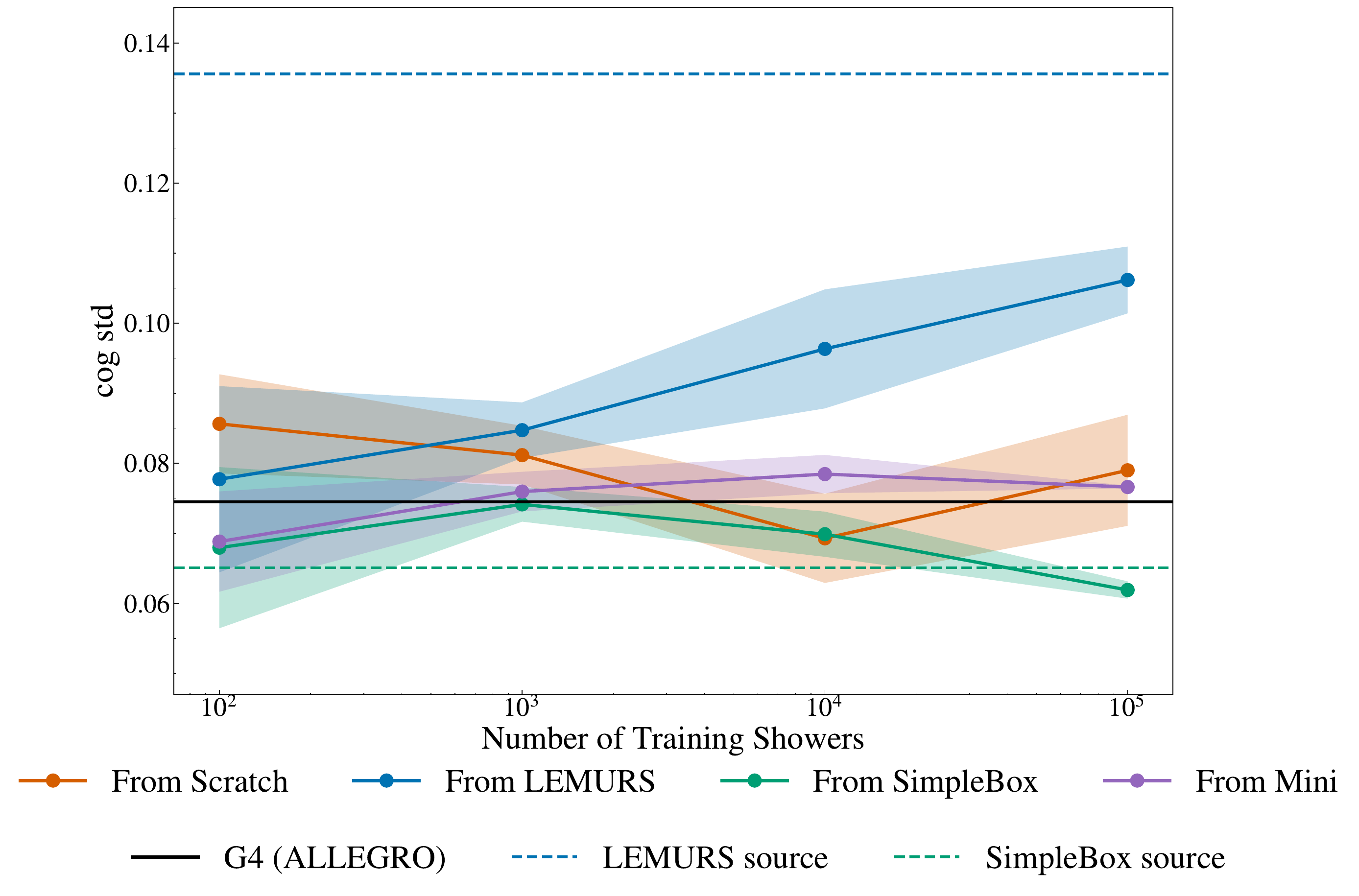}
    \caption{Drift of the generated multiplicity profile towards the pre-training prior. For each shower, the centre of gravity of the per-layer point counts is taken along the depth of the calorimeter, in units of fractional depth; the quantity plotted (cog std) is the standard deviation of that centre of gravity over the sample, that is, how much the depth of the occupancy profile varies from shower to shower. It is shown as a function of the fine-tuning dataset size $D$, with bands spanning the mean $\pm$ one standard deviation over five seeds. The solid black line is the \texttt{Geant4} FCCee-ALLEGRO reference; the dashed lines mark the same quantity for the LEMURS and \textsc{SimpleBox} pre-training pools. The model pre-trained on LEMURS moves away from the \texttt{Geant4} reference and towards its own pre-training value as $D$ grows. The \textsc{SimpleBox} arm moves the opposite way, towards its own narrower pool, and the from-scratch arm shows no trend.}
    \label{fig:cog_std_drift}
\end{figure}

ALLEGRO is the most demanding target: a liquid-argon calorimeter held out from both pre-training pools, with a coarser longitudinal segmentation ($N_{\rm layers}=11$) than any source detector and a sampling fraction of $0.162$, well above the range spanned by \textsc{SimpleBox}. Reaching it is extrapolation rather than interpolation.
We fine-tune on it from three priors (\textsc{SimpleBox}, LEMURS and \textsc{SimpleBox}-mini) and compare against training from scratch over the four budgets $D$, with five seeds each. \textsc{SimpleBox}-mini is a random $10^{5}$-shower subsample of the \textsc{SimpleBox} pool, corresponding to $2.5\%$ of the full sample (\cref{ssec:datasets-simplebox}).

\Cref{fig:observables_allegro} shows the three observables against \texttt{Geant4}.
At $D=10^{2}$ and $10^{3}$, the pre-trained models already reproduce them while the from-scratch baseline is visibly poorer, and the four arms (\cref{ssec:method-protocol}) become hard to separate by eye as $D$ grows; the quantitative comparison below shows that this visual agreement hides a residual difference. The \textsc{SimpleBox}-mini row shows that a prior pre-trained on $2.5\%$ of the pool still transfers, worse than the full one at the smallest budgets and matching it once $D$ reaches $10^{4}$ (\cref{app:simplebox-mini}), which \cref{ssec:results-breakeven} turns into a cost argument.

\Cref{fig:sample-efficiency} makes the comparison quantitative.
At $D=10^{3}$, LEMURS is the most sample-efficient prior, as one would expect from a pool of real detectors; at $D=10^{2}$, where every arm is data-starved, its longitudinal distance is comparable to from-scratch, and both \textsc{SimpleBox} priors are ahead. The trend is not monotonic: the LEMURS longitudinal distance falls to $0.016$ at $D=10^{3}$, then rises to $0.046$ at $10^{4}$ and $0.059$ at $10^{5}$, ending as the largest of the four arms. Beyond $D=10^{3}$, more target data makes this prior worse, the opposite of what the low-$D$ result alone would suggest. A pre-training imprint that extra target data fails to correct is not new in itself: it is reported as ossification in language model transfer~\cite{hernandez2021scalinglawstransfer}. Only one far-domain prior is available here, so we report the effect as a property of the LEMURS transfer and not of far-domain priors in general. The effect is largely invisible in the mean occupancy profile, so only a distributional metric exposes it (\cref{app:width-drift}).

\Cref{fig:g4cond_ablation} locates the effect. Repeating the measurement with the per-layer point counts taken from \texttt{Geant4} instead of from \textsc{PointCountFM} removes the rise entirely: the LEMURS longitudinal distance now falls monotonically, from $0.031$ at $D=10^{2}$ to $0.013$, $0.010$ and $0.0088$, against a statistical resolution of $0.0079$, and every other arm behaves the same way. Both sides use the same backbone checkpoints, so neither the coordinate model nor under-training accounts for the degradation. It belongs to the multiplicity stage, consistent with what \cref{ssec:results-sb-to-lemurs} already showed on the LEMURS detectors. Any residual difference between the mini and full \textsc{SimpleBox} priors at large $D$ also disappears once every arm is given the same occupancy, so it too came from the count model and not from the prior. Both observations survive a change of metric: a Kullback--Leibler divergence pooled over showers, in place of the per-shower sliced Wasserstein distance, reproduces the longitudinal LEMURS rise and its removal by the \texttt{Geant4}-count ablation (\cref{app:kl-allegro}).

To see what the count model does with the extra data, we work with the per-layer point counts directly, the quantity \textsc{PointCountFM} predicts. For shower $i$, let $n_{i,\ell}$ be the number of points in layer $\ell$, with $\ell\in\{0,\dots,N_{\rm layers}-1\}$ as in \cref{sec:datasets}, and let $z_\ell=(\ell+0.5)/N_{\rm layers}$ be the depth of that layer on a scale running from $0$ at the front face to $1$ at the back. The shower sits at the count-weighted mean depth
\begin{equation}
    \label{eq:cog}
    c_i = \frac{\sum_{\ell} n_{i,\ell}\, z_\ell}{\sum_{\ell} n_{i,\ell}},
\end{equation}
where dividing by the total number of points removes the overall size of the shower and leaves the shape of its count profile. The observable of \cref{fig:cog_std_drift}, the spread of this centre of gravity (cog std), is the standard deviation of $c_i$ over the shower sample: how much the depth of the shower fluctuates from one shower to the next. It is a second moment of that profile, and unlike the three observables of \cref{ssec:method-metrics} it is weighted by point counts rather than by energy.

\Cref{fig:cog_std_drift} shows this second moment as a function of $D$. For the model pre-trained on LEMURS, it drifts away from \texttt{Geant4} and towards the value of the LEMURS pool, from $0.078$ at $D=10^{2}$ to $0.106$ at $10^{5}$, against $0.0745$ for \texttt{Geant4} and $0.136$ for the pool itself. That is about half the distance between the two, and all five seeds move the same way. The \textsc{SimpleBox} arm is the control: it uses an identical fine-tuning recipe and differs only in its initialisation. Its pre-training pool is narrower than ALLEGRO, not wider. It drifts in the opposite direction, ending at $0.062$ at $D=10^{5}$, just below its own pool value of $0.065$. This is consistent with the source, not the recipe, setting the sign of the drift. The mini arm, whose prior comes from the same \textsc{SimpleBox} pool but from $2.5\%$ of it, shows no comparable drift and stays close to the \texttt{Geant4} value, consistent with a weaker prior leaving a weaker imprint (\cref{app:width-drift}). We find no corresponding movement in the first moments or the total point counts, so the change is in the second moments.

We analyse this width drift in more detail in \cref{app:width-drift}. Freezing part of the pre-trained core reduces the amplitude of the drift, but at a cost to the radial profile and with no benefit at the smallest $D$, so it is a mitigation rather than a solution. Freezing recipes of this kind are a standard response to the feature distortion that fine-tuning induces in discriminative transfer~\cite{kumar2022finetuningdistortpretrainedfeatures}.

The prior therefore buys sample efficiency in the low-$D$ regime that matters for a new detector, while at large $D$ the limiting component is the count model rather than the transfer itself.

\FloatBarrier

\subsection{Compute-budget analysis for an ALLEGRO-like downstream detector}
\label{ssec:results-breakeven}

A pre-trained prior is worthwhile only if its one-off cost is repaid across the detectors it is later applied to. We frame this as a simple break-even. Let $P$ be the one-off pre-training cost, $S$ the cost of training a generator from scratch for a single downstream detector, and $F$ the cost of fine-tuning the pre-trained model on that detector. Producing generators for $N$ detectors costs $N \times S$ from scratch, against $P + N \times F$ with pre-training, so the two are equal at
\begin{equation}
\label{eq:breakeven}
N^{\star} = \frac{P}{S - F},
\end{equation}
and pre-training is the cheaper route for $N > N^{\star}$.

\begin{figure}[htbp]
    \centering
    \includegraphics[width=\linewidth]{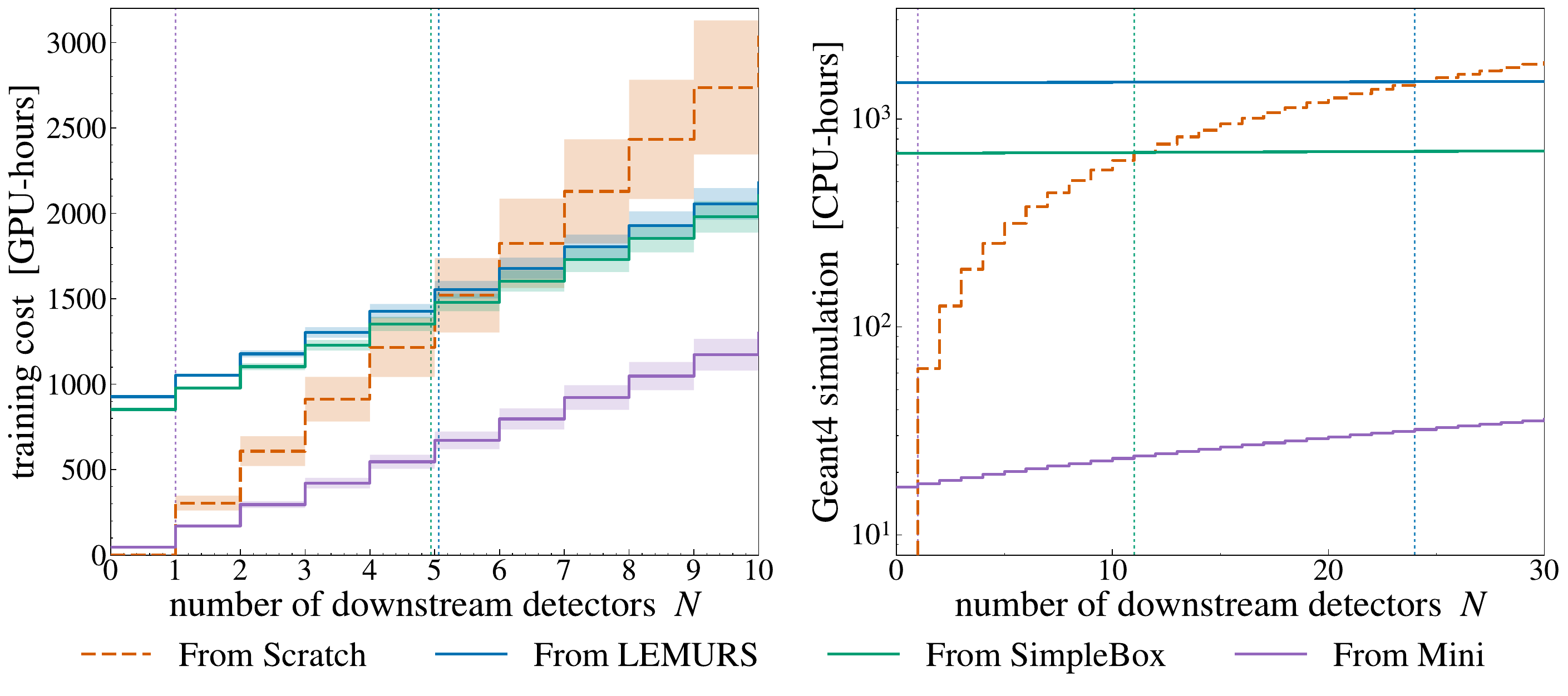}
    \caption{Cumulative cost of producing generators for $N$ ALLEGRO-like downstream detectors: generator training (left, GPU-hours) and \texttt{Geant4} simulation of the training data (right, CPU-hours, logarithmic scale). The left band is the per-detector five-seed standard deviation scaled linearly with $N$; the single-run pre-training cost carries no band. The right panel shows the as-produced simulation totals without a band, since \texttt{Geant4} throughput varies with the compute node and no single uncertainty applies. Dotted lines mark the break-even: the mini prior pays off within the first detector, the two full pools at about five detectors in GPU-hours (indistinguishable within the seed spread) and at eleven and twenty-four in CPU-hours. Per-detector costs are those measured on FCCee-ALLEGRO and taken equal for every subsequent detector.}
    \label{fig:breakeven-cost}
\end{figure}

Costs are measured on the production runs. A training cost in GPU-hours is the wall-clock training time up to the delivered checkpoint, multiplied by the number of A100 GPUs run in parallel (twelve for the full pre-trainings, four for the fine-tunings and the from-scratch runs). The delivered checkpoint is the one with the lowest validation loss. The reported $F$ and $S$ are the mean over five seeds. Data production is quoted in CPU-hours of single-threaded \texttt{Geant4} simulation. PCFM adds below one per cent and is included. All numbers refer to an ALLEGRO-like target: the per-detector cost depends on the detector geometry, the incident particle and the energy range. The absolute GPU-hours and CPU-hours are specific to the hardware used here and scale differently on other systems; what carries over is the break-even in the number of detectors, a ratio of costs measured on the same setup. \Cref{fig:breakeven-cost} shows the cumulative cost and \cref{tab:compute_budget} collects the budget.

\begin{table}[htbp]
    \centering
    \caption{Measured compute budget for an ALLEGRO-like downstream target. Training in GPU-hours (wall-clock $\times$ number of A100 GPUs); data production in CPU-hours of single-threaded \texttt{Geant4} simulation, summed over the production runs. PCFM is included (below one per cent of each total). Break-even counts $N^{\star}$ follow \cref{eq:breakeven} with the per-detector costs below; pre-training is the cheaper route for $N>N^{\star}$. Per-detector GPU-hour uncertainties are the standard deviation over five seeds; the CPU-hour totals are as-produced sums, since \texttt{Geant4} simulation throughput varies with detector geometry and, over a multi-month production campaign, with the software and cluster conditions at simulation time.}
    \label{tab:compute_budget}
    \small
    \begin{tabular}{lccccc}
    \toprule
    & \textbf{Showers} & \textbf{\texttt{Geant4} sim} & \textbf{Training} & \multicolumn{2}{c}{\textbf{Break-even $N^{\star}$}} \\
    & & [CPU-h] & [GPU-h] & GPU-h & CPU-h \\
    \midrule
    \multicolumn{6}{l}{\emph{One-off pre-training}} \\
    \textsc{SimpleBox}-mini & $10^{5}$        & 17   & 45  & 0.3 & 0.3 \\
    \textsc{SimpleBox}      & $4\times10^{6}$ & 683  & 852 & 4.8 & 10.9 \\
    \textsc{LEMURS}         & $4\times10^{6}$ & 1494 & 927 & 5.2 & 23.9 \\
    \midrule
    \multicolumn{6}{l}{\emph{Per downstream detector}} \\
    fine-tune ($D=10^{3}$)    & $10^{3}$ & 0.6 & $125 \pm 10$ & & \\
    from scratch ($D=10^{5}$) & $10^{5}$ & 63  & $304 \pm 44$ & & \\
    \bottomrule
    \end{tabular}
\end{table}

On the ALLEGRO target, fine-tuning matches the from-scratch agreement with a hundred times fewer target showers ($D=10^{3}$ against $D=10^{5}$, \cref{ssec:results-allegro-transfer}). In GPU-hours, the gap is a factor of $2.4$, $F = 125 \pm 10$ against $S = 304 \pm 44$ over five seeds: the fine-tune spends many inexpensive low-$D$ epochs rather than converging in fewer steps (\cref{app:hyperparams}). \Cref{eq:breakeven} then places the break-even at $N^{\star}\!\approx\!0.3$ for the mini prior and $\approx\!5$ for both full pools (\cref{fig:breakeven-cost}, left). The per-step cost, memory footprint, and convergence of the two full pre-trainings are comparable (\cref{tab:computational_cost_pretrain}).
In \texttt{Geant4} CPU time, a from-scratch training requires 63 CPU-hours of ALLEGRO simulation and a fine-tune 0.6; the one-off cost of a prior is the simulation of its own pool, which pushes the break-even to eleven and twenty-four detectors for the \textsc{SimpleBox} and LEMURS pools, respectively (\cref{tab:compute_budget}).
Two caveats follow from counting detectors this way. First, $F$ and $S$ are each measured once, on FCCee-ALLEGRO, and \cref{eq:breakeven} then assumes every further detector costs the same $F$. Detectors differ, and in both directions. A target inside the pre-training region can need little or no fine-tuning: the held-out \textsc{SimpleBox}-like configuration of \cref{fig:simplebox_pretrain_zeroshot} is already accurate zero-shot, so $F\!\to\!0$ and the prior is repaid immediately. A far-domain target such as ALLEGRO, whose sampling fraction sits outside the pre-training region, needs the full budget quoted here. The break-even is therefore a cost model for an ALLEGRO-like target, not a measurement across $N$ detectors. Second, the counting is most useful during detector development, where a single calorimeter is re-optimised many times, and each geometry variant needs its own generator: there $N$ counts design iterations rather than distinct experiments, with many fine-tuned showers produced per iteration.
The pre-training cost $P$ scales with the size of the pre-training set, so a smaller pool lowers both $P$ and the break-even $N^{\star}$, as long as the smaller prior still transfers. We probe this with \textsc{SimpleBox}-mini (\cref{ssec:datasets-simplebox}), pre-trained with the same architecture and protocol and then fine-tuned to FCCee-ALLEGRO. The \textsc{SimpleBox}-mini row of \cref{fig:observables_allegro} shows that the smaller prior does transfer: the longitudinal, radial, and cell-energy distributions track \texttt{Geant4} across the fine-tuning sizes $D$. It is not free, however: at $D=10^{3}$ the reduced pool performs worse than the full one, and \cref{app:simplebox-mini} sets the two side by side.

\FloatBarrier

\section{Conclusions}
\label{sec:conclusions}

We have studied whether a single point cloud calorimeter shower generator can be pre-trained once and transferred to detector geometries it never saw during training, and whether that pre-training can be built from synthetic geometric variation rather than real-detector data. Using a geometry-aware conditioning of the \textsc{AllShowers} backbone, we benchmarked a purely synthetic prior, the \textsc{SimpleBox} family of $10^{4}$ box calorimeters, against pre-training on the realistic \textsc{LEMURS} detectors at a matched $4\times10^{6}$-shower budget. On the held-out FCCee-ALLEGRO calorimeter, both priors beat training from scratch at $D=10^{3}$: the sliced Wasserstein distance to \texttt{Geant4}, aggregated over the three observables by the geometric mean $\bar{d}$ of \cref{eq:dbar}, drops by factors of $5.2$ and $8.0$ respectively.

Synthetic \textsc{SimpleBox} pre-training is competitive with the realistic LEMURS prior: it performs slightly worse at $D=10^{3}$ and better at larger target sizes, where the LEMURS prior degrades. On this target, geometric diversity is enough; detector realism is not what buys the transfer. For an ALLEGRO-like target, a compute-budget analysis places the break-even at about five downstream detectors in GPU-hours. A reduced $10^{5}$-shower pre-training ($2.5\%$ of the pool) still beats training from scratch at every fine-tuning size, though it performs worse than the full pool at $D=10^{3}$. Synthetic geometric variation is therefore a practical and inexpensive pre-training strategy for fast calorimeter simulation, and it needs no data from realistic detectors.

The study has clear limitations. 
It is restricted to electromagnetic, photon-induced showers. Hadronic showers, which dominate the simulation cost, are not addressed.
The geometry is encoded through only two scalars, the sampling fraction $f_{\rm s}$ and the number of layers $N_{\rm layers}$. They fix the energy-deposition scale and the longitudinal sampling, but not the material-dependent lateral (Moli\`ere) scale, which is recovered only during fine-tuning. The transfer is shown across five detectors of different materials and segmentations. The most extreme extrapolation, to a noble-liquid calorimeter whose sampling fraction lies far outside the pre-training region, is probed on one target only, FCCee-ALLEGRO.
Natural extensions are a pre-training prior covering all particle species and a more general conditioning on the full calorimeter description, towards a genuinely zero-shot generator.
Because the \textsc{SimpleBox} pool is sampled from a continuous range in sampling fraction rather than from a fixed set of geometries, it can be pushed to wider coverage to strengthen the prior and extend zero- and few-shot transfer to detectors further from the pre-training region. Reaching that goal will likely also require a synthetic pre-training set that spans several calorimeter geometry types, not only the box-shaped \textsc{SimpleBox} family, so that the prior is not tied to a single shape.

\acknowledgments
We thank Ivo Schulthess for valuable comments on the manuscript.

\paragraph*{Funding Information} 
This research was supported in part by the Maxwell computational resources operated at Deutsches
Elektronen-Synchrotron DESY, Hamburg, Germany.
This project has received funding from the
European Union’s Horizon 2020 Research and Innovation programme under Grant Agreement No
101004761. We acknowledge support by the Deutsche Forschungsgemeinschaft under Germany’s
Excellence Strategy – EXC 2121 Quantum Universe – 390833306 and via the KISS consortium
(05D23GU4, 13D22CH5) funded by the German Federal Ministry of Research, Technology, and
Space (BMFTR) in the ErUM-Data action plan.
P.M. has benefited from support by the CERN Strategic R$\&$D Programme on Technologies for Future Experiments~\cite{EPRD}.

\newpage
\appendix
\section{Code, weights and data availability}
\label{app:code-data}

The code and the pre-trained weights are collected in \cref{tab:code_release}: the \texttt{ddfastsim} simulation package~\cite{ddfastsim}, the \textsc{SimpleBox} and \textsc{LEMURS} dataset production~\cite{mgg_code_dataset}, the \textsc{AllShowers} backbone~\cite{mgg_code_allshowers}, and \textsc{PointCountFM}~\cite{mgg_code_pcfm}, the network that predicts the number of points in each layer. The pre-trained weights are under $2~\mathrm{MB}$ per model and are released on Hugging Face~\cite{mgg_weights_allshowers,mgg_weights_pcfm}, together with the fitted input transformations needed to sample from them.

\begin{table}[htbp]
    \centering
    \caption{Released code repositories and pre-trained model weights.}
    \label{tab:code_release}
    \small
    \begin{tabular}{lll}
    \toprule
    \textbf{Repository} & \textbf{Contents} & \textbf{Availability} \\
    \midrule
    \texttt{ddfastsim} & \texttt{Geant4}/DDG4 fast-simulation plugins & Ref.~\cite{ddfastsim} \\
    \texttt{multi-calorimeter-dataset} & dataset production and processing & Ref.~\cite{mgg_code_dataset} \\
    \texttt{AllShowers} & shower model training and generation & Ref.~\cite{mgg_code_allshowers} \\
    \texttt{PointCountFM} & per-layer point count model & Ref.~\cite{mgg_code_pcfm} \\
    \midrule
    \textsc{AllShowers} weights & \textsc{SimpleBox} and \textsc{LEMURS} pre-trainings & Ref.~\cite{mgg_weights_allshowers} \\
    \textsc{PointCountFM} weights & \textsc{SimpleBox} and \textsc{LEMURS} pre-trainings & Ref.~\cite{mgg_weights_pcfm} \\
    \bottomrule
    \end{tabular}
\end{table}

\Cref{tab:data_release} lists the released datasets, point clouds of \texttt{Geant4} steps in the schema of \cref{sec:datasets}.
All datasets are published as a single record in the Universit\"at Hamburg research data repository~\cite{mgg_datasets}, under \href{https://doi.org/10.25592/uhhfdm.19103}{DOI \texttt{10.25592/uhhfdm.19103}}. The record also carries a datasheet and a file manifest with per-file checksums, together with the held-out evaluation sets used in this work: a disjoint-seed $10^{4}$-shower test set for each LEMURS detector and for ALLEGRO, the eight held-out \textsc{SimpleBox} geometries ($8\times10^{4}$ showers in total), and the $10^{5}$-shower \textsc{SimpleBox} zero-shot grid sample. The record totals $465\,\mathrm{GB}$ over 16 files.
The synthetic \textsc{SimpleBox} pool and its $10^{5}$-shower subsample \textsc{SimpleBox-mini} are released here for the first time.
The public release of the \textsc{LEMURS} calorimeters~\cite{McKeown:2025gtw} is voxelised. The point cloud re-simulation used here is a separate artefact, released per detector. The optimal-transport coupling noise used at training time is regenerable and is not released.

\begin{table}[htbp]
    \centering
    \caption{Released datasets and their sizes, for the step-level point cloud HDF5 files of \cref{sec:datasets}; the \textsc{LEMURS} calorimeters are those of Ref.~\cite{McKeown:2025gtw}, re-simulated here in point cloud form. All datasets are published in the single record of Ref.~\cite{mgg_datasets}. The four entries total $452\,\mathrm{GB}$, with \textsc{SimpleBox-mini} a subsample of the \textsc{SimpleBox} dataset rather than additional data.}
    \label{tab:data_release}
    \small
    \begin{tabular}{llr}
    \toprule
    \textbf{Dataset} & \textbf{Showers} & \textbf{Size} \\
    \midrule
    \textsc{SimpleBox} dataset       & $4\times10^{6}$ & 164\,GB \\
    \textsc{SimpleBox-mini}          & $10^{5}$        & 4.1\,GB \\
    \textsc{LEMURS} (4 Si/Sci)       & $4\times10^{6}$ & 257\,GB \\
    FCCee-ALLEGRO                    & $10^{5}$        & 27\,GB  \\
    \bottomrule
    \end{tabular}
\end{table}

\FloatBarrier

\section{Training hyperparameters}
\label{app:hyperparams}

We summarise the optimisation and sampling settings of the two stages of
\textsc{AllShowers}~\cite{Buss:2026yrf}: the transformer that generates
the point coordinates
(\cref{tab:allshowers_hyperparameters,tab:allshowers_method_hyperparameters})
and the \textsc{PointCountFM} network that sets the number of points in
each layer (\cref{tab:pcfm_hyperparameters}). The architecture and the
geometry conditioning are described in \cref{sec:method} and are not
repeated here. Settings taken unchanged from Ref.~\cite{Buss:2026yrf}
are marked as inherited.

\begin{table}[htbp]
\centering
\caption{Coordinate (transformer) stage of \textsc{AllShowers}: training
and sampling settings, common to all runs. The learning rate and the
optimiser-step budget, which vary with the strategy and with the training-set
size $D$, are given in \cref{tab:allshowers_method_hyperparameters}.}
\label{tab:allshowers_hyperparameters}
\small
\begin{tabularx}{\linewidth}{lX}
\toprule
\textbf{Setting} & \textbf{Value} \\
\midrule
Optimiser & Ranger (RAdam with Lookahead), inherited \\
Learning-rate schedule & cosine to zero, stepped per epoch, no warmup \\
Batch size & $256$ global, divided across the GPUs ($64$ per GPU in the four-GPU downstream runs) \\
Weight decay & $10^{-2}$ \\
Gradient clipping & maximum norm $0.2$ \\
Moving average & none \\
Precision & single precision (\textsc{fp32}) \\
Device & NVIDIA A100, $4$ GPUs for fine-tuning, $12$ for pre-training \\
Training objective & conditional flow (velocity) matching with an exact optimal-transport coupling, inherited \\
Sampler & fixed-step midpoint solver (second order), $16$ steps \\
\bottomrule
\end{tabularx}
\end{table}

The coordinate stage is trained by conditional flow matching with an exact
optimal-transport coupling between the prior and the data points
(\cref{tab:allshowers_hyperparameters}). The lowest-validation-loss
checkpoint is kept, and a shower is generated by integrating the learned
velocity field with a fixed-step midpoint solver.

\begin{table}[htbp]
\centering
\caption{Coordinate-stage optimiser budget. The pre-trainings run on the
full pool. The downstream budget is the same for all four strategies, which
differ only in the starting learning rate: $10^{-3}$ for from-scratch
training and $10^{-4}$ for every fine-tuning, both annealed cosine to zero.
Each run stops at its epoch ceiling or at a $24$\,h wall-clock limit,
whichever comes first; the steps per epoch are $\lfloor n_{\rm
train}/256\rfloor$, with at least one batch per epoch. Downstream, the
training partition is $n_{\rm train}=D$, with the fixed
$10^{4}$-shower validation slice beyond it; at $D=10^{5}$ the
$10^{5}$-shower fine-tuning pool caps $n_{\rm train}$ at
$9\times10^{4}$.}
\label{tab:allshowers_method_hyperparameters}
\small
\begin{tabular}{lcc}
\toprule
\textbf{Run} & \textbf{Epochs} & \textbf{Optimiser steps} \\
\midrule
\textsc{SimpleBox} pre-training      & ${\sim}143$ & ${\sim}2.1\times10^{6}$ \\
LEMURS pre-training         & ${\sim}113$ & ${\sim}1.7\times10^{6}$ \\
\textsc{SimpleBox}-mini pre-training & $400$       & $1.4\times10^{5}$ \\
\midrule
Downstream, $D=10^{2}$ & $500$           & $500$ \\
Downstream, $D=10^{3}$ & $500$           & $1500$ \\
Downstream, $D=10^{4}$ & $140$ to $420$  & $6$ to $16\times10^{3}$ \\
Downstream, $D=10^{5}$ & $37$ to $72$    & $13$ to $25\times10^{3}$ \\
\bottomrule
\end{tabular}
\end{table}

The four downstream strategies, from-scratch training and the three
fine-tunings, share this budget and differ only in the starting learning
rate. The two smallest sizes complete the full $500$-epoch schedule, whereas $D=10^{4}$
and $10^{5}$ reach the $24$\,h limit first. The budget is the same across
the five downstream geometries, which vary only in the number of layers
$N_{\rm layers}$ and in the loaded checkpoint. The ALLEGRO results are means over five random seeds; the Si/Sci results are single runs.

\begin{table}[htbp]
\centering
\caption{\textsc{PointCountFM} stage of \textsc{AllShowers}: training
settings and the per-layer point count correction. The network sets the
number of points in each layer; several values vary with $D$ and with the
target detector.}
\label{tab:pcfm_hyperparameters}
\small
\begin{tabularx}{\linewidth}{lX}
\toprule
\textbf{Setting} & \textbf{Value} \\
\midrule
Optimiser & Adam for pre-training, AdamW downstream \\
Learning-rate schedule & one-cycle for pre-training, cosine (with warm restarts where used) downstream \\
Learning rate & $10^{-4}$ pre-training, $5\times10^{-5}$ to $10^{-3}$ downstream, set per $D$ and detector \\
Batch size & $16$, $64$, $256$, $1024$ for $D=10^{2}$, $10^{3}$, $10^{4}$, $10^{5}$ \\
Epochs & up to $5000$ (\textsc{SimpleBox}) or $400$ (LEMURS) pre-training and $3000$ downstream, with early stopping \\
Precision and device & single precision (\textsc{fp32}), one A100, no multi-GPU \\
\bottomrule
\end{tabularx}
\end{table}

The counts predicted by \textsc{PointCountFM} are corrected at generation time by a per-layer
multiplicative factor
$b_\ell=\langle n_\ell^{\rm real}\rangle/\langle n_\ell^{\rm gen}\rangle$,
the ratio of the mean true count to the mean generated count in layer
$\ell$. The factors are fitted on the training partition alone, separately
for each detector, strategy and training size, and averaged over five
incident-energy percentile bins. Inactive layers are set to $b_\ell=1$ and the factors are
clamped to $[0.5,2.0]$. Each predicted count is then rescaled by its factor
and rounded to the nearest non-negative integer. A per-energy-bin scalar
and a two-dimensional variant of the correction are implemented but are
not used for the production samples. Par04-SiW is generated without any
correction.

\Cref{tab:computational_cost_pretrain} reports the computational cost of the two full pre-trainings. The two backbones are similar in parameter count, per-step cost, memory footprint and steps to convergence. The choice of pre-training pool, synthetic \textsc{SimpleBox} or realistic \textsc{LEMURS}, therefore does not change the training budget.

\begin{table}[htbp]
    \centering
    \caption{Computational cost of the two pre-training strategies, measured on a single NVIDIA A100 80GB PCIe GPU with batch size 64.
    Per-step timings averaged over 3 epochs after 1 warmup epoch (epoch = 1{,}546 batches over a $10^{5}$-shower subsample of the pre-training dataset).
    Convergence steps reported under the production DDP protocol (effective global batch 256, 12 ranks of A100-SXM4-40GB).}
    \label{tab:computational_cost_pretrain}
    \small
    \begin{tabular}{lcccc}
    \toprule
    \textbf{Method} & \textbf{Params} & \textbf{ms/step} & \textbf{Peak VRAM (GB)} & \textbf{Conv. steps ($\times 10^3$)} \\
    \midrule
    \textsc{SimpleBox}-pretrain & 243\,846 & 221.4 $\pm$ 15.6$^{\dagger}$ & 6.7 & $\sim$1840 \\
    \textsc{LEMURS}-pretrain    & 269\,766 & 243.2 $\pm$ 12.3             & 8.2 & $\sim$1570 \\
    \bottomrule
    \end{tabular}
    \vspace{1mm}

    \footnotesize{$^{\dagger}$ Measured on a single A100 80GB PCIe; production pre-training ran on A100-SXM4-40GB under DDP.}
\end{table}

\FloatBarrier

\section{Dataset details}
\label{app:dataset-details}

This appendix gathers the construction and preprocessing details of the two datasets. \Cref{app:simplebox} covers the calibration that maps a target sampling fraction to the \textsc{SimpleBox} active fraction. \Cref{app:lemurs-preprocessing} covers the particle-centric local frame, and its acceptance windows, used to voxelise the LEMURS showers.

\subsection{SimpleBox}
\label{app:simplebox}

Building the \textsc{SimpleBox} family requires setting, for every geometry, the
active fraction $a$ that yields a chosen sampling fraction $f_{\rm s}$.
This map has no closed form: $f_{\rm s}$ rises with $a$, but two geometries that share the same $a$ and differ only in the number of layers $N_{\rm layers}$ reach different $f_{\rm s}$. 
The inversion must therefore depend on both quantities.

\begin{figure}[htbp]
\centering
\begin{minipage}[t]{0.47\linewidth}
\vspace{0pt}
\centering
\small
\captionof{table}{Five-fold cross-validated mean relative error on $a$ for
the candidate regressors of $\hat{a}(f_{\rm s},N_{\rm layers})$. The
retained model is highlighted.}
\label{tab:sb-modelsel}
\begin{tabular}{l S[table-format=2.2]}
\toprule
Regressor & {Mean rel.\ error [\%]} \\
\midrule
Polynomial, degree 1            & 2.26 \\
Polynomial, degree 2            & 0.59 \\
Polynomial, degree 3            & 0.30 \\
\textbf{Polynomial, degree 4}   & \bfseries 0.26 \\
Gradient boosting               & 2.51 \\
Random forest                   & 3.46 \\
SVM (RBF kernel)                & 7.64 \\
\bottomrule
\end{tabular}
\end{minipage}\hfill
\begin{minipage}[t]{0.51\linewidth}
\vspace{0pt}
\centering
\includegraphics[width=\linewidth]{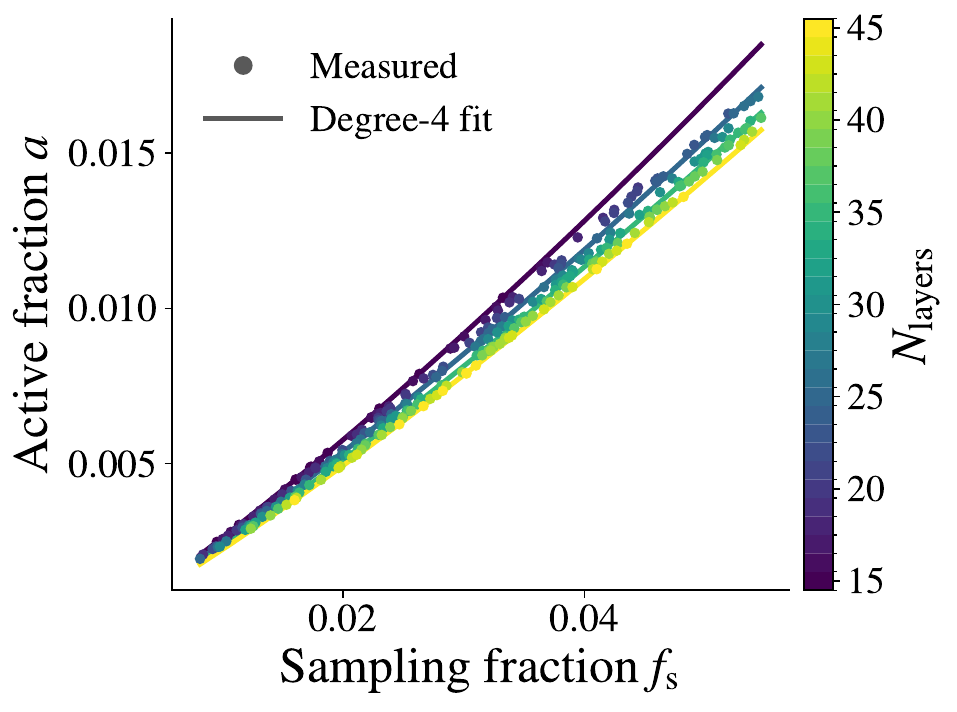}
\captionof{figure}{Retained degree-$4$ surface
$\hat{a}(f_{\rm s},N_{\rm layers})$ (lines, for
$N_{\rm layers}=15,25,35,45$) over the calibration points, coloured by
$N_{\rm layers}$.}
\label{fig:sb-deg4}
\end{minipage}
\end{figure}

\begingroup\emergencystretch=1em
We calibrate the inversion on $300$ geometries, drawing $a$ uniformly
for each $N_{\rm layers}\in\{15,\dots,45\}$ and measuring $f_{\rm s}$ with
\texttt{Geant4}. \Cref{tab:sb-modelsel} compares candidate regressors for
$\hat{a}(f_{\rm s},N_{\rm layers})$ under a common five-fold
cross-validation. Polynomials follow the smooth, monotone surface
more closely than the tree- and kernel-based models, and the
gain past degree 3 is small. We nonetheless keep the degree-$4$
polynomial, the most accurate of the candidates: it predicts $a$ to $0.26\%$, stays monotonic across the
calibration range, and is cheap to evaluate when generating the final geometries (\cref{fig:sb-deg4}).
\par\endgroup

When the final dataset is re-simulated with the predicted $a$, the
\texttt{Geant4} sampling fraction matches the target to a root-mean-square
error of $4\times10^{-5}$ ($\sim0.1\%$), with no visible bias over the full
$[0.01,0.05]$ range.

\subsection{LEMURS}
\label{app:lemurs-preprocessing}

Before clustering, every recorded \texttt{Geant4} step is mapped from the global detector frame into a particle-centric cylindrical frame attached to the incident photon.
For a step at global Cartesian position $(x,y,z)$ we form the detector radius $\rho$ and azimuth $\phi_{\rm hit}$, and from these the three local coordinates $(\mathrm{d}h_t,\mathrm{d}h_z,\ell)$:
\begin{align}
  \rho &= \sqrt{x^2+y^2}, &
  \phi_{\rm hit} &= \operatorname{atan2}(y,x), \\
  \mathrm{d}h_t &= \rho\,\operatorname{wrap}_\pi\!\left(\phi_{\rm hit}-\phi_{\rm gun}\right), &
  \mathrm{d}h_z &= z - z_{\rm gun}, \\
  \ell &= \left\lfloor \frac{\rho - R_{\min}}{t} \right\rfloor, &
  t &= \frac{R_{\max}-R_{\min}}{N_{\rm layers}},
  \label{eq:lemurs-frame}
\end{align}

where $\operatorname{wrap}_\pi$ folds an angle into $(-\pi,\pi]$, $\phi_{\rm gun}\equiv\phi$ is the gun azimuth, and $z_{\rm gun}\equiv R_{\rm g}\cot\theta$ is its longitudinal position, both read off the gun placement $\mathbf{x}_{\rm gun}$ of \cref{ssec:datasets-lemurs-calos} with $R_{\rm g}=R_{\min}-10^{-8}~\mathrm{mm}$. The active-stack extent $R_{\max}-R_{\min}$, where $R_{\max}$ is the outer radius of the active stack, is the ``Depth'' of \cref{tab:detectors}. It and the number of layers $N_{\rm layers}$ are the geometry inputs that fix the shell thickness $t$ in \cref{eq:lemurs-frame}.

\begin{figure}[htbp]
\centering
\includegraphics[width=\linewidth]{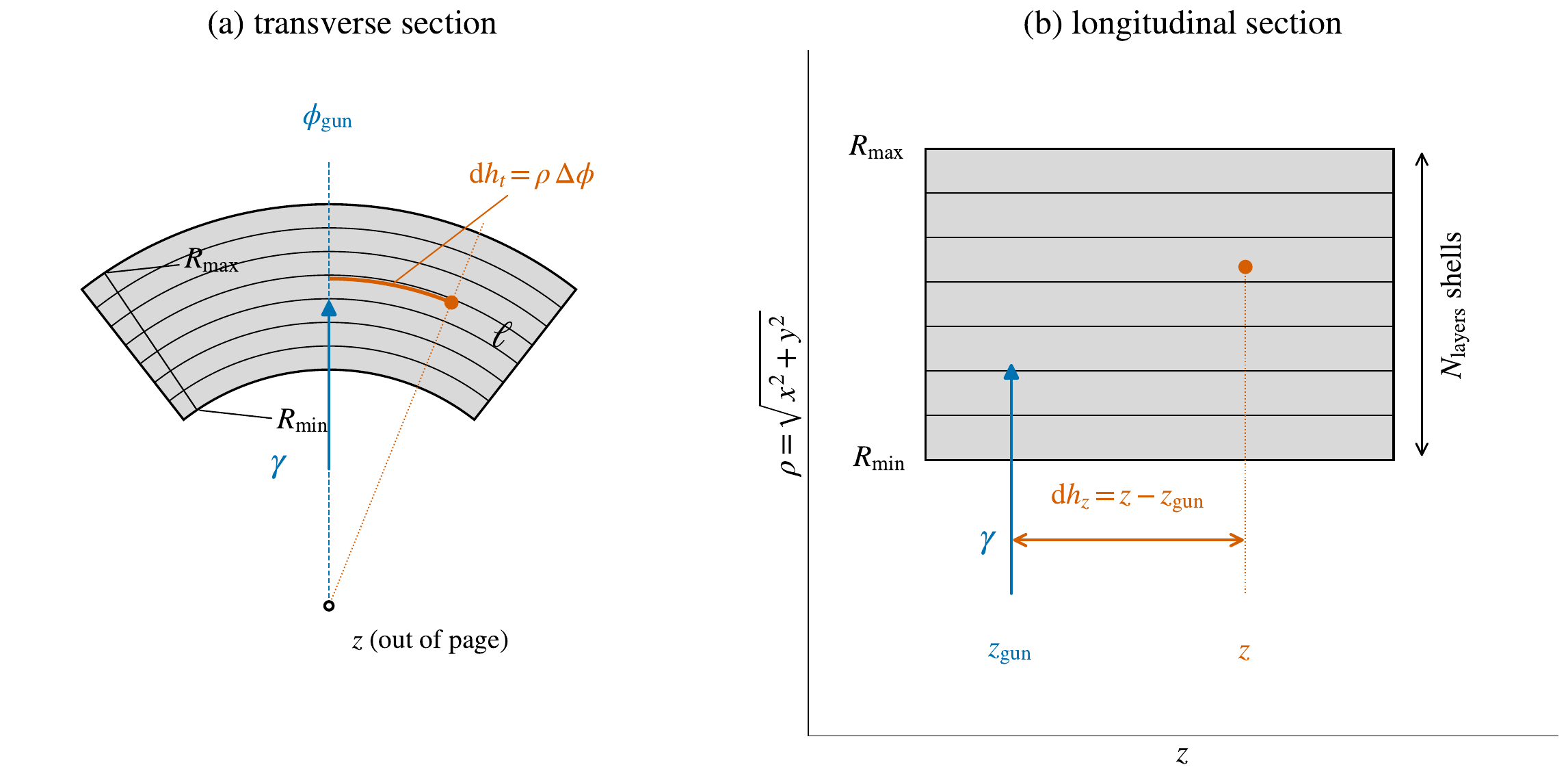}
\caption{Projection of a \texttt{Geant4} step into the particle-centric local frame. (a) Transverse section: equal-width radial shells (layer index $\ell$) and the azimuthal arc length $\mathrm{d}h_t=\rho\,\Delta\phi$ measured from the gun azimuth $\phi_{\rm gun}$. (b) Longitudinal section: the active stack between $R_{\min}$ and $R_{\max}$ and the axial displacement $\mathrm{d}h_z=z-z_{\rm gun}$ (transverse to the radial shower development).}
\label{fig:lemurs-frame}
\end{figure}

The barrel is a cylinder, so showers enter at the inner radius $R_{\min}$ and develop radially outward to $R_{\max}$: the radial direction $\rho$ is the depth axis.
The two coordinates transverse to the radial shower development are $\mathrm{d}h_t$ and $\mathrm{d}h_z$ (\cref{fig:lemurs-frame}): $\mathrm{d}h_t$ is the azimuthal arc length measured from the incident azimuth, and $\mathrm{d}h_z$ is the displacement along the detector (beam) axis. The shell index $\ell$, an equal-width radial bin, is the longitudinal (depth) coordinate.

We keep only steps inside an acceptance window, $|\mathrm{d}h_t|<500~\mathrm{mm}$ and $|\mathrm{d}h_z|<1000~\mathrm{mm}$, and drop any step whose radius falls outside the barrel envelope $[R_{\min},\,R_{\min}+N_{\rm layers}\,t]$ (equivalently $\ell\notin\{0,\dots,N_{\rm layers}-1\}$).
The surviving steps are clustered on a $1~\mathrm{mm}\times 1~\mathrm{mm}$ grid in $(\mathrm{d}h_t,\mathrm{d}h_z)$ within each shell; the step energies falling in a cell are summed, and cells below $E_{\rm thr}=10~\mathrm{keV}$ are discarded.

Two choices warrant justification.
The $500/1000~\mathrm{mm}$ windows extend well beyond the shower core while trimming the rare wide-angle backsplash that would otherwise scatter a handful of cells far from the core.
The shell index $\ell$ uses equal-width radial bins rather than each detector's native readout segmentation; this gives a uniform longitudinal schema across the five geometries. No step-level information is lost, since $\rho$ still enters the transverse clustering continuously.

\FloatBarrier

\section{SimpleBox-mini: a reduced-pool check}
\label{app:simplebox-mini}

The transfer results of the main text all use the full \textsc{SimpleBox} pre-training pool. This appendix asks how much of that pool is needed. It repeats the pre-training on \textsc{SimpleBox}-mini, a random $10^{5}$-shower subsample (about $2.5\%$) of the full pool, and follows the consequences through pre-training quality, transfer to the held-out FCCee-ALLEGRO calorimeter, and the compute break-even. Two pool sizes are a check, not a scan: they bound the question rather than resolve how performance varies with pool size in between.

\subsection{Setup and pre-training quality}
\label{ssec:mini-setup}

\textsc{SimpleBox}-mini draws $10^{5}$ showers uniformly at random from the full $4\times10^{6}$-shower \textsc{SimpleBox} pool, about $2.5\%$ of it, keeping the architecture, the geometry-aware conditioning and the optimiser settings of \cref{ssec:method-protocol} unchanged. One difference is unavoidable, because the learning rate anneals on a cosine tied to the epoch ceiling. The mini pre-training was budgeted $400$ epochs and reaches its lowest validation loss at epoch $145$, $36\%$ of the way through its schedule. The full pre-training was budgeted $143$ epochs and reaches its own minimum at epoch $123$, or $86\%$ of the way through. The two priors are therefore delivered at different points of their anneal, and the mini checkpoint is taken at a less annealed learning rate.

In optimiser steps the smaller pool is far cheaper. An epoch is $351$ steps for the mini pool against $1.5\times10^{4}$ for the full one. Reaching the minimum therefore takes $51\times10^{3}$ steps against $1.84\times10^{6}$, a factor of $36$ fewer, even though the mini needs slightly more epochs to get there. The per-step cost and the peak memory are unchanged, since the architecture is the same; \cref{tab:allshowers_method_hyperparameters} lists the budgeted epochs and steps of both runs, and \cref{tab:compute_budget} the resulting cost.

\begin{figure}[htbp]
    \centering
    \includegraphics[width=0.9\linewidth]{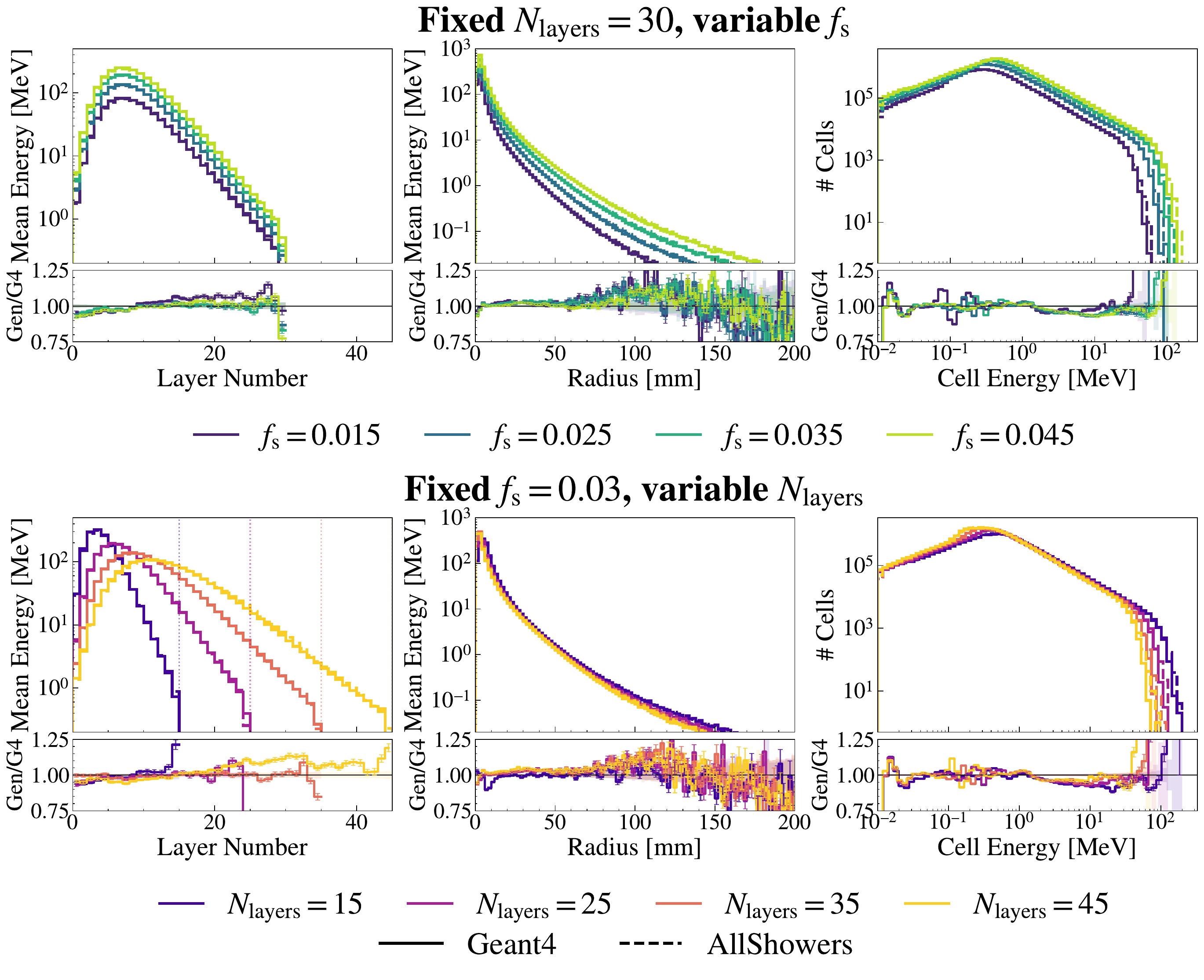}
    \caption{\textsc{SimpleBox}-mini pre-training quality on the same eight held-out geometries as \cref{fig:simplebox_pretrain_observables}: longitudinal profile (left), radial profile (middle), and cell-energy spectrum (right), with generated samples (dashed) over \texttt{Geant4} (solid) and the generated-to-\texttt{Geant4} ratio in the sub-panels.}
    \label{fig:simplebox_mini_pretrain_observables}
\end{figure}

The pre-training quality is assessed on the same eight held-out geometries used for the full pool in \cref{fig:simplebox_pretrain_observables}, so the two can be compared directly. \Cref{fig:simplebox_mini_pretrain_observables} shows that the mini prior follows the same conditioning trends along both axes of the grid and reproduces the three observables across all eight geometries. We do not claim the two pre-trainings are equivalent: the comparison here is a visual one on the held-out grid, and the transfer numbers of \cref{ssec:mini-allegro} show a residual gap at small fine-tuning sizes.

\subsection{Transfer to ALLEGRO across data scales}
\label{ssec:mini-allegro}

The mini prior is fine-tuned on the held-out FCCee-ALLEGRO target across the same budgets $D\in\{10^{2},10^{3},10^{4},10^{5}\}$ as in \cref{ssec:results-allegro-transfer}, with five seeds each; it is the fourth row of \cref{fig:observables_allegro} and the fourth arm of \cref{fig:sample-efficiency}. \Cref{tab:mini_vs_full} collects the aggregate distance $\bar{d}$ of \cref{eq:dbar} for training from scratch and for all three pre-trained priors, LEMURS included as the reference point already discussed in \cref{ssec:results-allegro-transfer}.

\begin{table}[htbp]
    \centering
    \caption{Aggregate sliced Wasserstein distance to \texttt{Geant4}, $\bar{d}$ of \cref{eq:dbar}, on FCCee-ALLEGRO as a function of the fine-tuning size $D$, for training from scratch and for the three pre-trained priors. Mean $\pm$ one standard deviation over five seeds. Lower is better; the smallest entry at each $D$ is set in bold.}
    \label{tab:mini_vs_full}
    \small
    \begin{tabular}{lcccc}
    \toprule
    $D$ & \textbf{from scratch} & \textbf{SimpleBox} & \textbf{LEMURS} & \textbf{SimpleBox-mini} \\
    \midrule
    $10^{2}$ & $0.1005 \pm 0.0042$ & $\mathbf{0.0390 \pm 0.0027}$ & $0.0734 \pm 0.0414$ & $0.0441 \pm 0.0024$ \\
    $10^{3}$ & $0.0845 \pm 0.0010$ & $0.0162 \pm 0.0020$ & $\mathbf{0.0105 \pm 0.0020}$ & $0.0252 \pm 0.0016$ \\
    $10^{4}$ & $0.0112 \pm 0.0039$ & $0.0085 \pm 0.0010$ & $0.0162 \pm 0.0029$ & $\mathbf{0.0081 \pm 0.0011}$ \\
    $10^{5}$ & $0.0135 \pm 0.0059$ & $0.0109 \pm 0.0008$ & $0.0220 \pm 0.0039$ & $\mathbf{0.0093 \pm 0.0022}$ \\
    \bottomrule
    \end{tabular}
\end{table}

The mini prior beats training from scratch at every budget, which is what makes it useful at all. It does not, however, match the full pool everywhere. At $D=10^{4}$ and $10^{5}$ the two priors agree within the seed spread, with the mini marginally ahead. At the two smallest budgets the full pool is ahead, slightly at $D=10^{2}$ and clearly at $D=10^{3}$. There $\bar{d}=0.0252$ against $0.0162$, so the mini is $56\%$ worse, with the seed bands well separated.

That gap matters more than its size suggests, because $D=10^{3}$ is the operating point the cost model is built on. It is the budget at which the fine-tuning cost $F$ is measured (\cref{ssec:results-breakeven}) and the one the abstract quotes. Measured against training from scratch at that budget, the full pool improves $\bar{d}$ by a factor of $5.2$ and the mini by $3.4$. The cheaper prior is therefore not free: it lowers the break-even from $N^{\star}\approx5$ to $N^{\star}\approx0.3$ (\cref{tab:compute_budget}) at the cost of part of the low-data advantage that motivates pre-training. Which of the two is preferable depends on how many detectors the prior is meant to serve and on how much target data each of them can afford.

\subsection{Computational cost and break-even}
\label{ssec:mini-cost}

The cost side is reported with the rest of the budget in \cref{ssec:results-breakeven}. The mini pre-training costs $P_{\rm mini}=45$ GPU-hours against $852$ for the full pool, while the per-detector cost of fine-tuning, $F=125\pm10$ GPU-hours, and of training from scratch, $S=304\pm44$ GPU-hours, are common to both priors. The break-even $N^{\star}=P/(S-F)$ therefore falls from about five detectors to about $0.3$, so the mini prior is repaid within the first downstream detector (\cref{tab:compute_budget}, \cref{fig:breakeven-cost}). The pre-training cost scales with the pool, while nothing else in the balance does, so the reduced pool moves $N^{\star}$ by the same factor as $P$.

\FloatBarrier

\section{Supplementary results}
\label{app:percalo-detailed}

This appendix gathers supplementary material: the per-detector distance metrics (\cref{app:percalo-kl}), individual pre-training showers, the transfer results with the multiplicity correction switched off (\cref{app:no-pcfm}), and a Kullback--Leibler counterpart of the FCCee-ALLEGRO transfer metrics (\cref{app:kl-allegro}).

\subsection{Per-detector distance metrics}
\label{app:percalo-kl}

\Cref{tab:sb-lemurs-w1} gives the per-observable $W_1/\sigma_{\rm ref}$ distance to \texttt{Geant4} for the four Si/Sci detectors, across the fine-tuning sizes $D$, for training from scratch and for \textsc{SimpleBox}-pretrain fine-tuning. It is the numerical counterpart of the transfer figures of \cref{ssec:results-sb-to-lemurs}.

\begin{table}[htbp]
\centering
\caption{Per-detector $W_1/\sigma_{\rm ref}$ distance to \texttt{Geant4}, by observable and fine-tuning size $D$, for training from scratch and for \textsc{SimpleBox}-pretrain fine-tuning. $\bar{d}$ (\cref{eq:dbar}) is the geometric mean over the three observables; the smaller of the two strategies at each $D$ is set in bold. Par04-SiW is generated without the per-layer bias correction. FCCee-ALLEGRO is covered in \cref{ssec:results-allegro-transfer}.}
\label{tab:sb-lemurs-w1}
\scriptsize
\setlength{\tabcolsep}{2pt}
\begin{tabular}{ll cccc !{\vrule} cccc}
\toprule
\multirow{2}{*}{\textbf{Detector}} & \multirow{2}{*}{\textbf{Observable}} & \multicolumn{4}{c}{\textbf{From scratch}} & \multicolumn{4}{c}{\textbf{SimpleBox fine-tuned}} \\
\cmidrule(lr){3-6}\cmidrule(lr){7-10}
 & & $10^{2}$ & $10^{3}$ & $10^{4}$ & $10^{5}$ & $10^{2}$ & $10^{3}$ & $10^{4}$ & $10^{5}$ \\
\midrule
\multirow{4}{*}{FCCee-CLD}   & Longitudinal & 0.153 & 0.206 & 0.0333 & 0.00626 & 0.394 & 0.0129 & 0.0174 & 0.00415 \\
                             & Radial       & 2.08 & 1.59 & 0.202 & 0.0741 & 0.462 & 0.0707 & 0.0392 & 0.0184 \\
                             & Cell energy  & 0.0995 & 0.0792 & 0.0249 & 0.00780 & 0.0323 & 0.0251 & 0.0188 & 0.00323 \\
                             & $\bar{d}$    & 0.316 & 0.296 & 0.0552 & 0.0154 & \textbf{0.181} & \textbf{0.0284} & \textbf{0.0234} & \textbf{0.00627} \\
\midrule
\multirow{4}{*}{ODD}         & Longitudinal & 0.303 & 0.310 & 0.0924 & 0.0194 & 0.659 & 0.0462 & 0.0318 & 0.0112 \\
                             & Radial       & 1.57 & 1.42 & 0.990 & 0.0933 & 0.890 & 0.108 & 0.0506 & 0.0106 \\
                             & Cell energy  & 0.0763 & 0.0757 & 0.0633 & 0.00615 & 0.0476 & 0.0340 & 0.0199 & 0.00492 \\
                             & $\bar{d}$    & 0.331 & 0.322 & 0.180 & 0.0223 & \textbf{0.303} & \textbf{0.0554} & \textbf{0.0318} & \textbf{0.00836} \\
\midrule
\multirow{4}{*}{Par04-SiW}   & Longitudinal & 0.431 & 0.333 & 0.307 & 0.0799 & 0.237 & 0.0949 & 0.347 & 0.554 \\
                             & Radial       & 1.64 & 1.28 & 1.02 & 0.0656 & 0.450 & 0.419 & 0.111 & 0.896 \\
                             & Cell energy  & 0.0789 & 0.0849 & 0.0310 & 0.00918 & 0.0848 & 0.0994 & 0.0216 & 0.0460 \\
                             & $\bar{d}$    & 0.382 & 0.331 & 0.214 & \textbf{0.0364} & \textbf{0.208} & \textbf{0.158} & \textbf{0.0940} & 0.284 \\
\midrule
\multirow{4}{*}{Par04-SciPb} & Longitudinal & 0.290 & 0.278 & 0.0468 & 0.00111 & 0.0880 & 0.0285 & 0.00651 & 0.00327 \\
                             & Radial       & 1.42 & 1.21 & 0.361 & 0.0411 & 0.194 & 0.0827 & 0.0221 & 0.0123 \\
                             & Cell energy  & 0.0709 & 0.0709 & 0.0213 & 0.00288 & 0.0371 & 0.0241 & 0.00852 & 0.00342 \\
                             & $\bar{d}$    & 0.308 & 0.288 & 0.0712 & \textbf{0.00508} & \textbf{0.0858} & \textbf{0.0384} & \textbf{0.0107} & 0.00516 \\
\bottomrule
\end{tabular}
\end{table}

\FloatBarrier

\subsection{Individual pre-training showers}
\begin{figure}[htbp]
    \centering
    \includegraphics[width=1\linewidth]{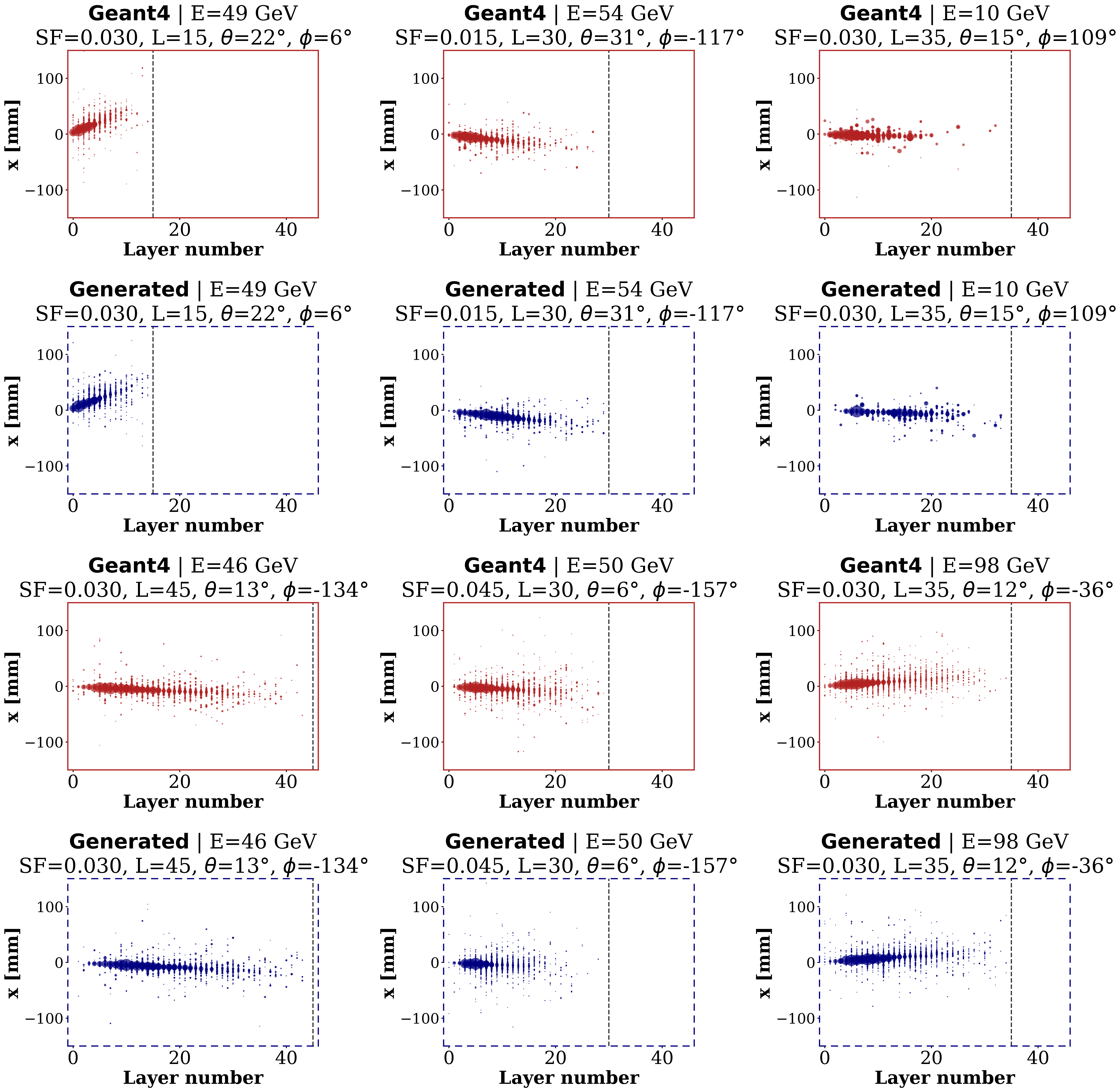}
    \caption{Individual \textsc{SimpleBox} showers, \texttt{Geant4} (red) versus generated (blue), shown as the transverse coordinate $x$ against layer number. Each panel corresponds to a different $(f_{\rm s},N_{\rm layers},E_{\rm inc},\theta,\phi)$ configuration drawn from the held-out set, with the dashed box marking the active depth $N_{\rm layers}$. The generator reproduces both the longitudinal containment fixed by $N_{\rm layers}$ and the angle-dependent lateral spread on a shower-by-shower basis.}
    \label{fig:simplebox_pretrain_showers}
\end{figure}

\begingroup\emergencystretch=1em
\Cref{ssec:results-pretrain} showed the distributional agreement of the \textsc{SimpleBox} pre-training across the $(f_{\rm s},N_{\rm layers})$ grid.
\Cref{fig:simplebox_pretrain_showers} complements this distributional view with individual showers; the per-shower longitudinal extent tracks $N_{\rm layers}$.
\par\endgroup

\subsection{Transfer without PointCountFM}
\label{app:no-pcfm}

These figures repeat the transfer results of \cref{ssec:results-sb-to-lemurs,ssec:results-allegro-transfer} with the \textsc{PointCountFM} (PCFM) correction of \cref{ssec:method-arch} switched off. The per-layer point counts are taken directly from \texttt{Geant4}, while the coordinates are still generated by the model.
This oracle-multiplicity setting isolates the coordinate-generation stage, so the residual discrepancies with \texttt{Geant4} here reflect the coordinate model alone.

\begin{figure}[htbp]
    \centering
    \includegraphics[width=0.82\linewidth]{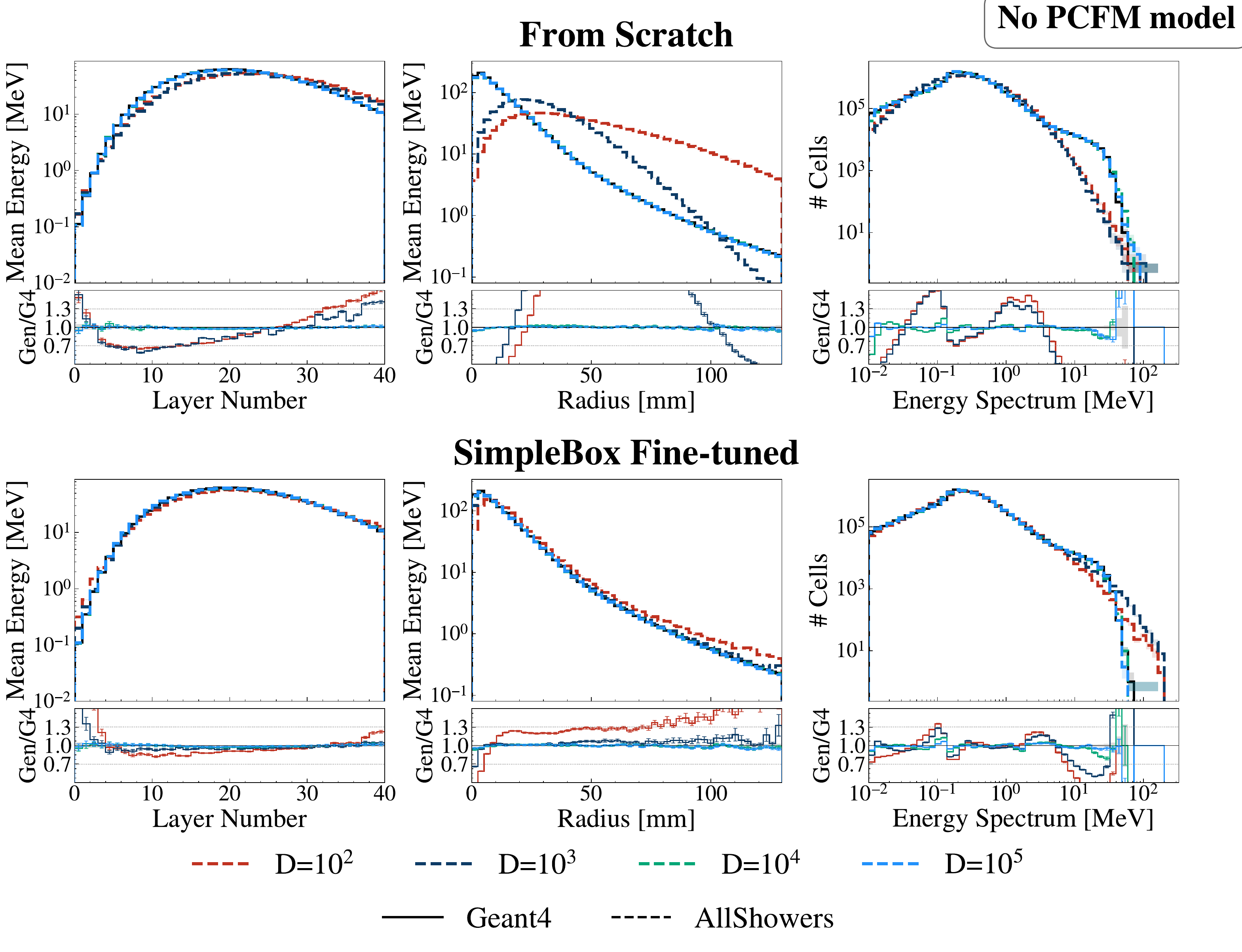}
    \caption{FCCee-CLD ($N_{\rm layers}=40$); PCFM counterpart in \cref{fig:simplebox_finetune_cld}.}
    \label{fig:simplebox_finetune_cld_g4}
\end{figure}

\begin{figure}[htbp]
    \centering
    \includegraphics[width=0.82\linewidth]{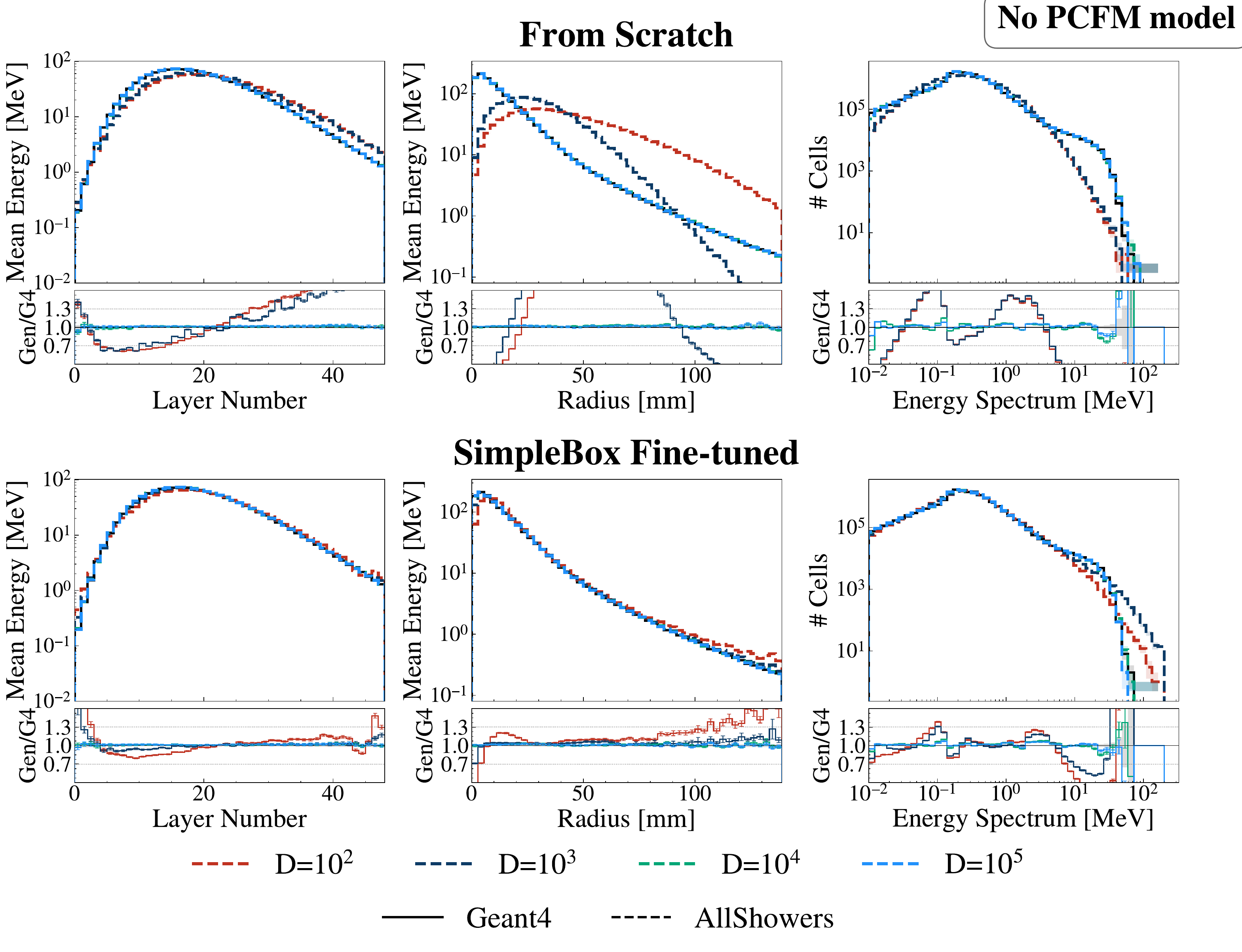}
    \caption{ODD ($N_{\rm layers}=48$); PCFM counterpart in \cref{fig:simplebox_finetune_odd}.}
    \label{fig:simplebox_finetune_odd_g4}
\end{figure}

\begin{figure}[htbp]
    \centering
    \includegraphics[width=0.82\linewidth]{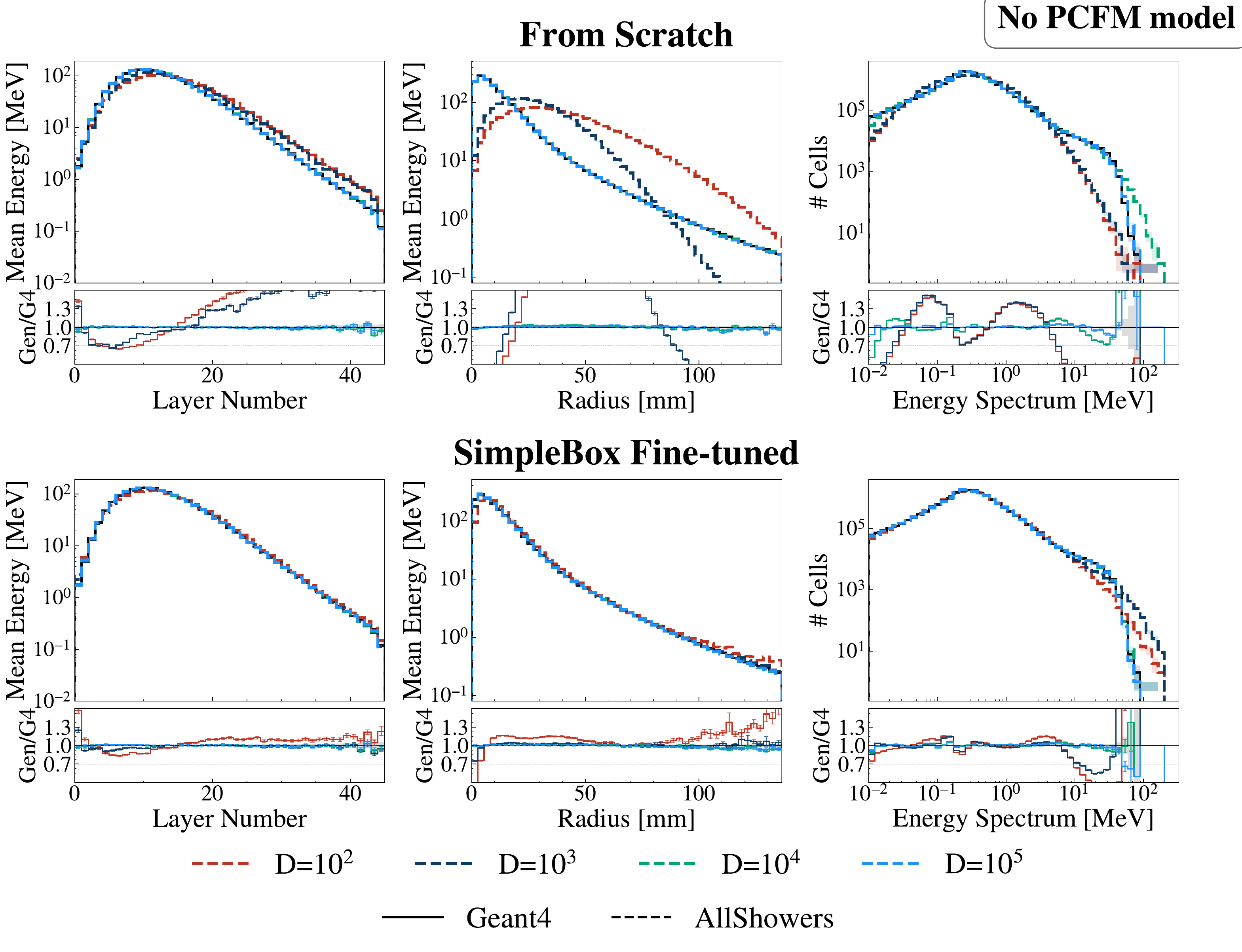}
    \caption{Par04-SciPb ($N_{\rm layers}=45$); PCFM counterpart in \cref{fig:simplebox_finetune_par04scipb}.} 
    \label{fig:simplebox_finetune_par04scipb_g4}
\end{figure}

\begin{figure}[htbp]
    \centering
    \includegraphics[width=0.82\linewidth]{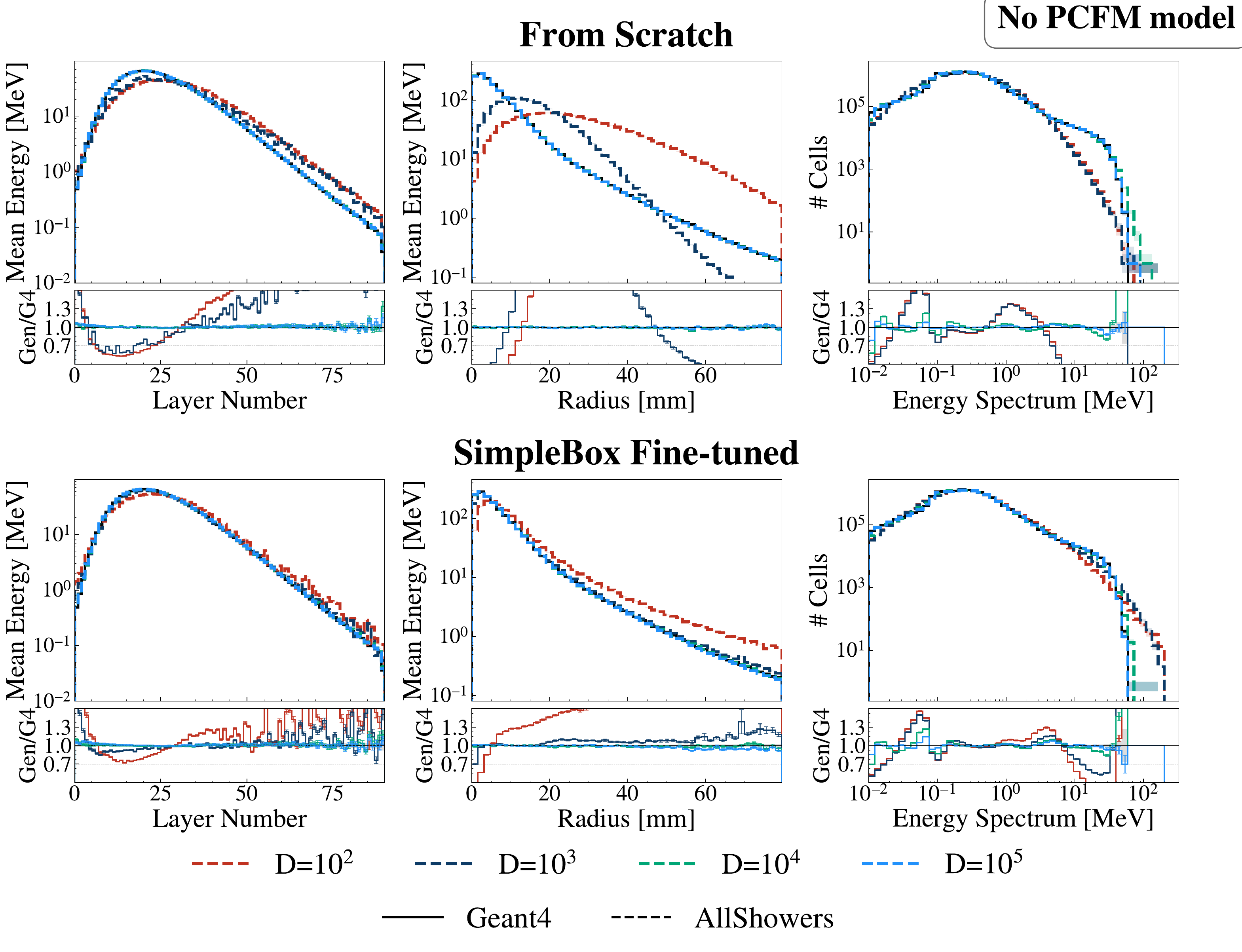}
    \caption{Par04-SiW ($N_{\rm layers}=90$); PCFM counterpart in \cref{fig:simplebox_finetune_par04siw}.}
    \label{fig:simplebox_finetune_par04siw_g4}
\end{figure}

\begin{figure}[htbp]
    \centering
    \includegraphics[width=0.98\linewidth]{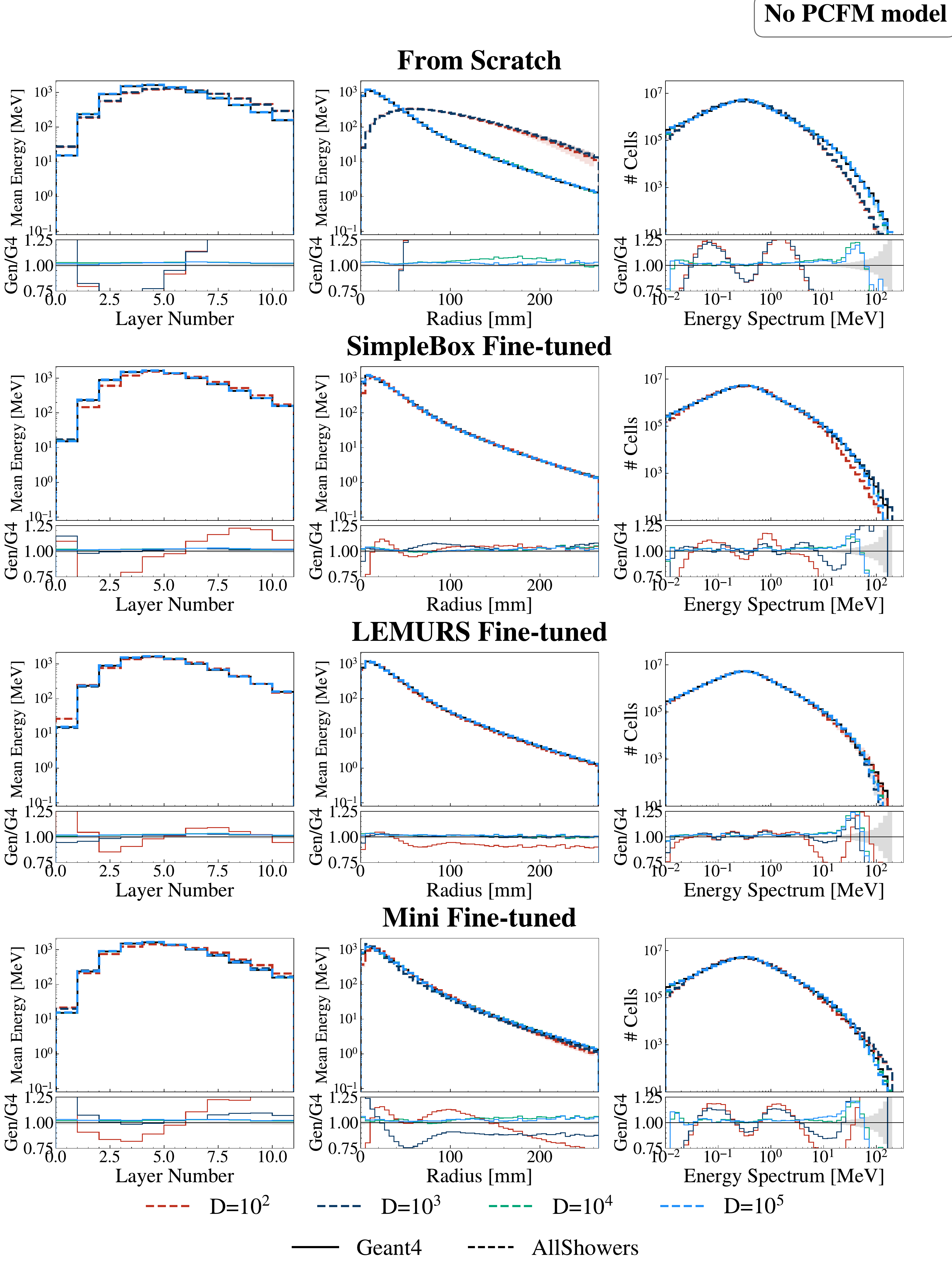}
    \caption{FCCee-ALLEGRO ($N_{\rm layers}=11$), one row per strategy, with the per-layer point counts taken from \texttt{Geant4} instead of \textsc{PointCountFM}; bands span the mean $\pm$ one standard deviation over five seeds. PCFM counterpart in \cref{fig:observables_allegro}.}
    \label{fig:observables_allegro_g4}
\end{figure}

\FloatBarrier
\subsection{Kullback--Leibler counterpart of the ALLEGRO transfer metrics}
\label{app:kl-allegro}

The sample-efficiency metric of \cref{ssec:results-allegro-transfer} is the sliced Wasserstein distance, which compares the two point clouds shower by shower (\cref{ssec:method-metrics}). A pooled information-theoretic divergence gives a complementary check with a different construction: it pools the hits of all showers into a single histogram per observable and measures the Kullback--Leibler (KL) divergence of the generated histogram from the \texttt{Geant4} one. We compute it for the same three observables. The longitudinal profile uses a categorical KL over the $N_{\rm layers}=11$ layers, weighted by deposited energy to match the energy-weighted longitudinal distance. The radial profile and the cell-energy spectrum use a KL on thirty \texttt{Geant4}-quantile bins of the pooled per-hit samples, matching the binning of \cref{ssec:method-metrics}. Their geometric mean aggregates the three, as $\bar{d}$ of \cref{eq:dbar} does for the sliced Wasserstein distance. As a statistical resolution we quote the KL between two disjoint halves of the \texttt{Geant4} reference, $1.3\times10^{-5}$ on the longitudinal profile; each comparison uses $5\times10^{3}$ showers per sample, as for the sliced Wasserstein distance. Unlike a distance, this pooled KL is a positively biased estimator whose same-distribution value shrinks as the sample grows. The dashed line therefore marks the resolution of the estimator rather than a lower bound: an arm that converges sits at it and can scatter below. The comparison between arms, not the absolute level, carries the result.

\begin{figure}[htbp]
    \centering
    \includegraphics[width=\linewidth]{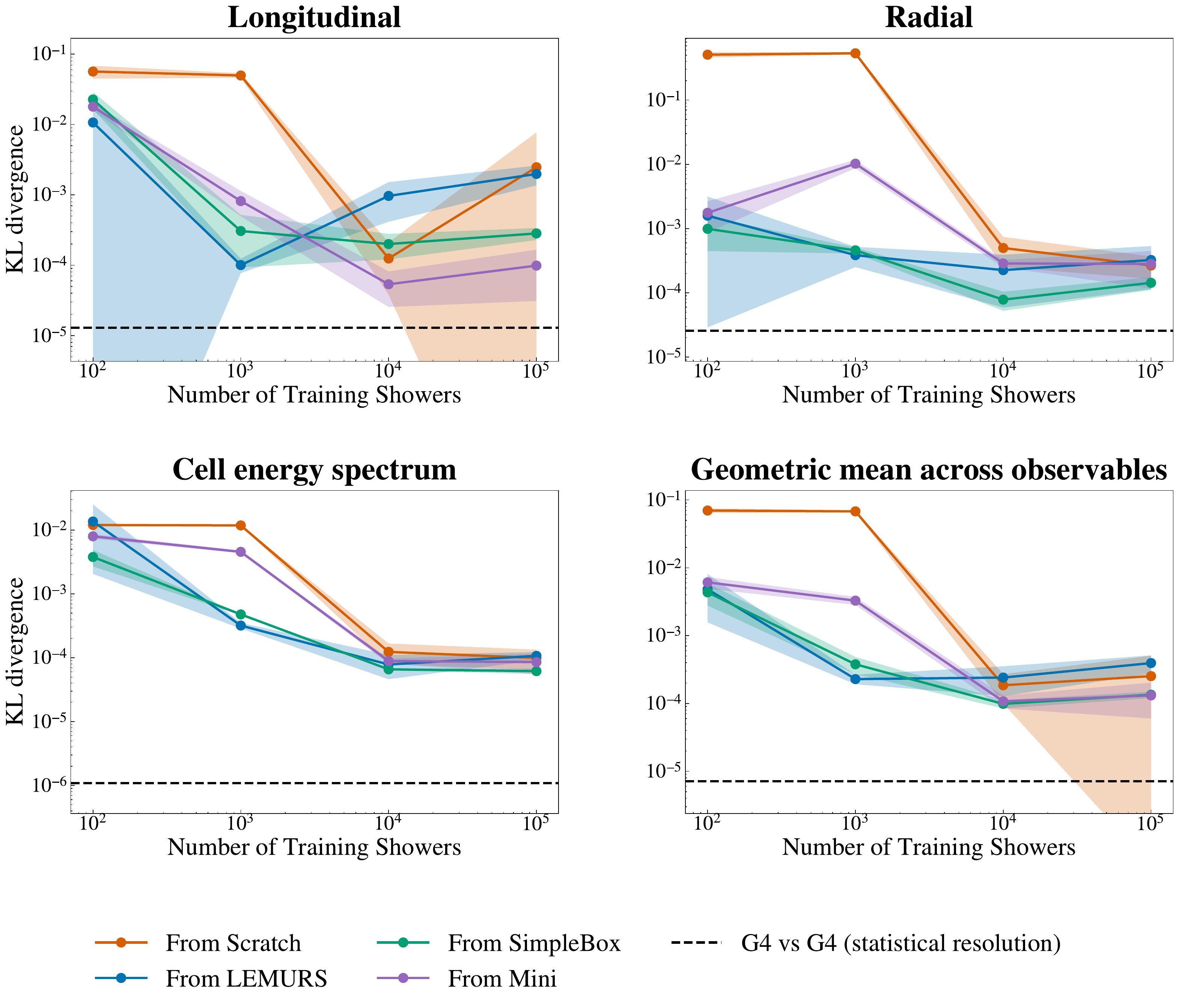}
    \caption{Kullback--Leibler counterpart of \cref{fig:sample-efficiency}: the KL divergence to \texttt{Geant4} on FCCee-ALLEGRO as a function of the fine-tuning size $D$, for the four strategies over five seeds. The longitudinal panel uses a categorical KL over the $N_{\rm layers}=11$ layers weighted by deposited energy; the radial and cell-energy panels use a thirty-bin quantile KL on the pooled per-hit samples; the fourth panel is their geometric mean. Bands span the mean $\pm$ one standard deviation over five seeds and the dashed line is the \texttt{Geant4}-versus-\texttt{Geant4} statistical resolution (\cref{ssec:method-metrics}). The vertical axis is clipped a factor of three below that line: the pooled KL is a biased estimator whose seed bands can extend far below the resolution near convergence, an artefact of the estimator's bias, magnified by the logarithmic axis, rather than a physical effect. The longitudinal LEMURS divergence is smallest at $D=10^{3}$ and rises at $10^{4}$ and $10^{5}$, reproducing the anomaly of \cref{fig:sample-efficiency}.}
    \label{fig:kl-allegro-pcfm}
\end{figure}

\Cref{fig:kl-allegro-pcfm} shows the result and \cref{tab:kl-allegro-long} lists the longitudinal values. The longitudinal panel reproduces the anomaly of \cref{fig:sample-efficiency}. The LEMURS divergence is smallest at $D=10^{3}$, $1.0\times10^{-4}$, and rises with more target data: $9.7\times10^{-4}$ at $10^{4}$ and $2.0\times10^{-3}$ at $10^{5}$, far above both \textsc{SimpleBox} priors. No other arm shows a comparable rise: both \textsc{SimpleBox} priors fall towards the resolution and stay there, and the from-scratch mean at $10^{5}$ is raised by a single high-KL seed (median $1.0\times10^{-4}$; \cref{tab:kl-allegro-long}). A pooled divergence built on a different principle therefore reproduces the rise, and it remains specific to the LEMURS transfer in this metric too.

\begin{figure}[htbp]
    \centering
    \includegraphics[width=\linewidth]{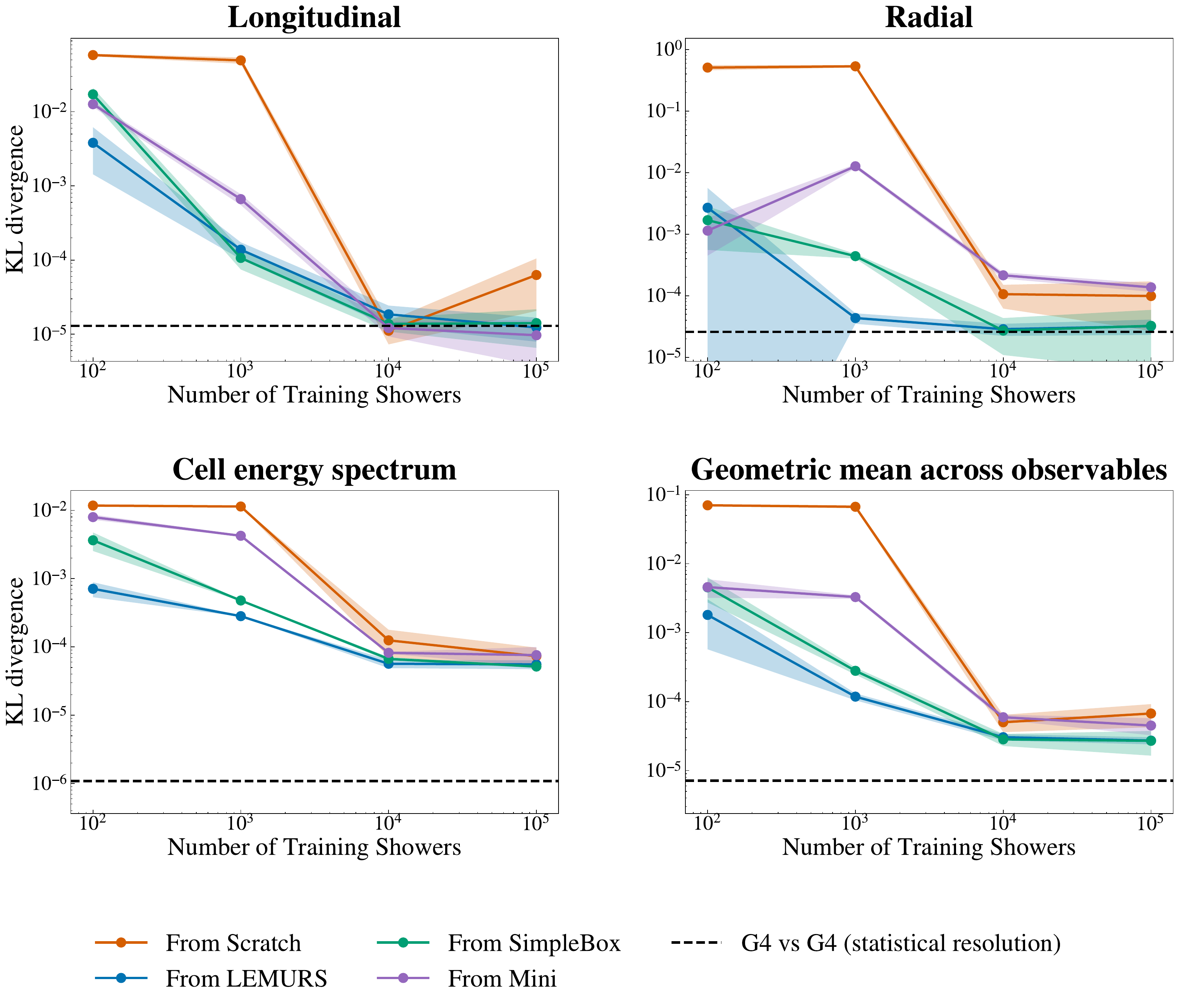}
    \caption{The KL divergences of \cref{fig:kl-allegro-pcfm}, recomputed with the per-layer point multiplicities taken from \texttt{Geant4} instead of from \textsc{PointCountFM} (the KL counterpart of \cref{fig:g4cond_ablation}). Panels, bands, normalisation and resolution line follow \cref{fig:kl-allegro-pcfm}. The longitudinal rise disappears and every arm falls monotonically to the statistical resolution.}
    \label{fig:kl-allegro-g4cond}
\end{figure}

\Cref{fig:kl-allegro-g4cond} repeats the measurement with the per-layer point counts taken from \texttt{Geant4} instead of from \textsc{PointCountFM}, the KL counterpart of \cref{fig:g4cond_ablation}. The longitudinal rise disappears: the LEMURS divergence falls monotonically, from $3.8\times10^{-3}$ at $D=10^{2}$ to $1.4\times10^{-4}$, $1.9\times10^{-5}$ and $1.2\times10^{-5}$, down to the statistical resolution, and every other arm does the same. As in \cref{ssec:results-allegro-transfer}, feeding the true occupancy removes the effect, placing it in the multiplicity stage rather than in the coordinate model, now with a second metric of a different construction.

The signal is clean on the longitudinal profile, where the multiplicity mechanism of \cref{ssec:results-allegro-transfer} acts, and weak on the geometric mean. From its $10^{3}$ minimum of $2.3\times10^{-4}$ the LEMURS geometric-mean KL grows only to $3.9\times10^{-4}$ at $10^{5}$, about a factor of two, against a factor of twenty on the longitudinal profile. The radial and cell-energy divergences of the pre-trained arms already lie within a small factor of the statistical resolution, which compresses the mean. We therefore read the KL counterpart on its longitudinal panel and do not draw a conclusion from the geometric mean. One display caveat: near the resolution the pooled KL of a single seed occasionally dominates the five-seed mean, which widens the corresponding bands in \cref{fig:kl-allegro-pcfm}. The LEMURS longitudinal rise is larger than this seed-to-seed dispersion at every point.

\begin{table}[htbp]
    \centering
    \caption{Longitudinal KL divergence to \texttt{Geant4} on FCCee-ALLEGRO, per strategy and fine-tuning size $D$, with the per-layer point counts from \textsc{PointCountFM}. Mean over five seeds; the per-seed dispersion is shown as the bands in \cref{fig:kl-allegro-pcfm}. Lower is better; the smallest entry at each $D$ is set in bold. The \texttt{Geant4}-versus-\texttt{Geant4} statistical resolution is $1.3\times10^{-5}$. The LEMURS entry is smallest at $D=10^{3}$ and then rises, the anomaly of \cref{ssec:results-allegro-transfer}; the from-scratch value at $D=10^{5}$ is raised by a single high-KL seed (median $1.0\times10^{-4}$).}
    \label{tab:kl-allegro-long}
    \small
    \begin{tabular}{lcccc}
    \toprule
    $D$ & \textbf{from scratch} & \textbf{SimpleBox} & \textbf{LEMURS} & \textbf{SimpleBox-mini} \\
    \midrule
    $10^{2}$ & $5.7\times10^{-2}$ & $2.3\times10^{-2}$ & $\mathbf{1.1\times10^{-2}}$ & $1.8\times10^{-2}$ \\
    $10^{3}$ & $5.0\times10^{-2}$ & $3.1\times10^{-4}$ & $\mathbf{1.0\times10^{-4}}$ & $8.2\times10^{-4}$ \\
    $10^{4}$ & $1.2\times10^{-4}$ & $2.0\times10^{-4}$ & $9.7\times10^{-4}$ & $\mathbf{5.4\times10^{-5}}$ \\
    $10^{5}$ & $2.5\times10^{-3}$ & $2.8\times10^{-4}$ & $2.0\times10^{-3}$ & $\mathbf{9.9\times10^{-5}}$ \\
    \bottomrule
    \end{tabular}
\end{table}

\FloatBarrier

\FloatBarrier

\clearpage
\section{Far-domain width drift: supporting analysis}
\label{app:width-drift}

\looseness=-1
This appendix supports the width-drift finding of \cref{ssec:results-allegro-transfer}. As the fine-tuning size $D$ grows, the model fine-tuned from the far-domain LEMURS prior generates a per-layer point-count profile whose depth distribution drifts towards the LEMURS pre-training value rather than towards \texttt{Geant4}. \Cref{ssec:results-allegro-transfer} locates this in the multiplicity stage predicted by \textsc{PointCountFM}, not in the coordinate model: feeding the per-layer counts from \texttt{Geant4} removes it (\cref{fig:g4cond_ablation}). There it is reported on one count-profile observable: the spread in depth of the per-shower centre of gravity, the cog std of \cref{eq:cog}. Here we reproduce it on a second count-profile observable, rule out four artefact explanations, and set out the limits of the directional attribution.

\begin{figure}[htb]
    \centering
    \includegraphics[width=0.42\linewidth]{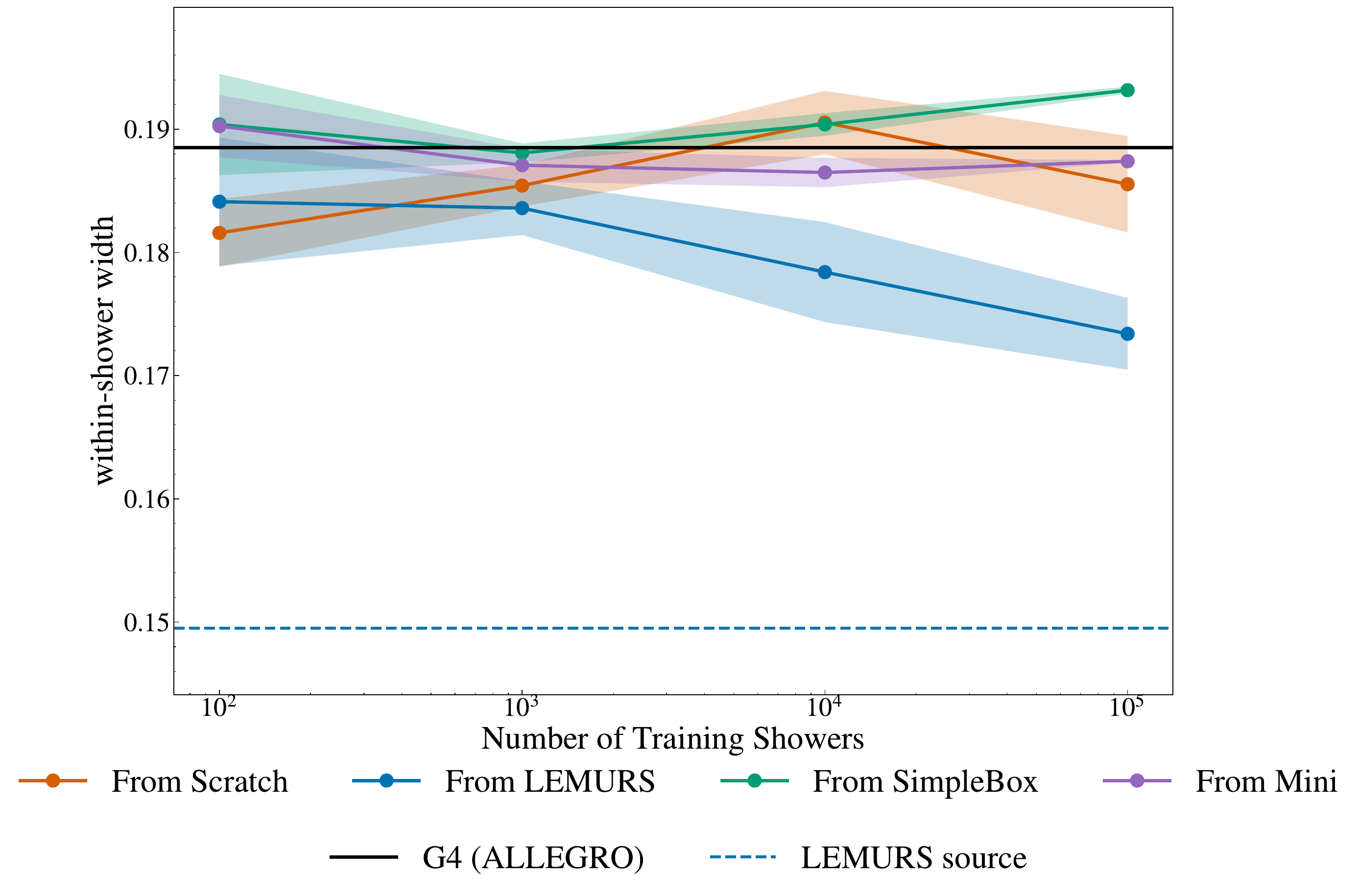}\\[0.5ex]
    \includegraphics[width=0.63\linewidth]{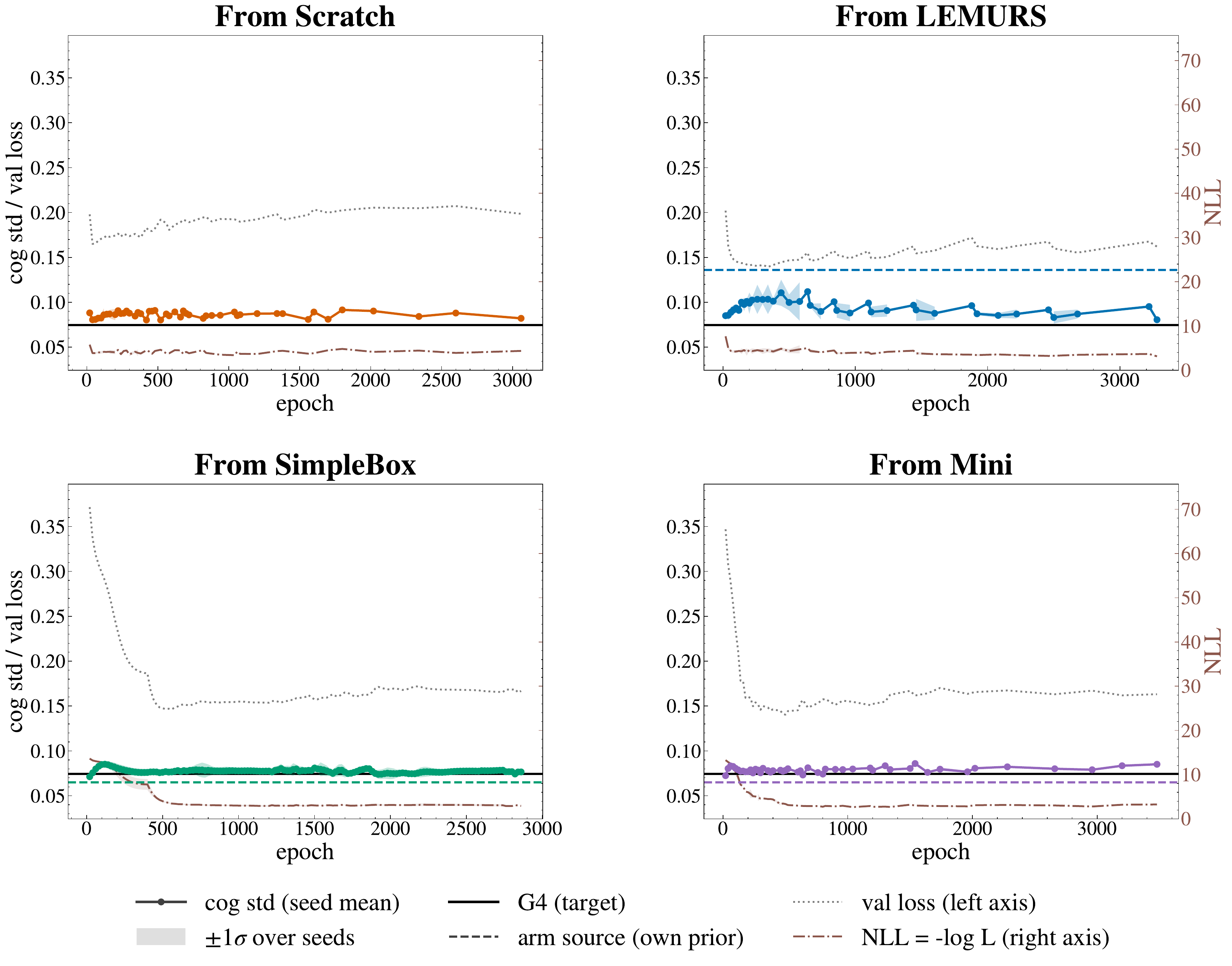}
    \caption{Top: within-shower width against the fine-tuning size $D$, bands the mean $\pm$ one standard deviation over five seeds; solid line \texttt{Geant4}, dashed line the LEMURS pool. Bottom: cog std (left axis, with the validation loss) and NLL (right axis) against training epoch at $D=10^{3}$, one panel per arm (the frozen-core variant of \cref{ssec:results-allegro-transfer} excluded); solid line \texttt{Geant4}, dashed line each arm's own pre-training source (the from-scratch arm has none). Both axes are shared with \cref{fig:cogstd-epoch-10k-100k} for direct comparison across $D$. At this budget the drift is present but weak.}
    \label{fig:width-within}
    \label{fig:cogstd-epoch-1k}
\end{figure}

\looseness=-1
The second observable is the within-shower width: for each shower, the count-weighted root-mean-square spread of the occupancy about its own mean depth, in the same fractional-depth units as \cref{eq:cog}, averaged over the sample. \Cref{fig:width-within} shows it against $D$. The LEMURS arm falls steadily in all five seeds, from about $0.184$ at $D=10^{2}$ to $0.173$ at $10^{5}$: away from the \texttt{Geant4} value near $0.189$ and towards the LEMURS pre-training width of $0.150$. It is the only arm to move that way on this observable: the from-scratch, \textsc{SimpleBox} and \textsc{SimpleBox}-mini arms all stay near the \texttt{Geant4} value. The \textsc{SimpleBox} cog std does move towards its own source (\cref{ssec:results-allegro-transfer}), but the two observables need not agree. The within-shower width therefore reproduces the LEMURS drift already seen in the cog std, so the drift is a property of the count profile rather than of a single summary statistic.

Whether the drift is aimed at the source, rather than a generic broadening, is harder to establish. The cog std does move towards the LEMURS pool value, but the control is weak. Only one far-domain prior is available, and the LEMURS and \textsc{SimpleBox} priors point in nearly parallel directions across the full set of count-profile features. A drift towards ``the source'' therefore cannot be fully distinguished from a drift towards any far-domain prior. We report the source-directedness as consistent with the data, not as established.

Four artefact explanations do not survive. Plotting the drift against cumulative optimiser steps rather than against $D$ does not collapse the arms onto one curve, so it scales with the fine-tuning size and not with training length. Re-selecting the delivered checkpoint on a width-aware criterion, rather than on validation loss, leaves it intact. It is already present in the raw \textsc{PointCountFM} prediction, before the per-layer bias correction, so it is not a post-processing artefact. Nor is it training instability: the validation loss falls smoothly throughout.

\begin{figure}[htbp]
    \centering
    \includegraphics[width=0.82\linewidth]{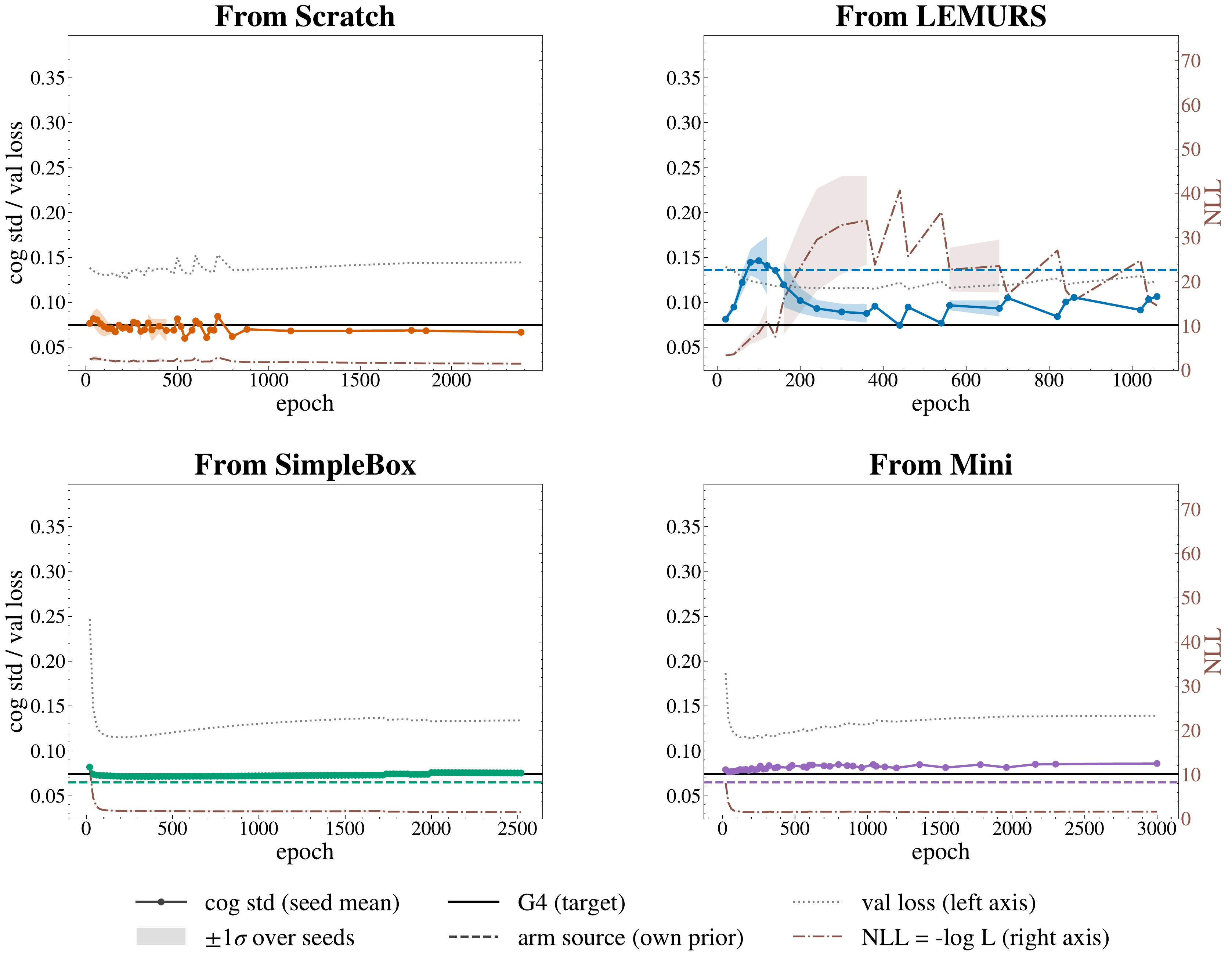}\\[1ex]
    \includegraphics[width=0.82\linewidth]{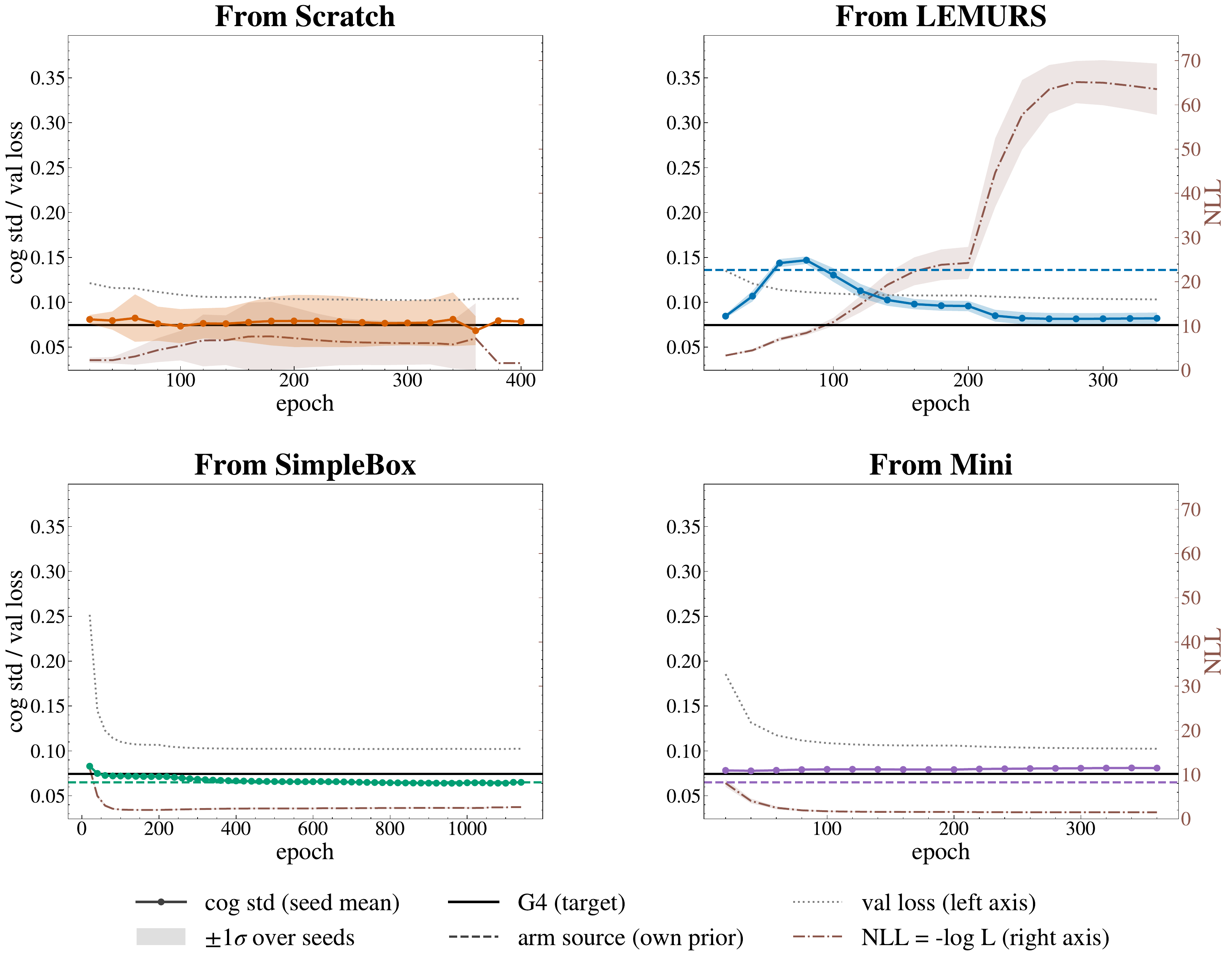}
    \caption{As \cref{fig:cogstd-epoch-1k}, at (top) $D=10^{4}$ and (bottom) $D=10^{5}$, the budget of the delivered checkpoints of \cref{ssec:results-allegro-transfer}. The LEMURS cog std rises above \texttt{Geant4} early in training while the validation loss is still falling; the negative log-likelihood degrades later and, at $D=10^{5}$, climbs to a plateau whose height grows with $D$. The two do not develop in step. The other three arms stay near \texttt{Geant4} throughout, with a flat, low negative log-likelihood.}
    \label{fig:cogstd-epoch-10k-100k}
\end{figure}

\looseness=-1
\Cref{fig:cogstd-epoch-1k,fig:cogstd-epoch-10k-100k} follow the LEMURS arm through training with a second diagnostic of a different construction alongside the cog std: the negative log-likelihood (NLL) of the \texttt{Geant4} reference counts under a Gaussian fitted to the model's own generated per-layer counts at that epoch. This checks whether the validation loss, the model's own training signal, stays a faithful proxy for how well the generated distribution still covers the true data as training continues. Because it uses the full covariance rather than one width scalar, it is also sensitive to the shape of the count distribution, not only its depth spread. The first moments stay calibrated (\cref{ssec:results-allegro-transfer}), so a collapse here is consistent with a second-moment effect. While the validation loss keeps falling, both diagnostics move against it, most strongly at $D=10^{5}$. They do not develop in step. The cog std rises early and partly settles back; the negative log-likelihood, on the shared right axis of the three epoch figures, degrades later and stays high. The degradation that the validation loss does not see therefore persists in the full count distribution after the single cog-std scalar has partly recovered. The from-scratch, \textsc{SimpleBox} and \textsc{SimpleBox}-mini arms show neither.

Freezing part of the pre-trained core, the mitigation used in \cref{ssec:results-allegro-transfer}, is partial. It suppresses the amplitude of the drift, but at a cost on the radial profile. It also brings no benefit at the smallest $D$, where the frozen features have not yet adapted. This trade-off bounds the effect rather than removing it.

\looseness=-1
A far-domain prior that grows more harmful as target data increases is not new. It is consistent with the ossification reported for pre-trained language models~\cite{hernandez2021scalinglawstransfer} and the feature distortion that linear-probe-then-fine-tune protocols are designed to counter~\cite{kumar2022finetuningdistortpretrainedfeatures}. It may also be the same effect behind the multi-detector adaptation behaviour reported for CaloDiT-2~\cite{Raikwar:2025fky}. What \cref{ssec:results-allegro-transfer} and this appendix add is that the effect is confined to the second moments of the count profile, so only a distributional metric exposes it.

\FloatBarrier

\clearpage
\bibliographystyle{JHEP}
\bibliography{biblio}

\end{document}